\documentclass[prb,aps,preprint,showkeys,showpacs]{revtex4}
\usepackage[utf8]{inputenc}
\usepackage{amsmath}
\usepackage{amsfonts}
\usepackage[dvips]{graphicx}
\usepackage{booktabs}
\usepackage[colorlinks=true, citecolor=red, linkcolor=blue, urlcolor=blue]{hyperref}
\begin{document}
\title{Exploration of free energy surface and thermal effects on relative population and infrared spectrum of the Be$_6$B$_{11}^{-}$ fluxional cluster.}

\author{Carlos Emiliano Buelna-Garc\'ia$^{1}$}
\author{C\'esar Castillo-Quevedo$^2$}
\author{Jesus Manuel Quiroz-Castillo$^{1}$}
\author{Gerardo Mart\'inez-Guajardo$^{3}$}
\author{Aned de-Leon-Flores$^{4}$}
\author{Gilberto Anzueto-S\'anchez$^{5}$}
\author{Martha Fabiola Martin-del-Campo-Solis$^{2}$}
\author{Jose Luis Cabellos$^{6\star}$}\email[email:]{sollebac@gmail.com}
\affiliation{$^1$Departamento de Investigaci\'on en Pol\'imeros y Materiales, Edificio 3G. Universidad de Sonora. Hermosillo, Sonora, M\'exico}
\affiliation{$^{2,2}$Departamento de Fundamentos del Conocimiento,
  Centro Universitario del Norte, Universidad de Guadalajara,
Carretera Federal No. 23, Km. 191, C.P. 46200, Colotl\'an, Jalisco, M\'exico}
\affiliation{$^3$Unidad Acad\'emica de Ciencias Qu\'imicas, \'Area de Ciencias de la Salud, Universidad Aut\'onoma de Zacatecas, Km. 6 carretera Zacatecas-Guadalajara s/n, Ejido La Escondida C. P. 98160, Zacatecas, Zac.}
\affiliation{$^4$Departamento de Ciencias Qu\'imico Biologicas, Edificio 5A.
  Universidad de Sonora. Hermosillo, Sonora, M\'exico}
\affiliation{$^5$ Centro de Investigaciones en \'Optica, A.C., 37150 Le\'on, Guanajuato, M\'exico}
\affiliation{$^6$Departamento de Investigaci\'on en F\'isica, Universidad de Sonora, Blvd. Luis Encinas y Rosales S/N, 83000 Hermosillo, Sonora, M\'exico}

\date{\today}
\affiliation{\textcolor{black}{$^*$corresponding author: sollebac@gmail.com, jose.cabellos@unison.mx}}
\begin{abstract}
The starting point to understanding cluster properties is the putative global minimum and all the nearby local energy minima; however, locating them is computationally expensive and challenging due to the degrees of liberty associated with the molecule rise as a function of the number of atoms. Therefore, the number of possible combinations increases exponentially, leading to a combinatorial explosion problem. 
The relative populations and spectroscopic properties of a molecule that are a function of temperature can be approximately computed by employing statistical thermodynamics.Here, we investigate entropy-driven isomers distribution on Be$_6$B$_{11}^-$ clusters and the effect of temperature on their infrared spectroscopy and relative populations. We identify the vibration modes possessed by the cluster that significantly contribute to the zero-point energy. A couple of steps are considered for computing the temperature-dependent relative population: First, using a genetic algorithm coupled to density functional theory, we performed an extensive and systematic exploration of the potential/free energy surface of  Be$_6$B$_{11}^-$ clusters to locate the putative global minimum and elucidate the low-energy structures. Second, the relative populations’ temperature effects are determined by considering the thermodynamic properties and Boltzmann factors. The temperature-dependent relative populations show that the entropies and temperature are essential for determining the global minimum. We compute the temperature-dependent total infrared spectra employing the Boltzmann factor weighted sums of each isomer’s infrared spectrum and find that at finite temperature, the total infrared spectrum is composed of an admixture of infrared spectra that corresponds to the spectrum of the lowest energy structure and its isomers located at higher energies. The methodology and results describe the thermal effects in the relative population and  the infrared spectra.
\end{abstract}
\pacs{61.46.-w,65.40.gd,65.,65.80.-g,67.25.bd,71.15.-m,71.15.Mb,74.20.Pq,74.25.Bt,74.25.Gz,74.25.Kc}
\keywords{global minimum; infrared spectrum; boron cluster; fluxional; density functional theory; temperature; Boltzmann factors; Gibbs free energy; entropy; lowest energy structure; Grimme's approach (DFT-D3); IR spectra, lowest energy estructure; genetic algorithm} 
\maketitle
\section{Introduction}
In recent years, the pure boron clusters, the metal, and non-metal doped boron clusters, have attracted considerable attention~\cite{Jalife,Wang,Sun,Jian,Teng-Teng,Gerardo,Dongliang,Peifang,Grande-Aztatzi,Dong} due to their unpredictable chemistry~\cite{Brothers,Axtell} and high potential to form novel structures.~\cite{Lai-Sheng5} Boron is the smallest and lightest semi-metal atom~\cite{Dongliang,Mannix} and a neighbor of carbon in the periodic table. It has electron deficiency~\cite{Vast,Peculiar,Hongxiao,Lai-Sheng5}  and a high capacity to combine and form novel atomic and molecular boron structures that are planar and quasi-planar.~\cite{Lai-Sheng,Lai-Sheng2,Lai-Sheng3} It can also form nanotubes,~\cite{Lai-Sheng4,Lai-Sheng5} borospherenes,~\cite{Qiang,Wang,Ying-Jin-Wang} borophene,~\cite{Wang} cages,~\cite{Lv} chiral helices,~\cite{Feng,Guo}, and nanosheets~\cite{Lai-Sheng,Mannix1513,Jimenez-Halla} consisting of triangle units of boron atoms. Boron can absorb neutrons that make it useful in nuclear and medical applications.~\cite{Barney,Lesnikowski,Ali,Lu}  Aromaticity, antiaromaticity, and conflicting aromaticity dominate the chemical bonding in boron-based clusters.~\cite{Ofelia,Alexandrova2,Feng,Boldyrev} The two most-used indices for quantifying aromaticity are the harmonic oscillator model of aromaticity, based on the geometric structure, and the nucleus-independent chemical shift, based on the magnetic response.
{{Aromaticity is not observable, cannot be directly measured~\cite{Poater}, and it correlates with electronic delocalization.~\cite{Mandado}}}
However, with applications in molecular devices, the dynamic structural fluxionality in boron and boron-doped based molecular systems is due to electronic delocalization.~\cite{Feng,Sudip2} Moreover, electronic localization/delocalization contributes significantly to stability, magnetic properties, and chemical reactivity.~\cite{Poater} Nowadays, dynamic structural fluxionality in boron nanoclusters is a topic of interest in nanotechnology.~\cite{Ying-Jin-Wang,Zhai} {{The fluxionality of an atomic cluster is highly relevant in terms of its catalytic activity ~\cite{Zhai2}, and in boron-based nanoscale rotors, it is a function of the atomic structure, size, bonding, and cluster charge.~\cite{Merino}}}
Moreover, doping a boron cluster with metals~\cite{Romanescu,Sun,Liang,Teng-Teng,Teng-Teng3,Dong2019,Popov2015} and non-metals~\cite{Duong} dramatically affects its structure, stability, and reactivity, as with the shut-down in the fluxionality of the boron-doped anion $B^{-}_{19}$~\cite{Cervantes-Navarro}.{{ It is important to mention, the emission of radiation as a competing cooling channel has to be considered in studying small cationic boron clusters' stabilities. Accordingly to Ferrari et al., this improving agreement between experiment and theory.}~\cite{Ferrari}} 

In this study, we consider that temperature and entropy are critical in elucidating the low-energy structures and highlight the importance of understanding the thermal and entropic effects in the Be$_6$B$_{11}^{-}$ fluxional cluster.
In the past years, a boron molecular Wankel motor\cite{Jimenez-Halla,Martinez-Guajardo,Moreno,Jalife,Truong,Fagiani,Yonggang} and subnanoscale tank treads have been reported;~\cite{Ying-Jin2015,Zhai2017} however, the entropic and temperature terms have not been considered.
In collaboration with Merino's group and Zhai's group, some of us studied and reported fluxionality in Be$_6$B$_{11}^{-}$.~\cite{Guo} The computations indicated that there were two competitive low-energy structures: a helix-type cluster and a fluxional coaxial multiple-layered cluster. More recently, another lowest-energy structure was found in the Be$_6$B$_{11}^{-}$ cluster by employing a cellular automaton algorithm.~\cite{Osvaldo} However, the putative global minimum energy structure and its molecular properties depend strongly on the {{temperature-entropy term}}.~\cite{Baletto,Grigoryan,Calvo1,Cahuzac,Foster}

In several previous works, one of the authors computed the barrier energy in a chemical reaction by taking into account the effect of temperature-entropy term~\cite{Dzib}, computed the temperature-dependent dipole moments for the HCl(H$_2$O)$_n$ clusters,~\cite{Vargas-Caamal}, computed the temperature-dependent linear optical properties of the Si(100) surface,~\cite{shkrebtii}  and more recently, it was considered in a study of {{gold clusters~\cite{Goldsmith,Ghiringhelli_2013,Schebarchov}}} and the thermochemical behavior of the sorghum molecule.~\cite{MENDOZAWILSON2020112912} Nevertheless, most theoretical density functional studies assume that the temperature is zero and neglect temperature-dependent and entropic contributions; consequently, their finite temperature properties remain unexplored,~\cite{Seitsonen,Prachi} whereas experimental studies are carried out in non-zero temperatures. Thus, it is necessary to understand the effect of the temperature on the cluster properties and the lowest energy structure's determination.~\cite{Jonathan,Seitsonen,Prachi} Herein, we investigate the effect of temperature-entropy term on the relative population and its infrared spectra, which need the putative elucidation global minimum and its low-energy isomers.~\cite{Zhen,Darby,Doye,Ohno,Baletto} The first starting point requires a minimum global search on the potential/free energy surfaces, which is a complicated task.

Taking temperature into account requires dealing with small systems' thermodynamics; The Gibbs free energy of classical thermodynamics also applies for small systems, known as thermodynamics of small systems.~\cite{Simon,Hill2,gibbs1961} or nanothermodynamics.~\cite{Li-Truhlar} The thermodynamics of clusters have been studied by various theoretical and simulation tools~\cite{Baletto,Hill,Simon,Calvo,Bixon,Kristensen,Wales925,Seitsonen,jena1992physics,Fox} like molecular-dynamics simulations,~\cite{Gerardo} Monte Carlo, and analytic methods. Under the harmonic superposition approximation, the temperature-entropy term can be computed with the vibrational frequencies on hand. The entropy effects have been considered for gold, copper, water, and sodium clusters.~\cite{Ghiringhelli_2013,Schebarchov,Goldsmith,C1FD00027F,Grigoryan,Zhen-Long,Malloum,Malloum2,Malloum3,Fifen}

As the second step for understanding cluster properties relies on the clusters’ spectroscopy, spectroscopy gives insight into the structure and detects structural transformations in clusters.~\cite{Gruene674,doi:10.1021/ja0509230,PhysRevLett.93.023401} The temperature effects on IR spectra have been studied experimentally and theoretically on small and neutral gold clusters~\cite{Ghiringhelli_2013,Goldsmith} and boron cluster.~\cite{Fagiani} In the same direction, the pristine Au$_{13}$ gold cluster's thermodynamical stability at finite temperature was studied using the replica-exchange method, which shows a fluxional behavior.~\cite{C1FD00027F} Au$_N$ clusters' thermodynamics properties (30 $<$ N $<$ 147) were studied  employing  the Gupta potential and DFT methodology~\cite{Schebarchov} The total absorption spectra were computed as the sum of the different spectra of different isomers.~\cite{Felix}

In this study will employ statistical thermodynamics to compute the Gibbs free energy entropic-temperature-dependency, evaluate relative populations as function of  temperature, and take into account the effects of temperature on the IR spectra. We also identify the vibration modes that make a significant contribution to the zero-point energy of the cluster that is strongly dominated at temperatures higher than 377 K, and also we show this structure poses the shortest B-B bond length. We investigate the effect of long-range van der Waals interactions on solid-solid transformation points; moreover, we found the vibrational modes responsible for the fluxionality of the cluster. Adicionally, we computed the relative population at single point CCSD(T) level of theory.  We believe that this yields useful information about which isomers will dominate at hot temperatures. No work has been attempted to investigate temperature-entropy driven isomers in the fluxional Be$_6$B$_{11}^{-}$ cluster as far as we know. The rest of the manuscript is distributed as follows: Section 2 briefly gives the theory and computational details. Section 3 discusses the lowest energy structures, energetic ordering at DFT/CCSD(T) level of theory, the relative population, and IR spectra taking into account the temperature-entropy term. Conclusions are given in Section 4.

\section{Theoretical Methods and Computational Details} 
\subsection{Global Minimum Search}
Despite advances in computing power, the minimum global search in molecular and atomic clusters remains a complicated task due to several factors. The exploration should be systematic and unbiased;~\cite{Min,Baletto} a molecule's degrees of freedom increase with the number of atoms;~\cite{Baletto,Wille_1985,Xu,Rossi_2009,Cheng} a molecule composed of N number of atoms possesses 3N degrees of freedom (i.e., a linear molecule has [3N-5]  degrees of vibrational modes, whereas a nonlinear molecule has [3N-6] degrees of vibrational modes); and, as a consequence, the potential/free energy surface depends on a large number of variables. The number of local minima increases exponentially as a function of the number of atoms in the molecule. Moreover, the total energy computation requires a quantum mechanical methodology to produce a realistic value for energy. In addition to that, there should be many initial structures. It is essential to sample a large region of the configuration space to ensure that we are not missing structures, making an incomplete sampling of the configurational space and introducing a significant problem to calculating the thermodynamic properties.~\cite{Zhen} A Complete sampling of the potential/free energy surface is nearly impossible, but a systematic exploration of the potential energy surface is extremely useful. Although searching for a global minimum in molecular systems is challenging, the design and use of algorithms dedicated to the search for global minima, such as simulated annealing,~\cite{kirkpatrick, metropolis, xiang, yang, vlachos, granville} kick method,~\cite{Saunders,Saunders2} genetic algorithms,~\cite{hsu, Wei, goldberg} Gradient Embedded Genetic Algorithm (GEGA),~\cite{alexa, alexandrova, alexan} and basin hopping,~\cite{harding, wales} has been accomplished over the years. In the past few years, one of us designed and employed genetic algorithms~\cite{Guo,Dong,Mondal,Ravell,Grande-Aztatzi} and kick methodology~\cite{Sudip,Cui,Vargas-Caamal2,Vargas-Caamal,Cui2,Vargas-Caamal2015,Florez,Ravell} coupled with density functional theory to explore atomic and molecular clusters' potential energy surfaces. They have led us to solve the minimum global search in a targeted way. In this paper, our computational procedure employs a recently developed and unbiased hybrid strategy for a search methodology that combines a modified-kick heuristic and genetic algorithm with density functional theory that has been implemented in the \emph{GALGOSON} code. \emph{GALGOSON} systematically and efficiently explores potential/free energy surfaces (PES/FES) of the atomic clusters to find the minimum energy structure. The methodology consists of a three-step search strategy where, in the first and second steps, we explore the PES, and in the third step, we explore the FES. First, the code builds a generation of random initial structures with an initial population of two hundred individuals per atom in the Be$_{6}$B$_{11}^{-}$ cluster using a kick methodology. The process to make 1D, 2D, and 3D structures is similar to that used in previous work.~\cite{Grande-Aztatzi,Osvaldo} and are restricted by two conditions~\cite{Grande-Aztatzi} that can be summarized as follows: a) All the atoms are confined inside a sphere with a radius determined by adding all atoms' covalent radii and multiplied by a factor established by the user, typically 0.9. b) The bond length between any two atoms are the sum of their covalent radii, modulated by a scale factor established by the user, typically close to 1.0; this allows us to compress/expand the bond length. These conditions avoid the high-energy local minima generated by poorly connected structures (too compact/ loose). Then, structures are optimized at the PBE0/3-21G level of theory employing Gaussian 09 code. As the second step, all energy structures lying in the energy range of 20 kcal/mol were re-optimized at the PBE0-GD3/LANL2DZ level of theory and joints with previously reported global minimum structures. Those structures comprised the initial population for the genetic algorithm. The optimization in this stage was at the PBE0-GD3/LANL2DZ level of theory. The criterion to stop the generation is if the lowest energy structure persists for 10 generations. In the third step, structures lying in 10 kcal/mol found in the previous step comprised the initial population for the genetic algorithm that uses Gibbs free energy extracted from the local optimizations at the PBE0-D3/def2-TZVP, taking into account the zero-point energy (ZPE) corrections. The criterion to stop is similar to that used in the previous stage. In the final step, the lowest energy structures are evaluated at a single point energy at the CCSD(T)/def2-TZVP//PBE0-D3/def2-TZVP level of theory. All the calculations were done employing the Gaussian 09 code.~\cite{gauss}
\subsection{Thermochemistry Properties}
All the information about a quantum system is contained in the wave function; similarly, the partition function provides all the information need to compute the thermodynamic properties, and it indicates the states accessible to the system at temperature T, so the thermodynamic functions are calculated using the temperature-dependent partition function Q shown in Equation~\ref{partition}.
\begin{equation}
\displaystyle
Q(T)=\sum_{i}g_i~e^{{-\Delta{E_i}}/{K_BT}}
\label{partition}
\end{equation}
In Eq.~\ref{partition}, the $g_i$ is the degeneracy or multiplicity, using degeneracy numbers is equivalent to take into account all degenerate states and the sum runs overall energy levels,  and $k_{\textup{B}}$ is the Boltzmann constant, $T$ is the temperature in Kelvins, and ${-\Delta{E_i}}$ is the total energy of a molecule.~\cite{Dzib,mcquarrie1975statistical}  An exact calculation of Q could be complicated due to the coupling of the internal modes, a way to decouple the electronic and nuclei modes is through the use of Born-Oppenheimer approximation. (BOA) This approach says that the electron movement is faster than the nuclei and assumes that the molecular wave function is the electronic and nuclear wavefunction product. $\psi=\psi_e\psi_n$
The vibrations change the momentum of inertia as a consequence, affect the rotations; this fact tightly couple the vibrational and rotational degrees of freedom; The separation of rotational and vibrational modes is called the rigid rotor, harmonic oscillator (RRHO) approximation, under this approximation, the molecule is treated rigidly,   this is generally good when vibrations are of small amplitude.
Here the vibration will be modeled in terms of harmonic oscillator and rotations in terms of the rigid rotor. Within BOA and RRHO approximations,  the partition function is factorized into electronic, translational, vibrational, and rotational energies. Consequently, the partition function, $q$, (Equation~\ref{qtqr}) can be given as a product of the corresponding contributions~\cite{hill1986introduction,Dzib}
\begin{equation}  
\displaystyle
q=q_{trans}~q_{rot}~q_{vib}~q_{elec}
\label{qtqr}
\end{equation}
Equations~\ref{q:all} are the contributions of electronic~(\ref{q:ele}), translational~(\ref{q:trans}),
vibrational~(\ref{q:vibra}),and rotational~(\ref{q:lineal},\ref{q:nonlineal})
to the canonical partition function.
\begin{subequations}
  \label{q:all}
Contributions to the canonical partition function
\begin{align}
      q_{trans}  &=\big((\frac{2\pi m k_{\textup{B}}T}{h^2}\big)^{\frac{3}{2}} \frac{k_{\textup{B}}T}{P},  \label{q:trans} \\
      q^{l}_{rot} &=\frac{1}{\sigma_r}\big(\frac{T}{\Theta_r}\big), \label{q:lineal}\\
      q^{nl}_{rot} &=\frac{\pi^{1/2}}{\sigma_r}\Big(\frac{T^{3/2}}{ (\Theta_{\textup{A}}{\Theta_{\textup{B}}\Theta_{\textup{C}}})^{1/2}}\Big),\Theta_{\text{i}}=\frac{\hbar}{2\textup{I}_{i}k_{\textup{B}}}, i=\textup{A},\textup{B},\textup{C}, \label{q:nonlineal}\\
      q_{vib} &=\prod_{i=1}^{n_{\nu}} \frac{e^{-\Theta_{{\textup{vib}_i}}/2T}}{1-{e^{-\Theta_{{\textup{vib}_i}}/T}}}, \Theta_{{\textup{vib}_i}}=\frac{h\nu_i}{k_{\textup{B}}},\label{q:vibra}\\ 
       q_{e} &=\omega_0,\label{q:ele}
\end{align}
\end{subequations}
We computed all partition functions at temperature T and a standard pressure of 1 bar (0.986923) atm.  The equations are equivalent to those given in the Ref.~\cite{Dzib}, and any standard text of thermodynamics~\cite{mcquarrie1975statistical,hill1986introduction} and applies for an ideal gas.  The implemented translational partition function in the Gaussian code~\cite{gauss} is the partition function, q$_{trans}$, given in Equation~\ref{q:trans}. In this study, the Equation~\ref{q:trans} (q$_{trans}$) is computed as a  function of T and is used to calculate the translational entropy.
In addition to using vibrational modes to identify true lowest energy structures from transition states, we also used them to compute the vibrational partition function. In this paper is considered vibrational modes, $\nu$, under the harmonic oscillator approximation, and total vibrational energy consists sum of the energies of each vibrational mode. In computing the electronic partition, we considerer that the energy gap between the first and higher excited states is more considerable than $k_{\textup{B}}$T, as consequence electronic partition function, q$_{elec}$, is given by q$_{elec}=\omega_0$ The Equations~\ref{q:all}, q$^{l}_{rot}$, q$_{elec}$, q$^{nl}_{rot}$, q$_{trans}$, are used to compute the internal energy (U), and entropy (S) contributions given in Equations~\ref{internal} 
\begin{equation}
\label{internal}
\displaystyle
\def\arraystretch{1.3}
\begin{array}{@{}lll@{}}
  \toprule 
  & \multicolumn{2}{c@{}}{}\\
  \cmidrule(l){1-3}
 \text{{\bf{Contribution}}}   & \text{{\bf{Internal energy}}} & \text{{\bf{Entropy}}} \\
\text{Translational}    & U_{trans}=\frac{3}{2}RT  
     &  S_{trans}=R\big(In~q_{trans}+\frac{5}{2}\big)\\
\text{Rotional Linear}    & U^{l}_{rot}=RT   
     &  S^{l}_{rot}=R\big(In~q^{l}_{rot}+1\big) \\
 \text{Rotational Nonlinear}   &   U^{nl}_{rot}=\frac{3}{2}RT   
   &  S^{nl}_{rot}=R\big(In~q^{nl}_{rot}+\frac{3}{2}\big) \\
\text{Vibrational}    & U_{vib}= R\sum_{i}^{\nu_i}\Theta_{{\textup{vib}_i}}\Big(\frac{1}{2}+\frac{1}{e^{\Theta_{{\textup{vib}_i}}/T}-1}\Big)    
     & S_{vib}=R\sum_i^{\nu_i}\Big[ \frac{e^{\Theta{{\textup{vib}_i}}/T}} {{e^{\Theta_{{\textup{vib}_i}}/T}}-1}-In\Big(1-{e^{\Theta_{{-\textup{vib}_i}}/T}}\Big)\Big]\\
\text{Electronic}     & U_{elec}=0    
     & S_{elec}=RIn~q_{elec}\\
\bottomrule
\end{array}
\end{equation}
The vibrational frequencies are calculated employing the Gaussian code,~\cite{gauss} and all the information needed to compute the total partition function is collected from the output.
The Gibbs free energy ($\Delta G$) and the enthalpy ($\Delta H$) are computed employing Equations~\ref{ag11} and ~\ref{ag22}), respectively. In
these equations, \emph{R} is the ideal gas constant and \emph{n} is the amount of substance, and \emph{T} is the
temperature in Kelvin.
\begin{equation}
  \centering
  \displaystyle
   \Delta H=U+nRT\label{ag11}
\end{equation}
\begin{equation}
  \centering 
  \displaystyle
  \Delta G=\Delta H-\Delta ST\label{ag22}
\end{equation}
\subsection{Boltzmann Population}
The properties observed in a molecule are statistical averages over the ensemble of geometrical conformations or isomers accessible to the cluster.~\cite{Teague} so the molecular properties are ruled by the Boltzmann distributions of isomers
that can change due to temperature-entropic term,~\cite{Vargas-Caamal,Bulusu,Grigoryan} and the soft vibrational modes that clusters possess make primary importance contributions to the entropy.~\cite{Malloum2} The relative populations of the low-energy isomers of the cluster Be$_6$B$_{11}^{-}$ are computed through the probabilities defined in Eq.~\ref{boltzman}
\begin{equation}
\centering 
\displaystyle
P(T)=\frac{e^{-\beta \Delta G^{k}}}{\sum e^{-\beta \Delta G^{k}}}\label{boltzman}, 
\end{equation}
where $\beta=1/k_{\textup{B}}T$, and $k_{\textup{B}}$ is the Boltzmann constant, $T$ is the temperature in Kelvin, $\Delta G^{k}$ is the Gibbs free energy of the $k^{th}$ isomer. Eq.~\ref{boltzman} establishes that the distribution of molecules will be among energy levels as a function of the energy and temperature.
It is worth mentioning that the energy of separation among isomers (energy gap between two isomers) is determinant in the computation of the solid-solid transition, T$_{ss}$ point.  T$_{ss}$ occurs when two competing structures are energetically equaled, and there is simultaneous coexistence of structural isomers at T. in other words, the T$_{ss}$ point is a function of the energy difference between two isomers and the
energy $\Delta G$  that the cluster posses. Boltzmann distribution finds a lot of applications as to native protein structures~\cite{Shortle} for a microscopy system, a temperature T or like method simulated annealing applied to the search of structures of minimum energy, rate of chemical reaction~\cite{Dzib}, sedimentation, among others.  For the calculation of the Gibbs free energies at temperature T and the relative populations, we used a homemade \emph{Python/Fortran} code called Boltzmann-Optics-Full-Ader (BOFA). 
\subsection{IR Spectra}
The vibrational spectra are useful for identifying phases and determining structures,~\cite{Kubicki_2019} among other applications mentioned above. In this study, the IR harmonic spectra for each isomer were calculated employing Gaussian code.~\cite{gauss} All isomers were characterized as minima because we found no negative frequencies in each isomer. The Lorentzian line shape, with a width at half maximum of 20 cm$^{-1}$, and a scaling factor of 0.98 were used to compute the IR spectra for each isomer. The most considerable contribution to total IR spectra is the putative global minimum atomic structure~\cite{Felix}, while the isomers located at high energies contribute little to the molecular properties. Therefore, the total IR spectrum is dependent on the temperature results from the contributions of all IR spectra weighted according to their relative populations. In this study, to obtain the total IR spectrum at temperature T, we weighted the IR spectrum of each isomer according to the probabilities computed in Equation~\ref{boltzman} and the sum of all of them; thus, we computed the total IR spectrum as a function of the temperature.
\begin{figure}[ht!]
\begin{center}  
  \includegraphics[scale=0.70]{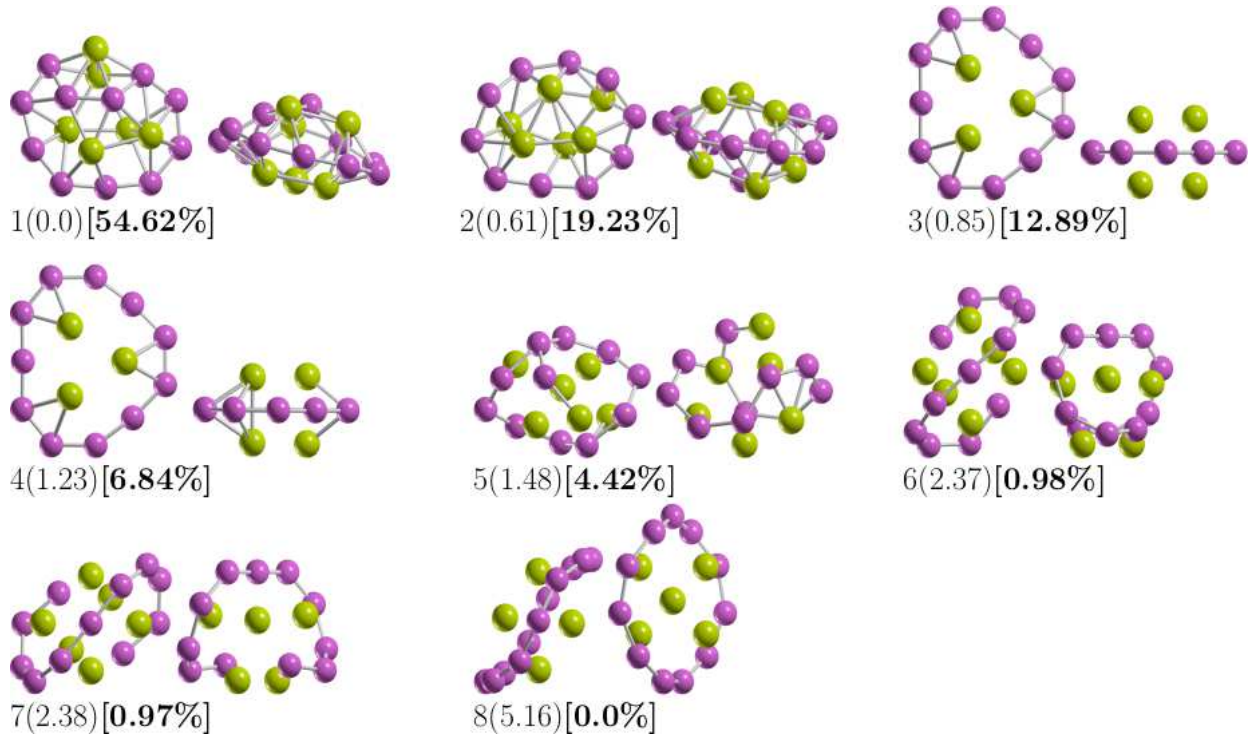}
\caption{(Color online). The optimized geometries of Be$_{6}$B$_{11}^{-}$ cluster. The most important energy isomers show in two orientations, front, and rotated 90 degrees up to plane paper.  Relative Gibbs free energies in kcal/mol (in round parenthesis) and the relative population [in square parenthesis], at PBE0-D3/def2-TZVP level of theory. The criterium to plot them is until the probability occupation is zero. The pink- and yellow-colored spheres represent the boron and beryllium atoms, respectively.}
\label{geometry_gibbs}
\end{center}    
\end{figure}
\subsection{Computational Details}
The global exploration of the potential and free energy surfaces of the Be$_6$B${^{-}_{11}}$ was done with a hybrid genetic algorithm written in Python. All local geometry optimization and vibrational frequencies were carried out employing the density functional theory (DFT) as implemented in the Gaussian 09~\cite{gauss} (Revision D.01) suite of programs, and no restrictions in the optimizations were imposed. Final equilibrium geometries and relative energies are reported at PBE0~\cite{Adamo}/def2-TZVP~\cite{Weigend} level of theory, taking into account the D3 version of Grimme’s dispersion corrections.~\cite{Grimme} and including the zero-point (ZPE) energy corrections. (PBE0-D3/def2-TZVP) As Pan et al.~\cite{Li-Li} reported, the computed relative energies with PBE0 functional are very close to the CCSD(T) values in  B$_9^{-}$ boron cluster. The def2-TZVP basis set from the \emph{Ahlrichs} can improve computations accuracy and describe the Be$_6$B${^{-}_{11}}$ cluster~\cite{Guo} To gain insight into its energetics, we evaluated the single point energy at CCSD(T)/def2TZVP//PBEO-D3/def2-TZVP level of theory for the putative global minima and the low-energy isomers. The total IR spectra dependent on temperature are computed employing Boltzmann weighted sum of the IR spectra of each isomer, and the relative populations using Boltzmann factors, both of them implemented in a made in home Python/Fortran code called BOFA. The BOFA code is employed in the computation of the relative population and weighted IR spectra. The code is available with the corresponding author.
\section{Results and Discussion}
\subsection{The lowest energy structures and energetics}
Figure~\ref{geometry_gibbs} shows the lowest energy structure of Be$_6$B${^{-}_{11}}$ clusters and seven low-energy competing isomers computed at the PBE0-D3/def2-TZVP level of theory. The criterion for drawing the structures is until the percentage of the relative population is zero. The relative Gibbs free energy is given in kcal/mol (round parenthesis) and computed at 298.15 K and 1 bar. In square parenthesis, and in  bold is given the percentage of the relative population computed employing Equation~\ref{boltzman} at 298.15 K. For the putative global minimum at the PBE0-D3/def2-TZVP level of theory, the optimized average B-B bond length is 1.64~\AA.In contrast, the optimized B-Be bond length is 2.01~\AA.
\begin{figure}[ht!]
\begin{center}  
\includegraphics[scale=1.0]{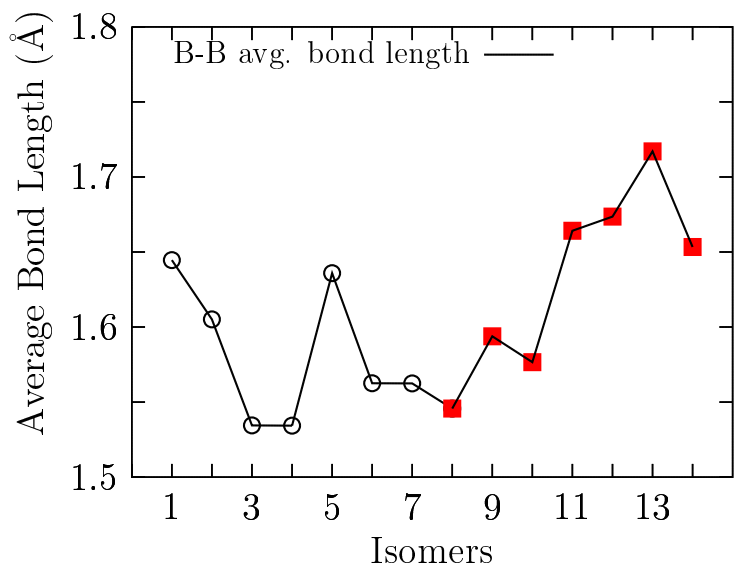}
\caption{(Color online). The average B-B bond length a function of the number of isomers. The isomers are energetically accommodated, from the most energetically favorable (1) to the less stable.(14). The coaxial Triple-layered structures with C$_s$ and C$_{2v}$ symmetries, isomers numbers 3 and 4, have the lowest average bond length of 1.53~\AA.~The black open-circles are the low-energy structures with a relative population different from zero at 298.15 K. The filled red squares are those isomers that the relative population is zero at 298.15 K.}
\label{lboro}
\end{center}
\end{figure}
To observe the trend in B-B bond length in the low-energy structures, Figure~\ref{lboro} shows the average bond length for B-B for the fourteen lowest energy isomers energetically accommodated, from the most energetically favorable, isomer number 1, to the least stable, isomer number 14. Our calculations indicated that the largest average value of the B-B bond length is 1.71~\AA~and belongs to isomer number 13, which is 25 kcal/mol less stable than the putative global minimum. The lowest average value of the B-B bond length is 1.53~\AA~ and corresponds to the isomers coaxial triple-layered structures with C$_s$ and C$_{2v}$ symmetries, located at energies of 0.85 and 1.23 kcal/mol above the putative global minima, respectively. The structures are depicted in Figure~\ref{geometry_gibbs}(3)(4). In these structures, the lowest average B-B bond length of 1.53~\AA~is considerably shorter compared with the: (a) length of a typical B-B single bond of 1.72~\AA~\cite{Shoji}, (b) the bond length of the B8 and B$_9^{-}$ molecular wheels,~\cite{Birch,Guo} and slightly shorter in 2.2{\%} than the B-B double bond length experimentally characterized in the range of 1.57–1.59~\AA.~\cite{Moezzi,Zhou} The average B-B bond length shortens from 1.64~\AA~to 1.53~\AA~, suggesting strong hyperconjugation in the coaxial triple-layered structures. The shortening of the B-B bond length is caused by orbital interaction, which is also a cause of C-C bond shortening in the Butyne molecule.~\cite{Feixas} Hyperconjugation has been shown in the shortening of B-B and C-C bond lengths~\cite{Szabo,Feixas}
 and which causes increases in the number of electrons shared between region. Figure~\ref{cubic} shows the average bond length for Be-B for the 14 low-energy isomers. The largest average value of the Be-B bond length is 2.0~\AA~ and 2.10~\AA,~ which correspond to the isomer coaxial triple-layered structures with C$_s$ and C$_{2v}$ symmetries, respectively. This suggests that if the shortening of the bond length increases the number of electrons shared in that region,~\cite{Feixas} the increase in bond length should decrease the number of electrons; consequently, the electron delocalization occurs in the ring of boron atoms. In Figure~\ref{geometry_gibbs}(1), is depicted the putative global minimum with 54{\%} of the relative population, and it has C$_1$ symmetry with a singlet electronic state $^{1}$A. It is a distorted, oblate spheroid with three berylliums atoms in one face and two in the other face. Nine boron and one beryllium atoms form a ring located around the spheroid's principal axes, and the remaining two boron atoms are located close to the boron ring on one of its faces. The second higher energy structure, at 298.15 K, lies at a Gibbs free energy of only 0.61 kcal/mol above the putative global minimum; it has C$_1$ symmetry with a singlet electronic state $^1$A. It is a prolate spheroid with 19{\%} of the relative population at a temperature of 298.15 K. The next two higher energy isomers, at 298.15 K, are located at 0.85 and 1.23 kcal/mol Gibbs energy above the putative global minimum. They are prolate, coaxial, triple-layered structures with C$_s$, and C$_{2v}$ symmetries and singlet electronic states $^1$A and $^1$A$_1$, respectively. This clearly shows that the low-symmetry structure C$_s$ becomes energetically preferred compared to the C$_{2v}$ symmetry, with a Gibbs free energy difference of 0.38 kcal/mol at 298.15 K due to the entropic effects. This is in agreement with a similar result found in Au$_{32}$.~\cite{Min} According to our computations, those structures are strongly dominant at temperatures higher than 377 K. The next structure shown in Figure~\ref{geometry_gibbs}(5) is located 1.48 kcal/mol above the global minimum; it is close to spherical in shape and corresponds to a prolate structure with C$_1$ symmetry and a singlet electronic state $^1$A. This structure makes up only 4.4{\%} of the relative population at 298.15 K. The next two structures, located at a Gibbs free energy of 2.37 kcal/mol above the global minimum, are the chiral helix-type structures. These were previously reported by Guo et al.~\cite{Guo} as the global minimum and also found with \emph{GALGOSON} code. They are prolate structures with C$_{2v}$ symmetries and their relative population is around only 1{\%}. We note that the chiral-helix structures are never the lowest energy structures throughout the entire temperature range. The relative population is zero for structures located at relative Gibbs free energies higher than 5.1 kcal/mol, and at 298.15 K, there is no contribution of these isomers to any total molecular property. A full understanding of the molecular properties requires the search for the global minimum and all its closest low-energy structures.~\cite{Zhen} The separation among isomers by energy difference is an important and critical characteristic that influences the relative population and, consequently, the overall molecular properties.  To gain insight into how the energy difference among isomers changes and how the energy ordering of the low-energy structures is affected.  We computed the putative global minima and the first seven low-energy structures a single point energy at the CCSD(T)/def2-TZVP level of theory corrected with the zero-point energy computed at the PBE0 D3/def2-TZVP level of theory. Figure~\ref{ccsdt_geo} shows the isomers’ energetic-ordering considering CCSD(T) energy in kcal/mol in parentheses, and the corrected  in kcal/mol in square brackets. At the CCSD(T) level of theory, the global minimum, the seven lowest energy isomers, and the energy order agree with those in a previous work,~\cite{Osvaldo} as shown in the first row of Table.~\ref{energia} The second row of Table~\ref{energia} shows the corrected  Interestingly, the energetic ordering of isomers does not change when considering the ZPE.
\begin{table}[!ht]\centering
\caption{The relative energies in kcal$\cdot$mol$^{-1}$, coupled cluster single-double and perturbative triple, CCSD(T), CCSD(T) with zero point energy ($\mathcal{E}_{\mathrm{ZPE}}$), CCSDT{\normalsize{$+\mathcal{E}_{\mathrm{ZPE}}$})}   Gibbs free energy ($\Delta G$) at 298.15 K, Electronic energy with $\mathrm{ZPE}$ {(\normalsize{$\mathcal{E}_0+\mathcal{E}_{\mathrm{ZPE}}$)}}, Electronic energy {(\normalsize{$\mathcal{E}_0$)}}, point group symmetry, electronic ground state, and the lowest frequency in cm$^{-1}$ for eight low-energy isomers. (Two isomers need to raise the frequencies below 100 cm$^{-1}$ to 100 cm$^{-1}$ ) }
\label{energia}
 \begin{tabular}{@{\extracolsep{0.2pt}} cl l l lll lll }
 \\[-1.8ex]\hline 
 \hline \\[-1.8ex] 
  &   &  \multicolumn{8}{c}{\normalsize{Isomers}}\\
\cline{3-10}\\ [-1.8ex] 
\multicolumn{1}{c} {{{\normalsize{Be$_6$B$^{-}_{11}$}}}}  & \multicolumn{1}{l}{Level} & \multicolumn{1}{c}{$i_1$} & \multicolumn{1}{c}{$i_2$ } & \multicolumn{1}{c}{$i_3$}& \multicolumn{1}{c}{$i_4$}& \multicolumn{1}{c}{$i_5$}  &\multicolumn{1}{c}{$i_6$} &\multicolumn{1}{c}{$i_7$}&\multicolumn{1}{c}{$i_8$}  \\
\hline \\[-1.8ex]
                             & CCSDT                                                &  0.0  &  1.75  &  1.84  & 1.84   & 4.10  & 4.13 & 2.64 & 2.42   \\
                             & CCSDT\normalsize{$+\mathcal{E}_{\mathrm{ZPE}}$}         &  0.0  &  0.58  &  0.85  & 0.86   & 1.19  & 1.23 & 1.68 & 1.81   \\
                             & $\Delta G$                                           &  0.0  & -1.48  &  0.89  & 0.88   & -0.63  & -0.25 & 4.14 & -0.87    \\
                             & \normalsize{$\mathcal{E}_0+\mathcal{E}_{\mathrm{ZPE}}$}  &  0.0  &  -0.29   &  1.51 & 1.52  & 2.41  & 2.42 & 5.0 & -0.08   \\
              {\normalsize{Be$_6$B$^{-}_{11}$}} & \normalsize{$\mathcal{E}_0$}       &  0.0  &  0.87   &  2.50  &  2.50  & 5.32  & 5.32 & 5.96  & 0.52    \\
                & {Point group symmetry}    &  {\normalsize{\emph{C$_{1}$}}} &  {\normalsize{\emph{C$_{1}$}}}   & {\normalsize{\emph{C$_{2}$}}}
              & {\normalsize{\emph{C$_{2}$}}}         &  {\normalsize{\emph{C$_{s}$}}} & {\normalsize{\emph{C$_{2v}$}}}   & {\normalsize{\emph{C$_{1}$}}} & {\normalsize{\emph{C$_{1}$}}}   \\
            
                             & {Electronic ground state}                 & $^1$A   &  $^1$A  &  $^1$A  &$^1$A & $^1$A$^{\text{'}}$   & $^1$A$_1$ & $^1$A & $^1$A   \\
              & {Frequencies }                            &  230  &  119  &  102 & 100  & 46  & 43 & 161 & 151   \\ 
\hline \\[-1.8ex]
\end{tabular}
\end{table}
Nevertheless, the energy-differenece among isomers were reduced drastically. For example, the energy difference between the first and second isomers was reduced by 66{\%}, from 1.75 to 0.58 kcal/mol; the energy difference between the second and third isomers was increased almost 300{\%}, from 0.1 to 0.27 kcal/mol, as shown in rows one and two of Table~\ref{energia} , respectively. This change (increase/decrease) in energy difference among isomers has an enormous impact on the relative population. Consequently, we deduced that the ZPE inclusion is essential to the isomers’ energy ordering and molecular properties. The third row of Table~\ref{energia}  shows the energy order considering the Gibbs free energy computed at 298.15 K; at this temperature, the isomers’ energy ordering changes: the second isomers are the putative global minima, and the first isomers have the fifth lowest energy. Interestingly, this energy ordering occurs at 298.15 K, and it is a  function of the temperature, which we discuss later in the relative population section. The fourth row in Table~\ref{energia} shows the electronic energy considering the ZPE. It follows the same trend in energy ordering when considering the Gibbs free energy, and it is the same putative global minima. The fifth row in Table~\ref{energia}  details the electronic energy. It almost follows the CCSD(T) energies trend, except isomer number 8 takes second place, located at 0.52 kcal/mol above the putative global minimum. The sixth, seventh, and eighth rows in Table~\ref{energia}  show the point group symmetry, electronic ground state, and the lowest vibrational frequency of each isomer, respectively. When we use the Gibbs free energy to energy order the structures, the second isomers change to first place, becoming the lowest energy structure; the energy ordering changes drastically, whereas the electronic energy follows a similar trend to that of CCSD(T) energy ordering. This shows us that the level of theory and the inclusion of entropy and temperature change the energy ordering and, therefore, the overall molecular properties.
\subsection{Relative population}
\begin{figure}[ht!]
  \begin{center}  
    \includegraphics[scale=1.0]{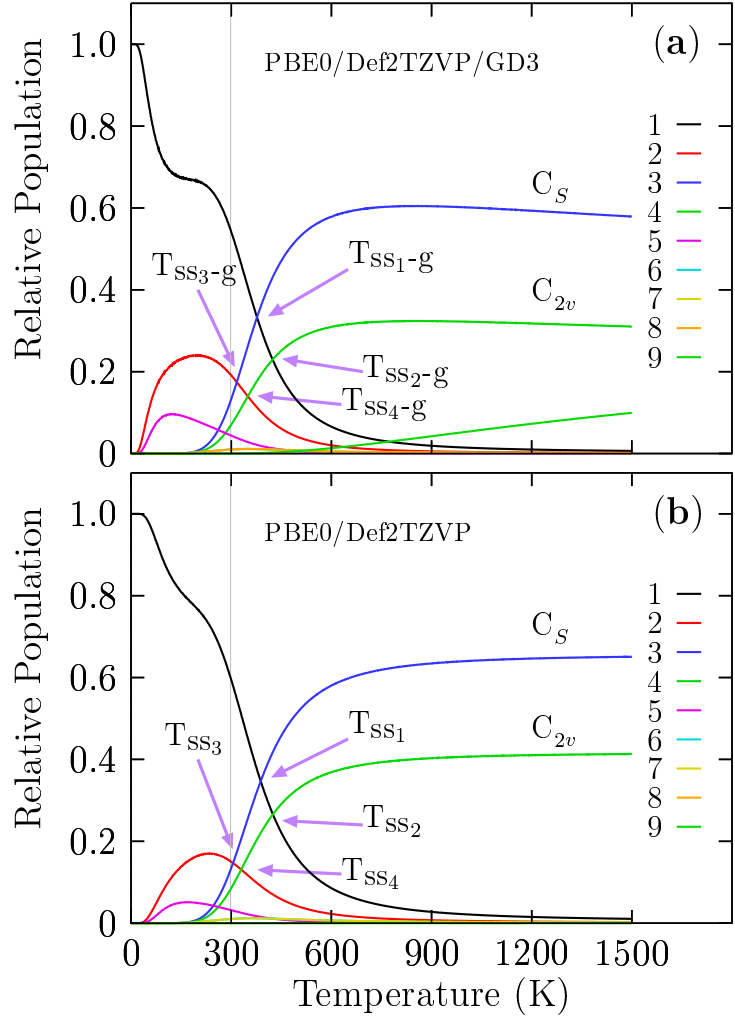}
    \caption{(Color online). The upper panel (a) shows the relative population for the temperatures ranging from 10 to 1500 K and at the PBE0-GD3/def2-TZVP level of theory. The lower panel (b) shows the relative population (without  Grimme's dispersion (D3)) for the temperatures ranging from 10 to 1500 K and at the PBE0/def2-TZVP level of theory. Notice that the Grimme's dispersion's effect shifts the transition solid-solid point ({\large{T$_{\textrm{ss$_1$-g}}$}}) to higher temperatures. The low-symmetry  C$_s$, and  C$_{2v}$ coaxial Triple-Layered structures become strongly dominating at high temperatures.}
    \label{popu}
  \end{center}
  \end{figure}

  \begin{table}[ht!]
  \caption{For easy comparison, the five points temperature solid-solid (T$_{\textrm{ss}}$), at the PBE0-D3/def2-TZVP taking into account the Grimme's dispersion (D3), ({\large{T$_{\textrm{ss$_i$-g}}$}}) and at  PBE0/def2-TZVP ({\large{T$_{\textrm{ss$_i$}}$}}) levels of theory without dispersion. T$_{\textrm{ss}}$ points are displayed in round parenthesis togheter with the probability of occurrence at that point in bold and square brackets.}
  \label{tss}
    \centering
    \begin{tabular}{l l l}   
      \hline\hline
      \rule{0pt}{4ex}    
      {\large{T$_{\textrm{ss$_i$-g}}$}}/{\large{{T$_{\textrm{ss$_i$}}$}}} & PBE0-D3/def2-TZVP  &  PBE0/def2-TZVP   
     \\[0.5ex]  
     \hline
     \rule{0pt}{4ex} 
     1        & (377)/[{\bf{33}}]    & (388)/[{\bf{34.5}}]     \\[1ex]
     2        & (424)/[{\bf{22.9}}]  & (444)/[{\bf{22.8}}]     \\[1ex]
     3        & (316.7)/[{\bf{14}}]  & (305.4)/[{\bf{14.7}}]   \\[1ex]
     4        & (349)/[{\bf{17.6}}]  & (343.6)/[{\bf{12.2}}]   \\[1ex]
     5        & (258)/[{\bf{5.7}}]   & (246.7)/[{\bf{4.2}}]   \\[1ex]
        \hline 
    \end{tabular}  
  \end{table}
Figure~\ref{popu} panel (a) shows the most important and strongly dominating {\large{T$_{\textrm{ss$_1$-g}}$}} point that is located at 377 K  temperature scale with a relative population of 33{\%}.
For temperatures ranging from 10 to 377 K, the relative population is strongly dominated by the putative global minima isomer distorted oblate spheroid with C$_1$ symmetry and this relative population is similar to -T$^{ -3}$ function with one point of inflection located at 180 K. After decreases monotonically up to 377 K.  At the {\large{T$_{\textrm{ss$_1$-g}}$}} point, the distorted oblate spheroid with C$_1$ symmetry co-exist and compete with the coaxial Triple-Layered structures with C$_s$ symmetry; This implies that the distorted oblate spheroid will be replaced with the coaxial Triple-Layered structures. Above temperature 377 K, the relative population is strongly dominated by the coaxial Triple-Layered structures with C$_s$ symmetry, located at 0.85 kcal$\cdot$mol$^{-1}$ above the global minimum at temperature of 298.15 K. This relative population depicted in blue-solid line in panel (a) has behavior as a  sigmoid function, from temperatures ranging from 377 to 600 K, it grows rapidally and from temperatures ranging from 600 to 1500 K, it almost keeps constant with 60{\%}. The second {\large{T$_{\textrm{ss$_2$-g}}$}} point is located at temperature 424 K with a relative population of 22.9{\%}, and this point the global minima distorted oblate spheroid with C$_1$ symmetry co-exist, and compete with the coaxial Triple-Layered structures with C$_{2v}$ symmetry, located at 1.23 kcal/mol above the global minima at 298.15 K.  The relative population of the coaxial Triple-Layered C$_{2v}$ symmetry depicted in green-solid line in panel (a) also has a behavior of a sigmoid function and up to 600 K it keeps constant with 32{\%} of relative population.
The {\large{T$_{\textrm{ss$_3$-g}}$}}, and  {\large{T$_{\textrm{ss$_4$-g}}$}} points, displayed in Figure~\ref{popu} panel (a), are located at 316.7 K, and 349 K  axis temperature with relative populations 14{\%} and 17{\%}, respectively. These relative populations correspond to the second isomer located just 0.61 kcal/mol at 298.15 K above the global minima, and co-existing at the temperatures 316.7 K and 349 K with the coaxial Triple-Layered structures with C$_s$, and C$_{2v}$ symmetries, respectively.
At low temperatures range, this isomer's relative population depicted in red-solid line of Figure~\ref{popu}a is around only 20{\%}, and up to room temperature, it decreases exponentially to zero. At temperatures up to 600 K, the relative population is zero; hence, at high temperatures these isomers do not contribute to the molecular properties. The relative population lower than 10{\%}, depicted in violet-solid line shows in Figure~\ref{popu}a, correspond to the isomers located at 1.48 kcal/mol above global minima at 298.15 K.
Interesting, this structure is the putative minimum global when the CCSD(T) energy is employed in the ordering energetic,
Despite that, this structure's relative population clearly shows that this structure does not contribute to molecular properties in all ranges of temperatures.  The average B-B bond length for this structure is 1.63~\AA,~distant for the lowest average B-B bond length of 1.53~\AA.~Moreover, this structure has the largest positive contribution to the relative zero-point energy. This suggests that not just the global minimum and its closest energy isomers of a potential/free energy surface are important, but also the contribution entropic effects and temperature are decisive in which isomers are going to contribute to molecular properties in a temperature ranging of interest. It should be pointed out that neither the helix-type structure reported by Guo et al.~\cite{Guo} nor the putative global minimum found in this study,  also reported by Ya\~nez et al.~\cite{Osvaldo} at a high level of theory, is the putative global minimum when we take into account the entropic term. Our results lead to that the entropic effect should be taken into account. So far, at this point, one may ask if there is a simple and easy method to elucidate which isomers have the largest entopic contributions. This question is going to be boarded in the relative zero-point energy decomposition section.  Another  interesting  question is: What is the effect of Grimme's dispersion (D3) on the relative population?. Figure~\ref{popu}, panel (b) shows four transitions solid-solid temperature points {\large{T$_{\textrm{ss$_1$}}$}}, {\large{T$_{\textrm{ss$_2$}}$}}, {\large{T$_{\textrm{ss$_3$}}$}}, and {\large{T$_{\textrm{ss$_4$}}$}}, without Grimme’s dispersion (D3), and for ease of comparison displayed in round parenthesis in Table.~\ref{tss} togheter with the probability of occurrence in bold and square brackets.  The {\large{T$_{\textrm{ss$_1$}}$}}, and  {\large{T$_{\textrm{ss$_2$}}$}}  point shifts in the temperature axis to a higeher temperature by 10 K, and 20K, whereas the relative population has a little variations not larger than 1.5{\%}. The {\large{T$_{\textrm{ss$_3$}}$}}, {\large{T$_{\textrm{ss$_4$}}$}}, and {\large{T$_{\textrm{ss$_5$}}$}} shifts in temperature axes to low tempearture, In the first galance, it suggest that the effect of the dispersion on the  reletive population is a little shift of  the two dominant {\large{T$_{\textrm{ss}}$}} points from low temperatures to higher temperatures, almost keeping the relative populations constant. In contrast with the {\large{T$_{\textrm{ss}}$}} points, with lower probability occurrence, shows a little  shift from high temperature to lower temperatures with little changes in the relative population. The total properties in a molecule are statistical averages over the ensemble of isomers. Thus, it is crucial as far as possible to make a complete sampling of the potential energy surface to consider all isomers. The search of the low-energy structures is not straightforward, and many times this could be lead to missing some low-energy isomers. In this respect, One might well ask what happens if we have a missing low-energy structure when we compute the relative populations and their consequence on on the computation of any molecular properties.
\begin{figure}[ht!]
  \begin{center}  
    \includegraphics[scale=1.0]{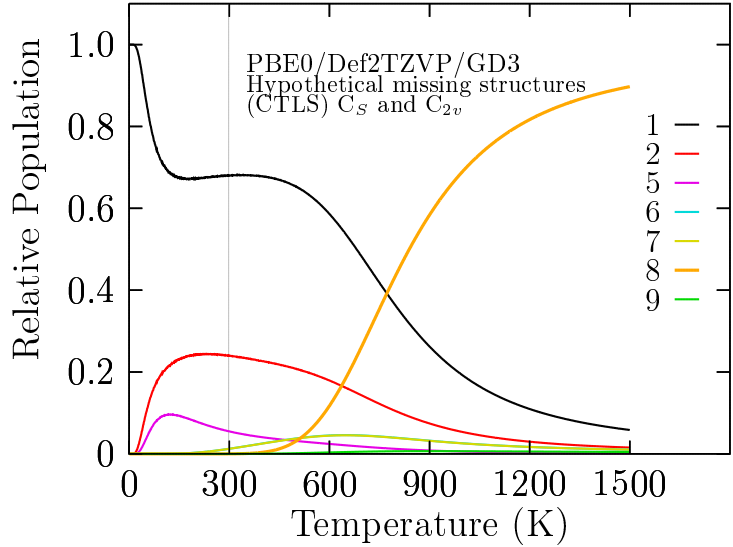}
    \caption{(Color online). The relative population of Be$_6$B$_{11}^{-}$ cluster for the temperatures ranging from 10 to 1500 K.
      It should be noted that in the abscence of C$_{2v}$ and C$_s$ symmetry coaxial Triple-Layered strutures (CTLS) in pool of isomers, the distorted-coaxial Triple-Layered structure with symmetry  C$_{2v}$ depicted in Figure~\ref{geometry_gibbs}(8), ~\ref{ccsdt_geo}(10), dominate for temperatures higher that 773 K. The lowest energy structure (at T=0), does not necessarily have the associated largest entropic effect.}
      \label{popu3}
  \end{center}
  \end{figure}
Figure~\ref{popu3} shows the computed relative population when the two coaxial Triple-Layered  C$_S$ and C$_{2v}$ structures have been taken out of the isomers pool database. Indeed,  in the temperature ranging from 773 to 1500 K, the relative population depicted in yellow-solid line in Figure~\ref{popu3} indicate   that the dominant structure is a distorted coaxial Triple-Layered depicted in Figure~\ref{ccsdt_geo}(10) and located at 9.20 kcal/mol above the putative global minimum at CCSD(T) level of theory.  Furthermore, analysis of results on the average B-B bond length shows in Figure~\ref{lboro} indicates that the structure with the second-lowest bond length is also the same distorted coaxial Triple-Layered structure.
This result leads to a couple of interesting observations in the case of Be$_6$B$_{11}^{-}$ cluster. Even at the high level of theory, the lowest energy structure (at T=0) does not necessarily have the associated largest entropic effect, and the structure with the lowest B-B bond is correlated with the largest entropic effects.
\begin{figure}[ht!]
  \begin{center}
  \includegraphics[scale=1.0]{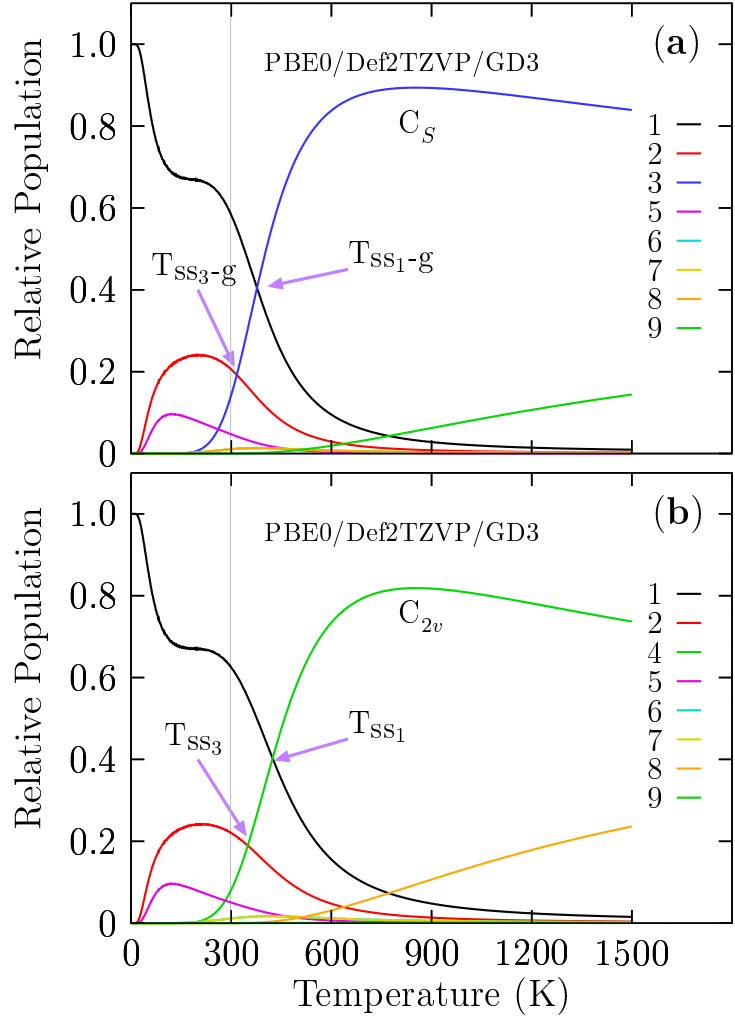}
    \caption{(Color online). The relative population of Be$_6$B$_{11}^{-}$ cluster for the temperatures ranging from 10 to 1500 K. The upper panel (a) shows (blue-soldi line) the relative population of the  of C$_{S}$ symmetry in the absence of C$_{2v}$ symmetry coaxial Triple-Layered structure in a pool of isomers, the low-symmetry  C$_s$, becomes strongly dominating for temperatures higher than 379 K ({\large{T$_{\textrm{ss$_1$-g}}$}}). The lower panel (b) shows (green-solid line) the relative population of the C$_{2v}$ symmetry in the absence of C$_{S}$ symmetry coaxial Triple-Layered structure in a pool of isomers, the high order-symmetry  C$_{2v}$, becomes strongly dominating for temperatures higher than 425 K ({\large{T$_{\textrm{ss$_1$}}$}}).  The absence of isomers with symmetries in the isomers-pool suggest important consequences in the molecular properties.}
    \label{popuSinC2v}
  \end{center}
  \end{figure}
It is worthwhile to note that in the temperature ranging from 377 to 1500 K, in Figure~\ref{popu} panel (a), the relative population depicted blue-solid line indicates that the coaxial Triple-Layered structures with C$_s$ symmetry is energetically more favorable than the coaxial Triple-Layered structures with C$_{2v}$ symmetry. Moreover, those two structures strongly dominate in this range of temperature. These results pointed out that we must consider more than one more isomer with point-group symmetries, ranging from the low-symmetry to high-symmetry. Figure~\ref{popuSinC2v} panel (a) display the relative population computed without taking into account the C$_{2v}$ symmetry coaxial Triple-Layered structure in the pool database, and panel (b) displays the relative population computed without taking into account the C$_{2v}$symmetry coaxial Triple-Layered symmetry in the pool database. A Comparison between the relative population shows in panel (a) and panel (b)    of  Figure~\ref{popuSinC2v} indicates, on the one hand, that that dominant T$_{ss}$ point does not shift when we do not consider the high-symmetry,   and on the other hand, the dominant T$_{ss}$ point shifts from 379 to 425 K when we no take into account the low-symmetry structure. This result leads to the observation that it is more important to calculate the relative population considering the low-symmetry structure than only structures with high symmetries. The reason is when we consider low-symmetry structures; the T$_{SS}$ point does not change. In contrast, when we take only higher -symmetries structures, the T$_{SS}$ shift with important consequences in the molecular properties when we compute the molecular properties as  statistical averages over the an ensemble of isomers.
\subsection{Relative population at CCSD(T) level of theory}
 \begin{figure}[ht!]
  \begin{center}  
    \includegraphics[scale=1.0]{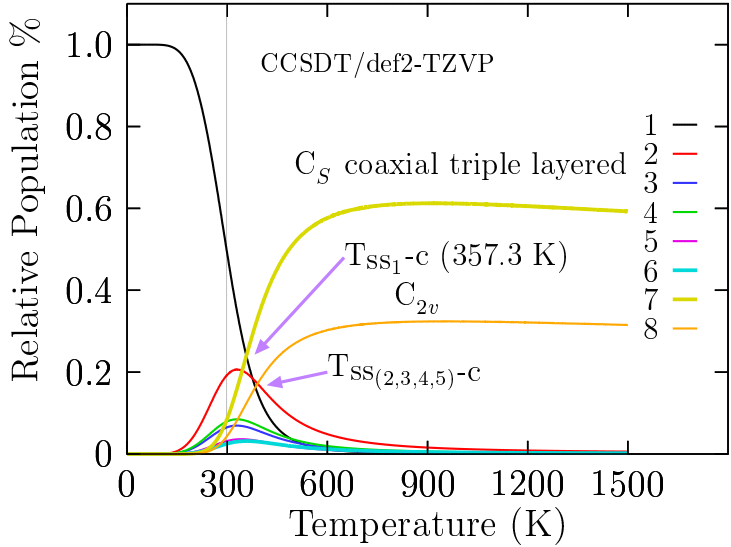}
    \caption{(Color online). The panel shows the relative population for temperatures ranging from 10 to 1500 K computed at single point CCSD(T)/def2-TZVP//PBE0-D3/def2-TZVP level of theory. The {\large{T$_{\textrm{ss$_1$-c}}$}} point is located at 357 K. The low-symmetry  C$_s$, and  C$_{2v}$ coaxial Triple-Layered structures become strongly dominating at temperatures higher than 357 K. The dominant structure at low tmeperatures  is the putative lowest energy structure depicted in Figure~\ref{ccsdt_geo}(1) with C$_1$ symmetry computed at SP CCSD(T) level of theory.}
    \label{popucc}
  \end{center}
  \end{figure}
 Figure~\ref{popucc} shows the relative population computed at single point  CCSD(T)/def2-TZVP//PBE0-D3/def2-TZVP level of theory.  For temperatures ranging from 10 to 357 K, the relative population is strongly dominated by the putative global minima isomer with C$_1$ symmetry and the relative population  decreases monotonically up to 357 K.  At the {\large{T$_{\textrm{ss$_1$-c}}$}} point, the distorted  C$_1$ symmetry co-exist and compete with the coaxial Triple-Layered structures with C$_s$ symmetry. The dominant structure at low tmeperatures  is the lowest energy structure depicted in Figure~\ref{ccsdt_geo}(1) with C$_1$ symmetry at SP CCSD(T) level of theory.
 \subsection{Molecular Dynamics}
 In this study to explore and gain insights into the dynamical behavior of Be$_6$B$_{11}^-$  a Born-Oppenheimer molecular dynamics (BOMD) was performed employing the  deMon2K program~\cite{demon2k} (deMon2k v. 6.01, Cinvestav, Mexico City 2011) at three different temperatures, 1600 K, 2000 K, and 2500 K,  and the  PBE/DZVP level of theory. We have chosen the temperatures from 1600K to 2500 K due to these temperatures are close to the melting points of boron (2349 K) and beryllium (1560 K). The BOMD's were started from the initial configuration of the coaxial-triple layered structure (the putative global minimum at a temperature of 1500 K ), employing a Hoover thermal bath with random initial velocities imposed to the atoms, and for a simulation time of 25 ps with a step size of 1 fs. As the temperature increases, Be$_6$B$^{-}_{11}$ cluster is subject to dissociation phenomena. Based on the BOMD simulation results, we found the dissociation processes of the Be$_6$B$^{-}_{11}$ cluster occurs at a temperature of 2000 K,  whereas there is no dissociation during the BOMD simulation at 1600 K; the cluster maintains its connectivity at this temperature.  At a temperature of 2500 K, the dissociation processes are stronger, and more beryllium atoms are escaped (see the movies in the Supplementary Information).
 Min Li et al.~\cite{Li-Min} noted that nanoparticles of tungsten dissociate when the temperature of tungsten nanoparticles is higher than the melting temperature. Our results make sense if we considered that the BOMD of Be$_6$B$^{-}_{11}$ cluster there is not dissociation at 1600 K, whereas at the temperature of 2000 K, there are dissociation phenomena. From the mentioned previously, we can infer that the melting point of the Be$_6$B$^{-}_{11}$ cluster is in the temperature ranging from 1600 K to 2000 K. 
  \begin{figure}[ht!]
  \begin{center}  
    \includegraphics[scale=1.0]{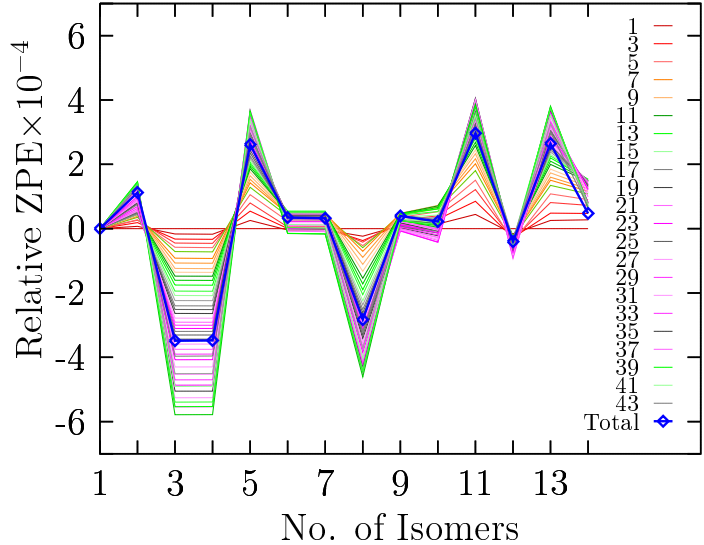}
    \caption{(Color online). Relative ZPE decomposition as a function of the vibrational modes (Hartree/particule), with reference ZPE of the lowest energy isomer. In axes, X is the number of isomers arranged from the lowest energy isomers (1) to higher energy isomer  (14). The lowest value of the total relative ZPE as a function of the number of isomers is correlated to the isomer that dominates as a global minimum at high range temperatures that correspond to the coaxial Triple-Layered structures with C$_s$ symmetry. The blue is the total ZPE taking all 45 vibrational modes the clusters pose. (3N-6 modes)}
    \label{zpe2}
  \end{center}
  \end{figure}
 \section{Contibutions of the vibrational modes to the ZPE energy.}
 At temperature zero, the lowest energy structure has the electronic energy plus zero-point energy computed as the sum of all vibration modes.  If one increases the system's temperature, entropic effects start to play an important role, and Gibbs's free energy determines the lowest energy structure. The significant contribution to entropy comes from low vibrational modes, and it is approximately proportional to the logarithmic sum of low frequencies,~\cite{spickermann2011entropies} in the other hand, high vibrational modes yield small contributions to vibrational entropy. The Equation~\ref{zpe} gives the zero-point energy (ZPE), where ~$\nu_i$~ are all 3N-6 vibrational modes that the cluster possesses. Figure~\ref{zpe2} shows the relative ZPE as a function of vibrational modes and isomers that are arranged in energy, from the lowest (1) to the highest energy isomer (14).  Remarkably,   the smallest value of the total relative ZPE (the minimum of ZPE) correlates with the lowest energy structure at high temperatures. The relative population displayed in Figure~\ref{popu} panel (a) shows that the isomer three and four, that correspond to coaxial triple-layered structure with C$_s$  and $C_{2v}$ symmetries respectively, strongly dominate in the temperature range up to 377 K.
 \begin{equation}
  \displaystyle
  \centering 
ZPE=\frac{1}{2}\sum_{i=1}^{3N-6} \nu_i
\label{zpe}
 \end{equation}
Figure~\ref{zpe2} shows the relative ZPE as a function of vibrational modes and isomers that are energetically ordered,  from the lowest (1) to the highest energy isomer (14).  Remarkably,   the smallest value of the total relative ZPE (the minimum of ZPE) correlates with the lowest energy structure at high temperatures. The relative population displayed in Figure~\ref{popu} panel (a) shows that the isomer three and four, that correspond to coaxial triple-layered structure with C$_s$  and $C_{2v}$ symmetries respectively, strongly dominate in the temperature range up to 377 K. In Figure~\ref{zpe2}, one can see that isomers three and four, the coaxial Triple-Layered structures with C$_s$  and $C_{2v}$ symmetries possess the lowest value of the relative ZPE. Interestingly, the structures with the lowest relative ZPE are correlated with the structures that strongly dominate the putative global minima at high temperatures.  This suggests that those structures possess the highest entropic effects. 
To understand which lowest vibrational modes contribute to the lowest ZPE, we decompose the relative ZPE as a function of the number of modes, adding the number of modes needed to build the smallest value of ZPE. The blue-solid line in Figure~\ref{zpe2} depicts the total relative ZPE employing the forty five vibrational modes, the red-solid lines depicts the relative ZPE with emplying modes from first to the sixth and so on. The Be$_6$ B$_{11}^{-}$ cluster posse forty five vibrational modes, we found that we have to add the lowest thirty eight vibrational modes to make the smallest value of relative ZPE. The frequency of mode thirty eight is  1026  cm$^{-1}$, which indicates that is the highest frequency (cutoff frequency)  that contributes to making minimum relative ZPE and, therefore, those vibrational frequencies that are in the range of 46 to 1026 cm $^{-1}$  makes the significant contribution to entropy. 
The vibrational modes number 39 to 45 (1036-1518  cm$^{-1}$) does not  make contributions to lowering the relative ZPE, as shown in Figure~\ref{zpe2}
\subsection{Infrared Spectroscopy}
In this section, each isomer’s IR spectra and how the relative stabilities contribute to the total IR spectra are discussed. In this study, each isomer’s IR spectra were computed using DFT as it is implemented in Gaussian 09 code; under the harmonic approximation, anharmonic effects are not considered. The effect of temperature on the total spectra and the total IR spectra were computed as a Boltzmann weighted sum of each isomer’s IR spectra, implemented in BOFA. As the Boltzmann factors depend on temperature, the total resulting IR spectra depend on temperature.
\begin{figure}[ht!]
  \begin{center}
  \includegraphics[scale=0.50]{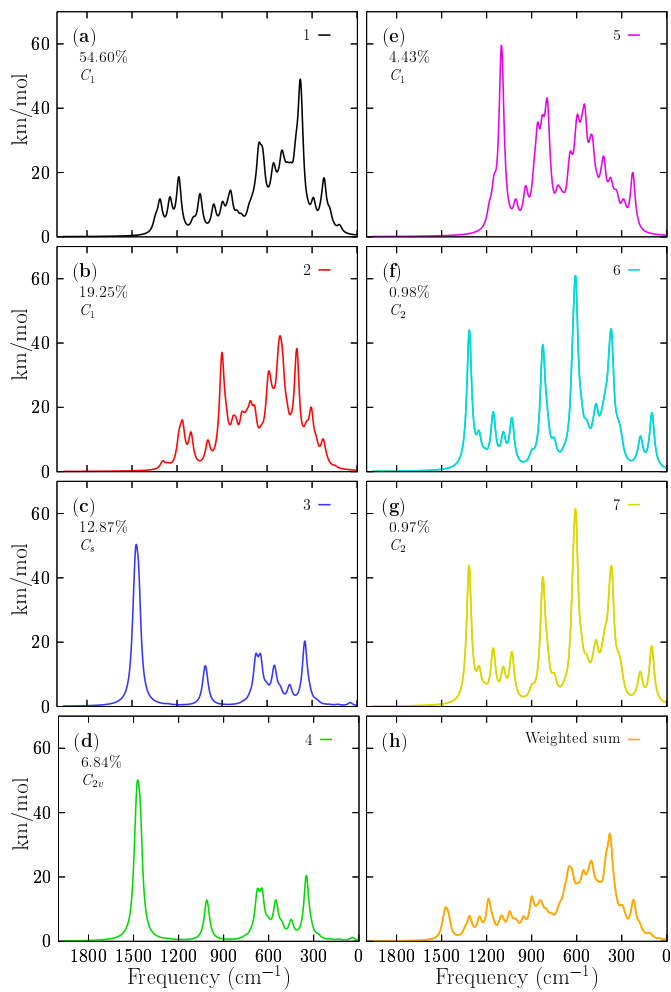} 
  \caption{(Color online). Panels (a) to (g) shows computed infrared spectra of boron clusters  in the range of 1 to 2000 $cm^{-1}$, based on the PBE0 functional with the def2-TZVP basis set, and considering version D3 Grimme's dispersion ( implemented in Gaussian 09 code)~\cite{gauss}. (h) The weighted IR spectrum Boltzmann sum of the IR spectra's results of the energetically competing structures, which provide different percentages to the entire IR spectrum. The total weighted IR spectrum shows a peak at 1500 cm$^{-1}$, which is not present in the IR spectrum that belongs to the putative global minimum at 298.15 K shown in (a). These IR bands are assigned to the 12.9{\%} contribution of the third isomer, which is coaxial triple layered structures with C$_s$ symmetry located at 0.85 kcal$\cdot$mol$^{-1}$ above of the global minimum.}
  \label{irs}
  \end{center}
\end{figure}
In a previous work,~\cite{Vargas-Caamal} one of the authors computed the total dipole moment as a dipole moment weighted by the Boltzmann factors and successfully compared it with experimental data.  From the experimental point of view,   Sieber et al.~\cite{Felix} compared the measured absorption spectrum of the Ag$_9$ cluster to a sum of different absorption spectra of the Ag$_9$ cluster computed by DFT.    Concerning boron clusters,  the vibrational spectrum of boron cluster B$_{13}^+$  was measured by infrared photodissociation spectroscopy and also compared with computed spectra.  Experimental spectroscopy studies employing anion photoelectron spectroscopy on boron anions cluster up to B$_{40}^{-}$ clusters have been done.  Additionally, the structure of neutral boron clusters B$_{11}$, B$_{16}$, and B$_{17}$ was also probed by IR.
The IR spectrum is related to vibrations that alter the dipole moment. These spectra are usually used to identify functional groups and chemical bond information, and are useful in organic/inorganic chemistry. However, from an experimental perspective, the assignment of IR bands to vibrational molecular modes can be somewhat difficult and requires \emph{ab-initio} calculations. In these computations, the temperature is generally not considered, and discrepancies between experimental and calculated IR spectra can result from finite temperature and anharmonic effects. It is also important to remember that the experiments are essential of multi-photon nature, whereas calculations IR spectra assume single-photon processes. Figure~\ref{irs}(a-g) display the individual IR spectra that belong to the putative global minima and the six lowest energy structures, respectively, located in the relative energy range up 0 to 2.38 kcal/mol at 298.15 K. Figure~\ref{irs}h shows the Boltzmann weighted spectrum at 298.15 K computed with BOFA. Interestingly, 93{\%} of the total weighted IR spectra was found to be composed of the individuals spectral contributions of the four lowest energy structures located at an energy scale-up of 0 to 1.23 kcal/mol. The other 7{\%} of the Boltzmann weighted spectra is composed of the IR spectra of the three structures located in the energy range from 1.48 to 2.38 kcal/mol. In the total weighted Boltzmann IR spectrum in Figure 7h, there are three segments on its frequency axis. The first segment is located in the frequency range of 0 to 700 cm$^{-1}$. The main bands observed in this range correspond to the IR vibrational modes of the global minimum. The highest peak is located in the 387 cm$^{-1}$ frequency axis, which corresponds to compression of the main ring formed by 10 boron atoms. It is located mainly on one side of the ring, accompanied by the vibrations of the two beryllium atoms. The second band is located at 669 cm$^{-1}$ in the frequency axis. This corresponds mainly to the 10-ring boron’s small asymmetric vibration and a minor vibration of the six beryllium atoms. The third peak is located at 225 cm$^{-1}$ on the frequency axis. It corresponds mainly to a stretching of the boron atom that does not form part of the boron ring, together with the two beryllium atoms located close to the boron. The second segment is located in the frequency range of 700 to 1400 cm$^{-1}$. In the Boltzmann weighted IR displayed in Figure~\ref{irs}h, the band observed at 900 cm$^{-1}$ is mainly composed of the 19.2{\%} contribution of the individual IR spectrum of the second isomer that lay 0.61 kcal/mol above the global minimum; this vibrational mode of the second isomer corresponds to the stretching of the three beryllium atoms located on one side along with a boron atom, together with the stretching of one of the boron atoms. The band observed at 1200 cm$^{-1}$ (Figure~\ref{irs}h) is mainly associated with the global minima’s IR spectrum, which corresponds to the boron atoms unique stretching. There is almost no vibration of the beryllium atoms. The band observed at 1500 cm$^{-1}$ (Figure~\ref{irs}h) is completely composed of the contribution of 12{\%} of the individual IR spectra of the third isomer, which has a coaxial triple-layered structure with C$_s$ symmetry. The fourth isomer contribution is the coaxial triple-layered structure with C$_{2v}$ symmetry located 1.23 kcal/mol above the global minimum. The different symmetries of the coaxial triple-layered structures ( C$_{2v}$ and C$_s$) are responsible for the different contributions to the total weighted IR spectrum. The low-symmetry isomers become more stable at high temperatures as a result of entropic effects. Interestingly, neither individual IR spectrum of the putative global minimum nor the individual IR spectrum of the second isomer, which was 0.61 kcal/mol above the putative global minimum, has any IR band in the range of 1500 cm$^{-1}$. Based on this, we assigned this band at 1500 cm$^{-1}$ in the total Boltzmann weighted IR spectrum to the third and four isomers, which have a coaxial triple-layered structure with two different symmetries. The helix-type structures proposed by Guo et al.~\cite{Guo} have a small contribution to the IR spectra in all ranges of temperature. The methodology employed in this paper for the assignment of the IR bands demonstrates that the total IR spectra are a mixture of many contributions from the low-energy structures. In this cluster, the total IR spectrum’s low-energy region is attributed to the putative minimum global contribution. In contrast, the high-energy region of the total IR spectrum is attributed to the isomers’ contribution on the high-energy axis. 
Figure~\ref{irs8} displays the IR spectra computed as a function of temperature. Figure~\ref{irs8}a shows the total Boltzmann weighted IR spectra in the temperature range of 10 to 300 K. Note that the IR spectrum at low temperatures is strongly dominated by the individual IR spectrum of the putative global minimum; this finding is in agreement with the relative population displayed in Figure~\ref{popu}. Below 377 K, the relative population is strongly dominated by the putative global minimum.
\begin{figure}[ht!]
  \begin{center}
  \includegraphics[scale=0.65]{figure_8.epsi} 
  \caption{(Color online). Panels (a) to (g) show the computed infrared spectra of boron clusters  in the range of 1 to 2000 $cm^{-1}$, based on the PBE0 functional with the def2-TZVP basis set, and considering version D3 Grimme’s dispersion ( implemented in Gaussian 09 code). (h) The weighted IR spectrum Boltzmann sum of the IR spectra’s results of the energetically competing structures, which provide different percentages to the entire IR spectrum. The total weighted IR spectrum shows a peak at 1500 cm$^{-1}$, which is not present in the IR spectrum that belongs to the putative global minimum at 298.15 K shown in (a). These IR bands are assigned to the 12.9{\%} contribution of the third isomer, which has coaxial triple-layered structures with C$_s$ symmetry.}
  \label{irs8}
  \end{center}
\end{figure}
The band observed at 1500 cm$^{-1}$ in Figure~\ref{irs8}a starts to increase at 200 K (pink line), increases again at 250 K (cyan line), and increases further at 300 K (yellow line). This IR band has contributions from the individual IR spectra of the coaxial triple-layered structures with Cs and C$_{2v}$ symmetries. It is in complete agreement with the relative population displayed in Figure~\ref{popu}a. The relative population of the coaxial triple-layered structures start to increase at 200 K. Figure~\ref{irs8}b shows the IR spectra in the range of 310 to 410 K. Within this temperature range, most transformation solid-solid point occur with different probabilities of occurrences as shown in Figure~\ref{popu}a; therefore, large changes in the total weighted IR spectra are also expected. In Figure~\ref{irs8}b, the IR band located at 1500 cm$^{-1}$ continues increasing at 310 K, and it persists, increasing to 430 K (cyan line). This vibrational mode pertains to an individual IR spectrum of the isomer with coaxial triple-layered structures displayed in Figure~\ref{irs}c. This is completely in agreement with the relative population displayed in Figure~\ref{popu}. From 377 to 1500 K, the relative population is strongly dominated by the coaxial contributions of the triple-layered structures with C$_s$ and C$_{2v}$ symmetries. The appearance and constant growth of the peak located at 1500 cm$^{-1}$ in the weighted total IR spectrum displayed in Figure~\ref{irs8}b, as a function of temperature, indicate the coexistence and competition of at least two strongly dominant structures at a specific finite temperature (377 K). Most importantly, the constant growth of the peak located at 1500 cm$^{-1}$ is indicative that putative global minimum interchange occurs as a function of temperature. This suggests that we must search exhaustively and systematically for the putative global minimum on the potential/free energy surface and its full distribution of all low-energy structures if we want to assign IR bands to specific vibrational modes. This paper shows how some IR bands in the Boltzmann weighted total IR spectrum belong to the IR spectra of isomers located on the higher energy axis. In summary, in the Boltzmann weighted total IR spectrum shown in Figure~\ref{irs8}b, the low-frequency range is dominated by the contributions of the putative global minimum, whereas the high-frequency range is dominated by geometric structures located at higher energies. Figure~\ref{irs8}b shows the IR spectra in the range of 310 to 410 K. Within this temperature range, most solid-solid transitions occur with different probabilities, as shown in Figure~\ref{popu}a; therefore, large changes in the total weighted IR spectra are also expected. In Figure~\ref{irs8}b, the IR band located at 1500 cm$^{-1}$ continues increasing at 310 K, and it persists, increasing up to 430 K (cyan line). This vibrational mode pertains to an individual IR spectrum of the coaxial triple-layered isomer displayed in Figure~\ref{irs}c. This is completely in agreement with the relative population displayed in Figure~\ref{popu}a. The peak located at 1500 cm$^{-1}$, shows in Figure~\ref{irs8}, panels (a-d), is present only in temperatures higher than 300 K, where the coaxial-triple layer structures start to be the lowest energy structures. The abovementioned reasons indicate that the vibrational modes located in the range of 1036 to 1518 cm$^{-1}$ are responsible for the cluster's fluxionality. The B-B stretching modes of the B$_{11}$ ring are located in the range of 1036 to 1518 cm$^{-1}$. This is correlated to hyperconjugation, delocalization, and fluxionality of the cluster.

\section{Conclusions}
In summary, we systematically explored the potential and free energy surface of the Be$_6$B$_{11}^{-}$ cluster using an unbiased hybrid, efficient, and multistep/multilevel algorithm implemented in Python and coupled to density functional theory. The temperature effects were considered employing Gibb’s free energy. If the system’s temperature is increased, entropic effects start to play an important role, and Gibbs’s free energy determines the lowest energy structure. 
We computed the relative population as a function of temperature using Boltzmann factors and the IR spectra dependent on temperature as a Boltzmann weighted sum of each IR spectrum’s isomer. Here, we demonstrate that the temperature and entropic effects produce several competing structures, so a mixture of isomers co-exist at a specific temperature. Our computations showed (with relative population) that the low-symmetry isomers have a higher stability than isomers with high symmetry at high temperatures as a result of the entropic effect. The coaxial triple-layered structures with C$_s$ symmetry are the putative global minima above 377 to 1500 K due to entropic effects. There are four T$_{ss}$ points in the relative population of the Be$_6$B$^{-}_{11}$ cluster; the most important and dominant of these is the T$_{ss}$ point located at 377 K with a relative population of 33{\%}. Additionally, our results give insight into the long-range \emph{van der Waals} interactions effects on the solid-solid transformation temperature points, hence the molecular properties. Indeed, the effect of dispersion shifts up in temperature the dominant T$_{ss}$ point, keeping the relative population almost invariant. The other T$_{ss}$ points shifted down on the temperature axis, so there is no clear trend in the up/downshifts in the  Be$_6$B$^{-}_{11}$ cluster. Remarkably, the coaxial triple-layered structures with C$_s$ and C$_{2v}$ symmetries have the lowest B-B bond length, and the same geometrical structures have the lowest relative zero-point energy. This suggests that both trends shortening of the B-B bond length and lowest relative zero-point energy are correlated with entropic effects. Analysis of our results leads to an interesting observation:  The strong dominant putative global minimum, under high-temperature conditions, has the shortest B-B bond length and the lowest relative zero-point energy. The low vibrational modes significantly contribute to entropy, whereas high vibrational modes provide small contributions to entropy. The Be$_6$B$^{-}_{11}$ cluster has 45 vibrational modes. We found the range of frequencies-the lowest to the highest vibrational modes that contribute to zero-point energy by computing the zero-point energy as a function of vibrational modes. We needed to sum the first 38 modes that contribute to zero-point energies; the frequency range was between 46 and 1026 cm$^1$. Vibrational modes outside of this range do not contribute to the zero-point energy. At the energy single-point CCSD(T) level of theory, the energetic ordering of isomers changes with respect to employing the electronic or Gibb’s free energies. The inclusion of the zero-point energy in CCSD(T) energies illustrates that the energy difference among isomers reduces drastically, which suggests that the dominant putative global minimum at zero temperature when we employ the CCSD(T) energies will change with the inclusion of temperature. 
The properties observed in a molecule are statistical averages over the ensemble of geometrical conformations or isomers accessible to the cluster, so the molecular properties are ruled by the Boltzmann distributions of isomers, which can change significantly with temperature, primarily due to entropic effects. We computed the IR spectra dependent on temperature as a Boltzmann weighted sum of each IR spectrum’s isomer. Our computations showed that the competing structures provide a different percentage to the entire molecular properties and IR spectra, in detail, the molecular properties can not be attributed to only the lowest energy structure. The structures located at high energy above the putative global minimum that have a significant energy difference among isomers on the potential/free energy surface do not contribute to the entire IR spectrum. Despite the number of isomers growing exponentially, the main contribution to the molecular properties comes from the low-energy structures close to the global minimum where the weighted Boltzmann factors’ temperature dependence are different from zero. (This depends strongly on the energy separation; if the energy separation is significant, the IR spectrum does not change.) The spectra that belong to the low-energy structure dominate the IR spectrum of the  Be$_6$B$^{-}_{11}$ cluster at low-temperature structures, whereas at high temperatures, it is strongly dominated by the spectra of structures located at high energy above the putative lowest energy structures. The increase/decrease in a peak/band in the IR spectra as a function of temperature is a clear signature of an interchange of the dominant lowest energy structure. With the IR spectra, we illustrated that the main contributions to the molecular properties are from the low-energy structures that are very close to the global minimum where the weighted Boltzmann factors’ temperature dependence is different from zero. 
The present study highlights the importance of entropy-temperature effects and what happens when some low-energy structures are not considered. We show that symmetry plays an important role in the definition of the global minimum and hence in molecular properties. We demonstrate that dispersion effects has a little the changes of the T$_{SS}$ points in temperature scale.  All of these effects have an impact on the spectroscopic and any other property of a molecular system. The Boltzmann-IR-spectra as a function of temperature were presented. The spectra unravel that the peak located at 1500 cm$^{-1}$, shows in Figure 8, panels (a-d), is present only in temperatures higher than  300 K,  where the coaxial-triple layered structures start to be the lowest energy structures.  The above mentioned reasons indicate that the vibrational modes located in the range of 1036 to 1518 cm$^{-1}$ are responsible for the cluster's fluxionality.  
An immediate future project is the computation of the optical spectra and other molecular properties employing the methodology described in this study and the computation of the relative population in many atomic and molecular clusters of interest employing higher levels of theory.

The Boltzmann Optics Full Adder (BOFA) \emph{Python} code supporting the findings of this study is available from the corresponding author upon reasonable request.

\section{Acknowledgments}
C.E.B.-G. thanks CONACyT for the Ph.D. scholarship (860052). We are grateful to Dra. Carmen Heras, and L.C.C. Daniel Mendoza for granting us access to their clusters and computational support. \emph{ACARUS} provided computational resources for this work through the High-Performance Computing Area of the University of Sonora. We also grateful to the computational chemistry laboratory (JLC) to provides computational resources, \emph{ELBAKYAN}, and \emph{PAKAL} supercomputers.
\section{Conflicts of Interest} The authors declare no conflict of interest.
\section{Funding} This research received no external funding.

\section{Supplementary Materials}
The following are available online, All XYZ atomic coordinates optimized of the Be$6$B$_{11}^{-}$ cluster at the PBE0-D3/def2-TZVP/Freq. (S1). Movies of the molecular dynamics simulation of Be$6$B$_{11}^{-}$ cluster at 1600 K, 2000 K, and 2500 K are shown in the supplementary material, MD1600.mp4, MD2000.mp4, MD2500.mp4, respectively. The Boltzmann Optics Full Adder (BOFA) Python code supporting the findings of this study is available from the corresponding author upon reasonable request. (sollebac@gmail.com)
\bibliography{bibliografia}

\begin{thebibliography}{146}
\expandafter\ifx\csname natexlab\endcsname\relax\def\natexlab#1{#1}\fi
\expandafter\ifx\csname bibnamefont\endcsname\relax
  \def\bibnamefont#1{#1}\fi
\expandafter\ifx\csname bibfnamefont\endcsname\relax
  \def\bibfnamefont#1{#1}\fi
\expandafter\ifx\csname citenamefont\endcsname\relax
  \def\citenamefont#1{#1}\fi
\expandafter\ifx\csname url\endcsname\relax
  \def\url#1{\texttt{#1}}\fi
\expandafter\ifx\csname urlprefix\endcsname\relax\def\urlprefix{URL }\fi
\providecommand{\bibinfo}[2]{#2}
\providecommand{\eprint}[2][]{\url{#2}}

\bibitem[{\citenamefont{Jalife et~al.}(2016)\citenamefont{Jalife, Liu, Pan,
  Cabellos, Osorio, Lu, Heine, Donald, and Merino}}]{Jalife}
\bibinfo{author}{\bibfnamefont{S.}~\bibnamefont{Jalife}},
  \bibinfo{author}{\bibfnamefont{L.}~\bibnamefont{Liu}},
  \bibinfo{author}{\bibfnamefont{S.}~\bibnamefont{Pan}},
  \bibinfo{author}{\bibfnamefont{J.~L.} \bibnamefont{Cabellos}},
  \bibinfo{author}{\bibfnamefont{E.}~\bibnamefont{Osorio}},
  \bibinfo{author}{\bibfnamefont{C.}~\bibnamefont{Lu}},
  \bibinfo{author}{\bibfnamefont{T.}~\bibnamefont{Heine}},
  \bibinfo{author}{\bibfnamefont{K.~J.} \bibnamefont{Donald}},
  \bibnamefont{and} \bibinfo{author}{\bibfnamefont{G.}~\bibnamefont{Merino}},
  \bibinfo{journal}{Nanoscale} \textbf{\bibinfo{volume}{8}},
  \bibinfo{pages}{17639} (\bibinfo{year}{2016}), \eprint{10.1039/C6NR06383G},
  \urlprefix\url{http://dx.doi.org/10.1039/C6NR06383G}.

\bibitem[{\citenamefont{Zhai et~al.}(2014)\citenamefont{Zhai, Zhao, Li, Chen,
  Bai, Hu, Piazza, Tian, Lu, Wu et~al.}}]{Wang}
\bibinfo{author}{\bibfnamefont{H.-J.} \bibnamefont{Zhai}},
  \bibinfo{author}{\bibfnamefont{Y.-F.} \bibnamefont{Zhao}},
  \bibinfo{author}{\bibfnamefont{W.-L.} \bibnamefont{Li}},
  \bibinfo{author}{\bibfnamefont{Q.}~\bibnamefont{Chen}},
  \bibinfo{author}{\bibfnamefont{H.}~\bibnamefont{Bai}},
  \bibinfo{author}{\bibfnamefont{H.-S.} \bibnamefont{Hu}},
  \bibinfo{author}{\bibfnamefont{Z.~A.} \bibnamefont{Piazza}},
  \bibinfo{author}{\bibfnamefont{W.-J.} \bibnamefont{Tian}},
  \bibinfo{author}{\bibfnamefont{H.-G.} \bibnamefont{Lu}},
  \bibinfo{author}{\bibfnamefont{Y.-B.} \bibnamefont{Wu}},
  \bibnamefont{et~al.}, \bibinfo{journal}{Nature Chemistry} pp.
  \bibinfo{pages}{727--731} (\bibinfo{year}{2014}),
  \urlprefix\url{https://doi.org/10.1038/nchem.1999}.

\bibitem[{\citenamefont{Sun et~al.}(2020)\citenamefont{Sun, Kang, Chen, Kuang,
  Ding, and Lu}}]{Sun}
\bibinfo{author}{\bibfnamefont{W.}~\bibnamefont{Sun}},
  \bibinfo{author}{\bibfnamefont{D.}~\bibnamefont{Kang}},
  \bibinfo{author}{\bibfnamefont{B.}~\bibnamefont{Chen}},
  \bibinfo{author}{\bibfnamefont{X.}~\bibnamefont{Kuang}},
  \bibinfo{author}{\bibfnamefont{K.}~\bibnamefont{Ding}}, \bibnamefont{and}
  \bibinfo{author}{\bibfnamefont{C.}~\bibnamefont{Lu}}, \bibinfo{journal}{The
  Journal of Physical Chemistry A} \textbf{\bibinfo{volume}{1}},
  \bibinfo{pages}{123} (\bibinfo{year}{2020}), \bibinfo{note}{pMID: 33085487},
  \eprint{doi:10.1021/acs.jpca.0c05197},
  \urlprefix\url{https://doi.org/10.1021/acs.jpca.0c05197}.

\bibitem[{\citenamefont{Jian et~al.}(2019)\citenamefont{Jian, Chen, Li,
  Boldyrev, Li, and Wang}}]{Jian}
\bibinfo{author}{\bibfnamefont{T.}~\bibnamefont{Jian}},
  \bibinfo{author}{\bibfnamefont{X.}~\bibnamefont{Chen}},
  \bibinfo{author}{\bibfnamefont{S.-D.} \bibnamefont{Li}},
  \bibinfo{author}{\bibfnamefont{A.~I.} \bibnamefont{Boldyrev}},
  \bibinfo{author}{\bibfnamefont{J.}~\bibnamefont{Li}}, \bibnamefont{and}
  \bibinfo{author}{\bibfnamefont{L.-S.} \bibnamefont{Wang}},
  \bibinfo{journal}{Chem. Soc. Rev.} \textbf{\bibinfo{volume}{48}},
  \bibinfo{pages}{3550} (\bibinfo{year}{2019}),
  \eprint{http://dx.doi.org/10.1039/C9CS00233B},
  \urlprefix\url{http://dx.doi.org/10.1039/C9CS00233B}.

\bibitem[{\citenamefont{Chen et~al.}(2019{\natexlab{a}})\citenamefont{Chen, Li,
  Chen, Li, and Wang}}]{Teng-Teng}
\bibinfo{author}{\bibfnamefont{T.-T.} \bibnamefont{Chen}},
  \bibinfo{author}{\bibfnamefont{W.-L.} \bibnamefont{Li}},
  \bibinfo{author}{\bibfnamefont{W.-J.} \bibnamefont{Chen}},
  \bibinfo{author}{\bibfnamefont{J.}~\bibnamefont{Li}}, \bibnamefont{and}
  \bibinfo{author}{\bibfnamefont{L.-S.} \bibnamefont{Wang}},
  \bibinfo{journal}{Chem. Commun.} \textbf{\bibinfo{volume}{55}},
  \bibinfo{pages}{7864} (\bibinfo{year}{2019}{\natexlab{a}}),
  \urlprefix\url{http://dx.doi.org/10.1039/C9CC03807H}.

\bibitem[{\citenamefont{Mart\'inez-Guajardo
  et~al.}(2015)\citenamefont{Mart\'inez-Guajardo, Cabellos, D\'iaz-Celaya, Pan,
  Islas, Chattaraj, Heine, and Merino}}]{Gerardo}
\bibinfo{author}{\bibfnamefont{G.}~\bibnamefont{Mart\'inez-Guajardo}},
  \bibinfo{author}{\bibfnamefont{J.~L.} \bibnamefont{Cabellos}},
  \bibinfo{author}{\bibfnamefont{A.}~\bibnamefont{D\'iaz-Celaya}},
  \bibinfo{author}{\bibfnamefont{S.}~\bibnamefont{Pan}},
  \bibinfo{author}{\bibfnamefont{R.}~\bibnamefont{Islas}},
  \bibinfo{author}{\bibfnamefont{P.~K.} \bibnamefont{Chattaraj}},
  \bibinfo{author}{\bibfnamefont{T.}~\bibnamefont{Heine}}, \bibnamefont{and}
  \bibinfo{author}{\bibfnamefont{G.}~\bibnamefont{Merino}},
  \bibinfo{journal}{Sci. Report} \textbf{\bibinfo{volume}{22}},
  \bibinfo{pages}{11287} (\bibinfo{year}{2015}),
  \urlprefix\url{https://www.nature.com/articles/srep11287}.

\bibitem[{\citenamefont{Dongliang et~al.}(2019)\citenamefont{Dongliang, Weiguo,
  Hongxiao, Cheng, and Xiaoyu}}]{Dongliang}
\bibinfo{author}{\bibfnamefont{K.}~\bibnamefont{Dongliang}},
  \bibinfo{author}{\bibfnamefont{S.}~\bibnamefont{Weiguo}},
  \bibinfo{author}{\bibfnamefont{S.}~\bibnamefont{Hongxiao}},
  \bibinfo{author}{\bibfnamefont{L.}~\bibnamefont{Cheng}}, \bibnamefont{and}
  \bibinfo{author}{\bibnamefont{Xiaoyu}}, \bibinfo{journal}{Sci. Rep.}
  \textbf{\bibinfo{volume}{9}}, \bibinfo{pages}{14367} (\bibinfo{year}{2019}),
  \urlprefix\url{https://www.nature.com/articles/s41598-019-50905-7}.

\bibitem[{\citenamefont{Li et~al.}(2018)\citenamefont{Li, Du, Wang, Lu, and
  Chen}}]{Peifang}
\bibinfo{author}{\bibfnamefont{P.}~\bibnamefont{Li}},
  \bibinfo{author}{\bibfnamefont{X.}~\bibnamefont{Du}},
  \bibinfo{author}{\bibfnamefont{J.~J.} \bibnamefont{Wang}},
  \bibinfo{author}{\bibfnamefont{C.}~\bibnamefont{Lu}}, \bibnamefont{and}
  \bibinfo{author}{\bibfnamefont{H.}~\bibnamefont{Chen}}, \bibinfo{journal}{The
  Journal of Physical Chemistry C} \textbf{\bibinfo{volume}{122}},
  \bibinfo{pages}{20000} (\bibinfo{year}{2018}),
  \eprint{https://doi.org/10.1021/acs.jpcc.8b05759},
  \urlprefix\url{https://doi.org/10.1021/acs.jpcc.8b05759}.

\bibitem[{\citenamefont{Grande-Aztatzi
  et~al.}(2014)\citenamefont{Grande-Aztatzi, Martínez-Alanis, Cabellos,
  Osorio, Martínez, and Merino}}]{Grande-Aztatzi}
\bibinfo{author}{\bibfnamefont{R.}~\bibnamefont{Grande-Aztatzi}},
  \bibinfo{author}{\bibfnamefont{P.~R.} \bibnamefont{Martínez-Alanis}},
  \bibinfo{author}{\bibfnamefont{J.~L.} \bibnamefont{Cabellos}},
  \bibinfo{author}{\bibfnamefont{E.}~\bibnamefont{Osorio}},
  \bibinfo{author}{\bibfnamefont{A.}~\bibnamefont{Martínez}},
  \bibnamefont{and} \bibinfo{author}{\bibfnamefont{G.}~\bibnamefont{Merino}},
  \bibinfo{journal}{Journal of Computational Chemistry}
  \textbf{\bibinfo{volume}{35}}, \bibinfo{pages}{2288} (\bibinfo{year}{2014}),
  \eprint{https://onlinelibrary.wiley.com/doi/pdf/10.1002/jcc.23748},
  \urlprefix\url{https://onlinelibrary.wiley.com/doi/abs/10.1002/jcc.23748}.

\bibitem[{\citenamefont{Dong et~al.}(2018)\citenamefont{Dong, Jalife,
  V\'asquez-Espinal, Ravell, Pan, Cabellos, Liang, Cui, and Merino}}]{Dong}
\bibinfo{author}{\bibfnamefont{X.}~\bibnamefont{Dong}},
  \bibinfo{author}{\bibfnamefont{S.}~\bibnamefont{Jalife}},
  \bibinfo{author}{\bibfnamefont{A.}~\bibnamefont{V\'asquez-Espinal}},
  \bibinfo{author}{\bibfnamefont{E.}~\bibnamefont{Ravell}},
  \bibinfo{author}{\bibfnamefont{S.}~\bibnamefont{Pan}},
  \bibinfo{author}{\bibfnamefont{J.~L.} \bibnamefont{Cabellos}},
  \bibinfo{author}{\bibfnamefont{W.-y.} \bibnamefont{Liang}},
  \bibinfo{author}{\bibfnamefont{Z.-h.} \bibnamefont{Cui}}, \bibnamefont{and}
  \bibinfo{author}{\bibfnamefont{G.}~\bibnamefont{Merino}},
  \bibinfo{journal}{Angewandte Chemie International Edition}
  \textbf{\bibinfo{volume}{57}}, \bibinfo{pages}{4627} (\bibinfo{year}{2018}),
  \eprint{https://onlinelibrary.wiley.com/doi/pdf/10.1002/anie.201800976},
  \urlprefix\url{https://onlinelibrary.wiley.com/doi/abs/10.1002/anie.201800976}.

\bibitem[{\citenamefont{Brothers}(2008)}]{Brothers}
\bibinfo{author}{\bibfnamefont{P.~J.} \bibnamefont{Brothers}},
  \bibinfo{journal}{Chem. Commun.} pp. \bibinfo{pages}{2090--2102}
  (\bibinfo{year}{2008}), \urlprefix\url{http://dx.doi.org/10.1039/B714894A}.

\bibitem[{\citenamefont{Axtell et~al.}(2018)\citenamefont{Axtell, Saleh, Qian,
  Wixtrom, and Spokoyny}}]{Axtell}
\bibinfo{author}{\bibfnamefont{J.~C.} \bibnamefont{Axtell}},
  \bibinfo{author}{\bibfnamefont{L.~M.~A.} \bibnamefont{Saleh}},
  \bibinfo{author}{\bibfnamefont{E.~A.} \bibnamefont{Qian}},
  \bibinfo{author}{\bibfnamefont{A.~I.} \bibnamefont{Wixtrom}},
  \bibnamefont{and} \bibinfo{author}{\bibfnamefont{A.~M.}
  \bibnamefont{Spokoyny}}, \bibinfo{journal}{Inorganic Chemistry}
  \textbf{\bibinfo{volume}{57}}, \bibinfo{pages}{2333} (\bibinfo{year}{2018}),
  \bibinfo{note}{pMID: 29465227},
  \eprint{https://doi.org/10.1021/acs.inorgchem.7b02912},
  \urlprefix\url{https://doi.org/10.1021/acs.inorgchem.7b02912}.

\bibitem[{\citenamefont{Piazza et~al.}(2017)\citenamefont{Piazza, Hu, Li, Zhao,
  Li, and Wang}}]{Lai-Sheng5}
\bibinfo{author}{\bibfnamefont{Z.~A.} \bibnamefont{Piazza}},
  \bibinfo{author}{\bibfnamefont{H.-S.} \bibnamefont{Hu}},
  \bibinfo{author}{\bibfnamefont{W.-L.} \bibnamefont{Li}},
  \bibinfo{author}{\bibfnamefont{Y.-F.} \bibnamefont{Zhao}},
  \bibinfo{author}{\bibfnamefont{J.}~\bibnamefont{Li}}, \bibnamefont{and}
  \bibinfo{author}{\bibfnamefont{L.-S.} \bibnamefont{Wang}},
  \bibinfo{journal}{Nature Reviews Chemistry} \textbf{\bibinfo{volume}{1}},
  \bibinfo{pages}{0071} (\bibinfo{year}{2017}),
  \eprint{https://doi.org/10.1038/s41570-017-0071},
  \urlprefix\url{https://doi.org/10.1038/s41570-017-0071}.

\bibitem[{\citenamefont{Mannix et~al.}(2018)\citenamefont{Mannix, Zhang,
  Guisinger, Yakobson, and Hersam}}]{Mannix}
\bibinfo{author}{\bibfnamefont{A.~J.} \bibnamefont{Mannix}},
  \bibinfo{author}{\bibfnamefont{Z.}~\bibnamefont{Zhang}},
  \bibinfo{author}{\bibfnamefont{N.~P.} \bibnamefont{Guisinger}},
  \bibinfo{author}{\bibfnamefont{B.~I.} \bibnamefont{Yakobson}},
  \bibnamefont{and} \bibinfo{author}{\bibfnamefont{M.~C.}
  \bibnamefont{Hersam}}, \bibinfo{journal}{Nature Nanotechnology}
  \textbf{\bibinfo{volume}{13}}, \bibinfo{pages}{444} (\bibinfo{year}{2018}),
  \urlprefix\url{https://doi.org/10.1038/s41565-018-0157-4}.

\bibitem[{\citenamefont{Vast et~al.}(1997)\citenamefont{Vast, Baroni, Zerah,
  Besson, Polian, Grimsditch, and Chervin}}]{Vast}
\bibinfo{author}{\bibfnamefont{N.}~\bibnamefont{Vast}},
  \bibinfo{author}{\bibfnamefont{S.}~\bibnamefont{Baroni}},
  \bibinfo{author}{\bibfnamefont{G.}~\bibnamefont{Zerah}},
  \bibinfo{author}{\bibfnamefont{J.~M.} \bibnamefont{Besson}},
  \bibinfo{author}{\bibfnamefont{A.}~\bibnamefont{Polian}},
  \bibinfo{author}{\bibfnamefont{M.}~\bibnamefont{Grimsditch}},
  \bibnamefont{and} \bibinfo{author}{\bibfnamefont{J.~C.}
  \bibnamefont{Chervin}}, \bibinfo{journal}{Phys. Rev. Lett.}
  \textbf{\bibinfo{volume}{78}}, \bibinfo{pages}{693} (\bibinfo{year}{1997}),
  \urlprefix\url{https://link.aps.org/doi/10.1103/PhysRevLett.78.693}.

\bibitem[{\citenamefont{Fujimori et~al.}(1999)\citenamefont{Fujimori, Nakata,
  Nakayama, Nishibori, Kimura, Takata, and Sakata}}]{Peculiar}
\bibinfo{author}{\bibfnamefont{M.}~\bibnamefont{Fujimori}},
  \bibinfo{author}{\bibfnamefont{T.}~\bibnamefont{Nakata}},
  \bibinfo{author}{\bibfnamefont{T.}~\bibnamefont{Nakayama}},
  \bibinfo{author}{\bibfnamefont{E.}~\bibnamefont{Nishibori}},
  \bibinfo{author}{\bibfnamefont{K.}~\bibnamefont{Kimura}},
  \bibinfo{author}{\bibfnamefont{M.}~\bibnamefont{Takata}}, \bibnamefont{and}
  \bibinfo{author}{\bibfnamefont{M.}~\bibnamefont{Sakata}},
  \bibinfo{journal}{Phys. Rev. Lett.} \textbf{\bibinfo{volume}{82}},
  \bibinfo{pages}{4452} (\bibinfo{year}{1999}),
  \urlprefix\url{https://link.aps.org/doi/10.1103/PhysRevLett.82.4452}.

\bibitem[{\citenamefont{Shi et~al.}(2020)\citenamefont{Shi, Kuang, and
  Lu}}]{Hongxiao}
\bibinfo{author}{\bibfnamefont{H.}~\bibnamefont{Shi}},
  \bibinfo{author}{\bibfnamefont{X.}~\bibnamefont{Kuang}}, \bibnamefont{and}
  \bibinfo{author}{\bibfnamefont{C.}~\bibnamefont{Lu}},
  \bibinfo{journal}{Scientific Reports} \textbf{\bibinfo{volume}{10}},
  \bibinfo{pages}{1642} (\bibinfo{year}{2020}),
  \urlprefix\url{https://doi.org/10.1038/s41598-020-57769-2}.

\bibitem[{\citenamefont{Piazza et~al.}(2014)\citenamefont{Piazza, Hu, Li, Zhao,
  Li, and Wang}}]{Lai-Sheng}
\bibinfo{author}{\bibfnamefont{Z.~A.} \bibnamefont{Piazza}},
  \bibinfo{author}{\bibfnamefont{H.-S.} \bibnamefont{Hu}},
  \bibinfo{author}{\bibfnamefont{W.-L.} \bibnamefont{Li}},
  \bibinfo{author}{\bibfnamefont{Y.-F.} \bibnamefont{Zhao}},
  \bibinfo{author}{\bibfnamefont{J.}~\bibnamefont{Li}}, \bibnamefont{and}
  \bibinfo{author}{\bibfnamefont{L.-S.} \bibnamefont{Wang}},
  \bibinfo{journal}{Nature Communications} \textbf{\bibinfo{volume}{1}},
  \bibinfo{pages}{3113} (\bibinfo{year}{2014}),
  \eprint{https://doi.org/10.1038/ncomms4113},
  \urlprefix\url{https://doi.org/10.1038/ncomms4113}.

\bibitem[{\citenamefont{Li et~al.}(2014)\citenamefont{Li, Zhao, Hu, Li, and
  Wang}}]{Lai-Sheng2}
\bibinfo{author}{\bibfnamefont{W.-L.} \bibnamefont{Li}},
  \bibinfo{author}{\bibfnamefont{Y.-F.} \bibnamefont{Zhao}},
  \bibinfo{author}{\bibfnamefont{H.-S.} \bibnamefont{Hu}},
  \bibinfo{author}{\bibfnamefont{J.}~\bibnamefont{Li}}, \bibnamefont{and}
  \bibinfo{author}{\bibfnamefont{L.-S.} \bibnamefont{Wang}},
  \bibinfo{journal}{Angewandte Chemie International Edition}
  \textbf{\bibinfo{volume}{53}}, \bibinfo{pages}{5540} (\bibinfo{year}{2014}),
  \eprint{https://onlinelibrary.wiley.com/doi/pdf/10.1002/anie.201402488},
  \urlprefix\url{https://onlinelibrary.wiley.com/doi/abs/10.1002/anie.201402488}.

\bibitem[{\citenamefont{Chen et~al.}(2019{\natexlab{b}})\citenamefont{Chen,
  Chen, Li, Zhao, Chen, Zhai, Li, and Wang}}]{Lai-Sheng3}
\bibinfo{author}{\bibfnamefont{Q.}~\bibnamefont{Chen}},
  \bibinfo{author}{\bibfnamefont{T.-T.} \bibnamefont{Chen}},
  \bibinfo{author}{\bibfnamefont{H.-R.} \bibnamefont{Li}},
  \bibinfo{author}{\bibfnamefont{X.-Y.} \bibnamefont{Zhao}},
  \bibinfo{author}{\bibfnamefont{W.-J.} \bibnamefont{Chen}},
  \bibinfo{author}{\bibfnamefont{H.-J.} \bibnamefont{Zhai}},
  \bibinfo{author}{\bibfnamefont{S.-D.} \bibnamefont{Li}}, \bibnamefont{and}
  \bibinfo{author}{\bibfnamefont{L.-S.} \bibnamefont{Wang}},
  \bibinfo{journal}{Nanoscale} \textbf{\bibinfo{volume}{11}},
  \bibinfo{pages}{9698} (\bibinfo{year}{2019}{\natexlab{b}}),
  \urlprefix\url{http://dx.doi.org/10.1039/C9NR01524H}.

\bibitem[{\citenamefont{Kiran et~al.}(2005)\citenamefont{Kiran, Bulusu, Zhai,
  Yoo, Zeng, and Wang}}]{Lai-Sheng4}
\bibinfo{author}{\bibfnamefont{B.}~\bibnamefont{Kiran}},
  \bibinfo{author}{\bibfnamefont{S.}~\bibnamefont{Bulusu}},
  \bibinfo{author}{\bibfnamefont{H.-J.} \bibnamefont{Zhai}},
  \bibinfo{author}{\bibfnamefont{S.}~\bibnamefont{Yoo}},
  \bibinfo{author}{\bibfnamefont{X.~C.} \bibnamefont{Zeng}}, \bibnamefont{and}
  \bibinfo{author}{\bibfnamefont{L.-S.} \bibnamefont{Wang}},
  \bibinfo{journal}{Proceedings of the National Academy of Sciences}
  \textbf{\bibinfo{volume}{102}}, \bibinfo{pages}{961} (\bibinfo{year}{2005}),
  ISSN \bibinfo{issn}{0027-8424},
  \eprint{https://www.pnas.org/content/102/4/961.full.pdf},
  \urlprefix\url{https://www.pnas.org/content/102/4/961}.

\bibitem[{\citenamefont{Chen et~al.}(2015)\citenamefont{Chen, Li, Zhao, Zhang,
  Hu, Bai, Li, Tian, Lu, Zhai et~al.}}]{Qiang}
\bibinfo{author}{\bibfnamefont{Q.}~\bibnamefont{Chen}},
  \bibinfo{author}{\bibfnamefont{W.-L.} \bibnamefont{Li}},
  \bibinfo{author}{\bibfnamefont{Y.-F.} \bibnamefont{Zhao}},
  \bibinfo{author}{\bibfnamefont{S.-Y.} \bibnamefont{Zhang}},
  \bibinfo{author}{\bibfnamefont{H.-S.} \bibnamefont{Hu}},
  \bibinfo{author}{\bibfnamefont{H.}~\bibnamefont{Bai}},
  \bibinfo{author}{\bibfnamefont{H.-R.} \bibnamefont{Li}},
  \bibinfo{author}{\bibfnamefont{W.-J.} \bibnamefont{Tian}},
  \bibinfo{author}{\bibfnamefont{H.-G.} \bibnamefont{Lu}},
  \bibinfo{author}{\bibfnamefont{H.-J.} \bibnamefont{Zhai}},
  \bibnamefont{et~al.}, \bibinfo{journal}{ACS Nano}
  \textbf{\bibinfo{volume}{9}}, \bibinfo{pages}{754} (\bibinfo{year}{2015}),
  \bibinfo{note}{pMID: 25517915}, \eprint{https://doi.org/10.1021/nn506262c},
  \urlprefix\url{https://doi.org/10.1021/nn506262c}.

\bibitem[{\citenamefont{Wang et~al.}(2016)\citenamefont{Wang, Zhao, Li, Jian,
  Chen, You, Ou, Zhao, Zhai, Li et~al.}}]{Ying-Jin-Wang}
\bibinfo{author}{\bibfnamefont{Y.-J.} \bibnamefont{Wang}},
  \bibinfo{author}{\bibfnamefont{Y.-F.} \bibnamefont{Zhao}},
  \bibinfo{author}{\bibfnamefont{W.-L.} \bibnamefont{Li}},
  \bibinfo{author}{\bibfnamefont{T.}~\bibnamefont{Jian}},
  \bibinfo{author}{\bibfnamefont{Q.}~\bibnamefont{Chen}},
  \bibinfo{author}{\bibfnamefont{X.-R.} \bibnamefont{You}},
  \bibinfo{author}{\bibfnamefont{T.}~\bibnamefont{Ou}},
  \bibinfo{author}{\bibfnamefont{X.-Y.} \bibnamefont{Zhao}},
  \bibinfo{author}{\bibfnamefont{H.-J.} \bibnamefont{Zhai}},
  \bibinfo{author}{\bibfnamefont{S.-D.} \bibnamefont{Li}},
  \bibnamefont{et~al.}, \bibinfo{journal}{The Journal of Chemical Physics}
  \textbf{\bibinfo{volume}{144}}, \bibinfo{pages}{064307}
  (\bibinfo{year}{2016}), \eprint{https://doi.org/10.1063/1.4941380},
  \urlprefix\url{https://doi.org/10.1063/1.4941380}.

\bibitem[{\citenamefont{Lv et~al.}(2015)\citenamefont{Lv, Wang, Zhang, Lin,
  Zhao, and Ma}}]{Lv}
\bibinfo{author}{\bibfnamefont{J.}~\bibnamefont{Lv}},
  \bibinfo{author}{\bibfnamefont{Y.}~\bibnamefont{Wang}},
  \bibinfo{author}{\bibfnamefont{L.}~\bibnamefont{Zhang}},
  \bibinfo{author}{\bibfnamefont{H.}~\bibnamefont{Lin}},
  \bibinfo{author}{\bibfnamefont{J.}~\bibnamefont{Zhao}}, \bibnamefont{and}
  \bibinfo{author}{\bibfnamefont{Y.}~\bibnamefont{Ma}},
  \bibinfo{journal}{Nanoscale} \textbf{\bibinfo{volume}{7}},
  \bibinfo{pages}{10482} (\bibinfo{year}{2015}),
  \urlprefix\url{http://dx.doi.org/10.1039/C5NR01659B}.

\bibitem[{\citenamefont{Feng et~al.}(2020)\citenamefont{Feng, Guo, Li, and
  Zhai}}]{Feng}
\bibinfo{author}{\bibfnamefont{L.-Y.} \bibnamefont{Feng}},
  \bibinfo{author}{\bibfnamefont{J.-C.} \bibnamefont{Guo}},
  \bibinfo{author}{\bibfnamefont{P.-F.} \bibnamefont{Li}}, \bibnamefont{and}
  \bibinfo{author}{\bibfnamefont{H.-J.} \bibnamefont{Zhai}},
  \bibinfo{journal}{Chemistry - An Asian Journal}
  \textbf{\bibinfo{volume}{15}}, \bibinfo{pages}{1094} (\bibinfo{year}{2020}),
  \eprint{https://onlinelibrary.wiley.com/doi/pdf/10.1002/asia.201901640},
  \urlprefix\url{https://onlinelibrary.wiley.com/doi/abs/10.1002/asia.201901640}.

\bibitem[{\citenamefont{Guo et~al.}(2017)\citenamefont{Guo, Feng, Wang, Jalife,
  Vásquez-Espinal, Cabellos, Pan, Merino, and Zhai}}]{Guo}
\bibinfo{author}{\bibfnamefont{J.-C.} \bibnamefont{Guo}},
  \bibinfo{author}{\bibfnamefont{L.-Y.} \bibnamefont{Feng}},
  \bibinfo{author}{\bibfnamefont{Y.-J.} \bibnamefont{Wang}},
  \bibinfo{author}{\bibfnamefont{S.}~\bibnamefont{Jalife}},
  \bibinfo{author}{\bibfnamefont{A.}~\bibnamefont{Vásquez-Espinal}},
  \bibinfo{author}{\bibfnamefont{J.~L.} \bibnamefont{Cabellos}},
  \bibinfo{author}{\bibfnamefont{S.}~\bibnamefont{Pan}},
  \bibinfo{author}{\bibfnamefont{G.}~\bibnamefont{Merino}}, \bibnamefont{and}
  \bibinfo{author}{\bibfnamefont{H.-J.} \bibnamefont{Zhai}},
  \bibinfo{journal}{Angewandte Chemie International Edition}
  \textbf{\bibinfo{volume}{56}}, \bibinfo{pages}{10174} (\bibinfo{year}{2017}),
  \eprint{https://onlinelibrary.wiley.com/doi/pdf/10.1002/anie.201703979},
  \urlprefix\url{https://onlinelibrary.wiley.com/doi/abs/10.1002/anie.201703979}.

\bibitem[{\citenamefont{Mannix et~al.}(2015)\citenamefont{Mannix, Zhou, Kiraly,
  Wood, Alducin, Myers, Liu, Fisher, Santiago, Guest et~al.}}]{Mannix1513}
\bibinfo{author}{\bibfnamefont{A.~J.} \bibnamefont{Mannix}},
  \bibinfo{author}{\bibfnamefont{X.-F.} \bibnamefont{Zhou}},
  \bibinfo{author}{\bibfnamefont{B.}~\bibnamefont{Kiraly}},
  \bibinfo{author}{\bibfnamefont{J.~D.} \bibnamefont{Wood}},
  \bibinfo{author}{\bibfnamefont{D.}~\bibnamefont{Alducin}},
  \bibinfo{author}{\bibfnamefont{B.~D.} \bibnamefont{Myers}},
  \bibinfo{author}{\bibfnamefont{X.}~\bibnamefont{Liu}},
  \bibinfo{author}{\bibfnamefont{B.~L.} \bibnamefont{Fisher}},
  \bibinfo{author}{\bibfnamefont{U.}~\bibnamefont{Santiago}},
  \bibinfo{author}{\bibfnamefont{J.~R.} \bibnamefont{Guest}},
  \bibnamefont{et~al.}, \bibinfo{journal}{Science}
  \textbf{\bibinfo{volume}{350}}, \bibinfo{pages}{1513} (\bibinfo{year}{2015}),
  ISSN \bibinfo{issn}{0036-8075},
  \eprint{https://science.sciencemag.org/content/350/6267/1513.full.pdf},
  \urlprefix\url{https://science.sciencemag.org/content/350/6267/1513}.

\bibitem[{\citenamefont{Jimenez-Halla et~al.}(2010)\citenamefont{Jimenez-Halla,
  Islas, Heine, and Merino}}]{Jimenez-Halla}
\bibinfo{author}{\bibfnamefont{J.}~\bibnamefont{Jimenez-Halla}},
  \bibinfo{author}{\bibfnamefont{R.}~\bibnamefont{Islas}},
  \bibinfo{author}{\bibfnamefont{T.}~\bibnamefont{Heine}}, \bibnamefont{and}
  \bibinfo{author}{\bibfnamefont{G.}~\bibnamefont{Merino}},
  \bibinfo{journal}{Angewandte Chemie International Edition}
  \textbf{\bibinfo{volume}{49}}, \bibinfo{pages}{5668} (\bibinfo{year}{2010}),
  \eprint{https://onlinelibrary.wiley.com/doi/pdf/10.1002/anie.201001275},
  \urlprefix\url{https://onlinelibrary.wiley.com/doi/abs/10.1002/anie.201001275}.

\bibitem[{\citenamefont{Barney et~al.}(1958)\citenamefont{Barney, Sehmel, and
  Seymour}}]{Barney}
\bibinfo{author}{\bibfnamefont{W.~K.} \bibnamefont{Barney}},
  \bibinfo{author}{\bibfnamefont{G.~A.} \bibnamefont{Sehmel}},
  \bibnamefont{and} \bibinfo{author}{\bibfnamefont{W.~E.}
  \bibnamefont{Seymour}}, \bibinfo{journal}{Nuclear Science and Engineering}
  \textbf{\bibinfo{volume}{4}}, \bibinfo{pages}{439} (\bibinfo{year}{1958}),
  \eprint{https://doi.org/10.13182/NSE58-A25540},
  \urlprefix\url{https://doi.org/10.13182/NSE58-A25540}.

\bibitem[{\citenamefont{Le\'snikowski}(2016)}]{Lesnikowski}
\bibinfo{author}{\bibfnamefont{Z.~J.} \bibnamefont{Le\'snikowski}},
  \bibinfo{journal}{Journal of Medicinal Chemistry}
  \textbf{\bibinfo{volume}{59}}, \bibinfo{pages}{7738} (\bibinfo{year}{2016}),
  \bibinfo{note}{pMID: 27124656},
  \eprint{https://doi.org/10.1021/acs.jmedchem.5b01932},
  \urlprefix\url{https://doi.org/10.1021/acs.jmedchem.5b01932}.

\bibitem[{\citenamefont{Ali et~al.}(2020)\citenamefont{Ali, S~Hosmane, and
  Zhu}}]{Ali}
\bibinfo{author}{\bibfnamefont{F.}~\bibnamefont{Ali}},
  \bibinfo{author}{\bibfnamefont{N.}~\bibnamefont{S~Hosmane}},
  \bibnamefont{and} \bibinfo{author}{\bibfnamefont{Y.}~\bibnamefont{Zhu}},
  \bibinfo{journal}{Molecules} \textbf{\bibinfo{volume}{25}},
  \bibinfo{pages}{828} (\bibinfo{year}{2020}),
  \eprint{https://pubmed.ncbi.nlm.nih.gov/32070043},
  \urlprefix\url{https://pubmed.ncbi.nlm.nih.gov/32070043}.

\bibitem[{\citenamefont{Lu et~al.}(2016)\citenamefont{Lu, Wang, Jiang, liu, and
  Huang}}]{Lu}
\bibinfo{author}{\bibfnamefont{T.}~\bibnamefont{Lu}},
  \bibinfo{author}{\bibfnamefont{L.}~\bibnamefont{Wang}},
  \bibinfo{author}{\bibfnamefont{Y.}~\bibnamefont{Jiang}},
  \bibinfo{author}{\bibfnamefont{Q.}~\bibnamefont{liu}}, \bibnamefont{and}
  \bibinfo{author}{\bibfnamefont{C.}~\bibnamefont{Huang}}, \bibinfo{journal}{J.
  Mater. Chem. B} \textbf{\bibinfo{volume}{4}}, \bibinfo{pages}{6103}
  (\bibinfo{year}{2016}), \urlprefix\url{http://dx.doi.org/10.1039/C6TB01481J}.

\bibitem[{\citenamefont{O\~na et~al.}(2015)\citenamefont{O\~na, Torres-Vega,
  Torre, Lain, Alcoba, V\'asquez-Espinal, and Tiznado}}]{Ofelia}
\bibinfo{author}{\bibfnamefont{O.~B.} \bibnamefont{O\~na}},
  \bibinfo{author}{\bibfnamefont{J.~J.} \bibnamefont{Torres-Vega}},
  \bibinfo{author}{\bibfnamefont{A.}~\bibnamefont{Torre}},
  \bibinfo{author}{\bibfnamefont{L.}~\bibnamefont{Lain}},
  \bibinfo{author}{\bibfnamefont{D.~R.} \bibnamefont{Alcoba}},
  \bibinfo{author}{\bibfnamefont{A.}~\bibnamefont{V\'asquez-Espinal}},
  \bibnamefont{and} \bibinfo{author}{\bibfnamefont{W.}~\bibnamefont{Tiznado}},
  \bibinfo{journal}{Theoretical Chemistry Accounts}
  \textbf{\bibinfo{volume}{134}}, \bibinfo{pages}{28} (\bibinfo{year}{2015}),
  \eprint{https://doi.org/10.1007/s00214-015-1627-5},
  \urlprefix\url{https://doi.org/10.1007/s00214-015-1627-5}.

\bibitem[{\citenamefont{Alexandrova et~al.}(2006)\citenamefont{Alexandrova,
  Boldyrev, Zhai, and Wang}}]{Alexandrova2}
\bibinfo{author}{\bibfnamefont{A.~N.} \bibnamefont{Alexandrova}},
  \bibinfo{author}{\bibfnamefont{A.~I.} \bibnamefont{Boldyrev}},
  \bibinfo{author}{\bibfnamefont{H.-J.} \bibnamefont{Zhai}}, \bibnamefont{and}
  \bibinfo{author}{\bibfnamefont{L.-S.} \bibnamefont{Wang}},
  \bibinfo{journal}{Coordination Chemistry Reviews}
  \textbf{\bibinfo{volume}{250}}, \bibinfo{pages}{2811 }
  (\bibinfo{year}{2006}), ISSN \bibinfo{issn}{0010-8545}, \bibinfo{note}{18th
  Main Group Chemistry},
  \urlprefix\url{http://www.sciencedirect.com/science/article/pii/S0010854506001408}.

\bibitem[{\citenamefont{Zubarev and Boldyrev}(2007)}]{Boldyrev}
\bibinfo{author}{\bibfnamefont{D.~Y.} \bibnamefont{Zubarev}} \bibnamefont{and}
  \bibinfo{author}{\bibfnamefont{A.~I.} \bibnamefont{Boldyrev}},
  \bibinfo{journal}{Journal of Computational Chemistry}
  \textbf{\bibinfo{volume}{28}}, \bibinfo{pages}{251} (\bibinfo{year}{2007}),
  \eprint{https://onlinelibrary.wiley.com/doi/pdf/10.1002/jcc.20518},
  \urlprefix\url{https://onlinelibrary.wiley.com/doi/abs/10.1002/jcc.20518}.

\bibitem[{\citenamefont{Poater et~al.}(2005)\citenamefont{Poater, Duran, Solà,
  and Silvi}}]{Poater}
\bibinfo{author}{\bibfnamefont{J.}~\bibnamefont{Poater}},
  \bibinfo{author}{\bibfnamefont{M.}~\bibnamefont{Duran}},
  \bibinfo{author}{\bibfnamefont{M.}~\bibnamefont{Solà}}, \bibnamefont{and}
  \bibinfo{author}{\bibfnamefont{B.}~\bibnamefont{Silvi}},
  \bibinfo{journal}{Chemical Reviews} \textbf{\bibinfo{volume}{105}},
  \bibinfo{pages}{3911} (\bibinfo{year}{2005}), \bibinfo{note}{pMID: 16218571},
  \eprint{https://doi.org/10.1021/cr030085x},
  \urlprefix\url{https://doi.org/10.1021/cr030085x}.

\bibitem[{\citenamefont{Mandado et~al.}(2007)\citenamefont{Mandado,
  Gonz\'alez-Moa, and Mosquera}}]{Mandado}
\bibinfo{author}{\bibfnamefont{M.}~\bibnamefont{Mandado}},
  \bibinfo{author}{\bibfnamefont{M.~J.} \bibnamefont{Gonz\'alez-Moa}},
  \bibnamefont{and} \bibinfo{author}{\bibfnamefont{R.~A.}
  \bibnamefont{Mosquera}}, \bibinfo{journal}{Journal of Computational
  Chemistry} \textbf{\bibinfo{volume}{28}}, \bibinfo{pages}{127}
  (\bibinfo{year}{2007}),
  \eprint{https://onlinelibrary.wiley.com/doi/pdf/10.1002/jcc.20468},
  \urlprefix\url{https://onlinelibrary.wiley.com/doi/abs/10.1002/jcc.20468}.

\bibitem[{\citenamefont{Pan et~al.}(2019)\citenamefont{Pan, Barroso, Jalife,
  Heine, Asmis, and Merino}}]{Sudip2}
\bibinfo{author}{\bibfnamefont{S.}~\bibnamefont{Pan}},
  \bibinfo{author}{\bibfnamefont{J.}~\bibnamefont{Barroso}},
  \bibinfo{author}{\bibfnamefont{S.}~\bibnamefont{Jalife}},
  \bibinfo{author}{\bibfnamefont{T.}~\bibnamefont{Heine}},
  \bibinfo{author}{\bibfnamefont{K.~R.} \bibnamefont{Asmis}}, \bibnamefont{and}
  \bibinfo{author}{\bibfnamefont{G.}~\bibnamefont{Merino}},
  \bibinfo{journal}{Accounts of Chemical Research}
  \textbf{\bibinfo{volume}{52}}, \bibinfo{pages}{2732} (\bibinfo{year}{2019}),
  \bibinfo{note}{pMID: 31487150},
  \eprint{https://doi.org/10.1021/acs.accounts.9b00336},
  \urlprefix\url{https://doi.org/10.1021/acs.accounts.9b00336}.

\bibitem[{\citenamefont{Wang et~al.}(2017{\natexlab{a}})\citenamefont{Wang,
  Feng, Guo, and Zhai}}]{Zhai}
\bibinfo{author}{\bibfnamefont{Y.-J.} \bibnamefont{Wang}},
  \bibinfo{author}{\bibfnamefont{L.-Y.} \bibnamefont{Feng}},
  \bibinfo{author}{\bibfnamefont{J.-C.} \bibnamefont{Guo}}, \bibnamefont{and}
  \bibinfo{author}{\bibfnamefont{H.-J.} \bibnamefont{Zhai}},
  \bibinfo{journal}{Chemistry - An Asian Journal}
  \textbf{\bibinfo{volume}{12}}, \bibinfo{pages}{2899}
  (\bibinfo{year}{2017}{\natexlab{a}}),
  \eprint{https://onlinelibrary.wiley.com/doi/pdf/10.1002/asia.201701310},
  \urlprefix\url{https://onlinelibrary.wiley.com/doi/abs/10.1002/asia.201701310}.

\bibitem[{\citenamefont{Zhai and Alexandrova}(2017)}]{Zhai2}
\bibinfo{author}{\bibfnamefont{H.}~\bibnamefont{Zhai}} \bibnamefont{and}
  \bibinfo{author}{\bibfnamefont{A.~N.} \bibnamefont{Alexandrova}},
  \bibinfo{journal}{ACS Catalysis} \textbf{\bibinfo{volume}{7}},
  \bibinfo{pages}{1905} (\bibinfo{year}{2017}),
  \eprint{https://doi.org/10.1021/acscatal.6b03243},
  \urlprefix\url{https://doi.org/10.1021/acscatal.6b03243}.

\bibitem[{\citenamefont{Merino and Heine}(2012)}]{Merino}
\bibinfo{author}{\bibfnamefont{G.}~\bibnamefont{Merino}} \bibnamefont{and}
  \bibinfo{author}{\bibfnamefont{T.}~\bibnamefont{Heine}},
  \bibinfo{journal}{Angewandte Chemie International Edition}
  \textbf{\bibinfo{volume}{51}}, \bibinfo{pages}{10226} (\bibinfo{year}{2012}),
  \eprint{https://onlinelibrary.wiley.com/doi/pdf/10.1002/anie.201206188},
  \urlprefix\url{https://onlinelibrary.wiley.com/doi/abs/10.1002/anie.201206188}.

\bibitem[{\citenamefont{Romanescu et~al.}(2013)\citenamefont{Romanescu, Galeev,
  Li, Boldyrev, and Wang}}]{Romanescu}
\bibinfo{author}{\bibfnamefont{C.}~\bibnamefont{Romanescu}},
  \bibinfo{author}{\bibfnamefont{T.~R.} \bibnamefont{Galeev}},
  \bibinfo{author}{\bibfnamefont{W.-L.} \bibnamefont{Li}},
  \bibinfo{author}{\bibfnamefont{A.~I.} \bibnamefont{Boldyrev}},
  \bibnamefont{and} \bibinfo{author}{\bibfnamefont{L.-S.} \bibnamefont{Wang}},
  \bibinfo{journal}{Accounts of Chemical Research}
  \textbf{\bibinfo{volume}{46}}, \bibinfo{pages}{350} (\bibinfo{year}{2013}),
  \bibinfo{note}{pMID: 23210660}, \eprint{https://doi.org/10.1021/ar300149a},
  \urlprefix\url{https://doi.org/10.1021/ar300149a}.

\bibitem[{\citenamefont{Liang et~al.}(2019)\citenamefont{Liang, Barroso,
  Jalife, Orozco-Ic, Zarate, Dong, Cui, and Merino}}]{Liang}
\bibinfo{author}{\bibfnamefont{W.-y.} \bibnamefont{Liang}},
  \bibinfo{author}{\bibfnamefont{J.}~\bibnamefont{Barroso}},
  \bibinfo{author}{\bibfnamefont{S.}~\bibnamefont{Jalife}},
  \bibinfo{author}{\bibfnamefont{M.}~\bibnamefont{Orozco-Ic}},
  \bibinfo{author}{\bibfnamefont{X.}~\bibnamefont{Zarate}},
  \bibinfo{author}{\bibfnamefont{X.}~\bibnamefont{Dong}},
  \bibinfo{author}{\bibfnamefont{Z.-h.} \bibnamefont{Cui}}, \bibnamefont{and}
  \bibinfo{author}{\bibfnamefont{G.}~\bibnamefont{Merino}},
  \bibinfo{journal}{Chem. Commun.} \textbf{\bibinfo{volume}{55}},
  \bibinfo{pages}{7490} (\bibinfo{year}{2019}),
  \urlprefix\url{http://dx.doi.org/10.1039/C9CC03732B}.

\bibitem[{\citenamefont{Chen et~al.}(2019{\natexlab{c}})\citenamefont{Chen, Li,
  Bai, Chen, Dong, Li, and Wang}}]{Teng-Teng3}
\bibinfo{author}{\bibfnamefont{T.-T.} \bibnamefont{Chen}},
  \bibinfo{author}{\bibfnamefont{W.-L.} \bibnamefont{Li}},
  \bibinfo{author}{\bibfnamefont{H.}~\bibnamefont{Bai}},
  \bibinfo{author}{\bibfnamefont{W.-J.} \bibnamefont{Chen}},
  \bibinfo{author}{\bibfnamefont{X.-R.} \bibnamefont{Dong}},
  \bibinfo{author}{\bibfnamefont{J.}~\bibnamefont{Li}}, \bibnamefont{and}
  \bibinfo{author}{\bibfnamefont{L.-S.} \bibnamefont{Wang}},
  \bibinfo{journal}{The Journal of Physical Chemistry A}
  \textbf{\bibinfo{volume}{123}}, \bibinfo{pages}{5317}
  (\bibinfo{year}{2019}{\natexlab{c}}),
  \eprint{https://doi.org/10.1021/acs.jpca.9b03942},
  \urlprefix\url{https://doi.org/10.1021/acs.jpca.9b03942}.

\bibitem[{\citenamefont{Dong et~al.}(2019)\citenamefont{Dong, Jalife,
  V\'asquez-Espinal, Barroso, Orozco-Ic, Ravell, Cabellos, Liang, Cui, and
  Merino}}]{Dong2019}
\bibinfo{author}{\bibfnamefont{X.}~\bibnamefont{Dong}},
  \bibinfo{author}{\bibfnamefont{S.}~\bibnamefont{Jalife}},
  \bibinfo{author}{\bibfnamefont{A.}~\bibnamefont{V\'asquez-Espinal}},
  \bibinfo{author}{\bibfnamefont{J.}~\bibnamefont{Barroso}},
  \bibinfo{author}{\bibfnamefont{M.}~\bibnamefont{Orozco-Ic}},
  \bibinfo{author}{\bibfnamefont{E.}~\bibnamefont{Ravell}},
  \bibinfo{author}{\bibfnamefont{J.~L.} \bibnamefont{Cabellos}},
  \bibinfo{author}{\bibfnamefont{W.-y.} \bibnamefont{Liang}},
  \bibinfo{author}{\bibfnamefont{Z.-h.} \bibnamefont{Cui}}, \bibnamefont{and}
  \bibinfo{author}{\bibfnamefont{G.}~\bibnamefont{Merino}},
  \bibinfo{journal}{Nanoscale} \textbf{\bibinfo{volume}{11}},
  \bibinfo{pages}{2143} (\bibinfo{year}{2019}),
  \urlprefix\url{http://dx.doi.org/10.1039/C8NR09173K}.

\bibitem[{\citenamefont{Popov et~al.}(2015)\citenamefont{Popov, Jian, Lopez,
  Boldyrev, and Wang}}]{Popov2015}
\bibinfo{author}{\bibfnamefont{I.~A.} \bibnamefont{Popov}},
  \bibinfo{author}{\bibfnamefont{T.}~\bibnamefont{Jian}},
  \bibinfo{author}{\bibfnamefont{G.~V.} \bibnamefont{Lopez}},
  \bibinfo{author}{\bibfnamefont{A.~I.} \bibnamefont{Boldyrev}},
  \bibnamefont{and} \bibinfo{author}{\bibfnamefont{L.-S.} \bibnamefont{Wang}},
  \bibinfo{journal}{Nature Communications} \textbf{\bibinfo{volume}{6}},
  \bibinfo{pages}{8654} (\bibinfo{year}{2015}),
  \urlprefix\url{https://doi.org/10.1038/ncomms9654}.

\bibitem[{\citenamefont{Van~Duong and Tho~Nguyen}(2017)}]{Duong}
\bibinfo{author}{\bibfnamefont{L.}~\bibnamefont{Van~Duong}} \bibnamefont{and}
  \bibinfo{author}{\bibfnamefont{M.}~\bibnamefont{Tho~Nguyen}},
  \bibinfo{journal}{Phys. Chem. Chem. Phys.} \textbf{\bibinfo{volume}{19}},
  \bibinfo{pages}{14913} (\bibinfo{year}{2017}),
  \urlprefix\url{http://dx.doi.org/10.1039/C7CP01740E}.

\bibitem[{\citenamefont{Cervantes-Navarro
  et~al.}(2014)\citenamefont{Cervantes-Navarro, Mart\'inez-Guajardo, Osorio,
  Moreno, Tiznado, Islas, Donald, and Merino}}]{Cervantes-Navarro}
\bibinfo{author}{\bibfnamefont{F.}~\bibnamefont{Cervantes-Navarro}},
  \bibinfo{author}{\bibfnamefont{G.}~\bibnamefont{Mart\'inez-Guajardo}},
  \bibinfo{author}{\bibfnamefont{E.}~\bibnamefont{Osorio}},
  \bibinfo{author}{\bibfnamefont{D.}~\bibnamefont{Moreno}},
  \bibinfo{author}{\bibfnamefont{W.}~\bibnamefont{Tiznado}},
  \bibinfo{author}{\bibfnamefont{R.}~\bibnamefont{Islas}},
  \bibinfo{author}{\bibfnamefont{K.~J.} \bibnamefont{Donald}},
  \bibnamefont{and} \bibinfo{author}{\bibfnamefont{G.}~\bibnamefont{Merino}},
  \bibinfo{journal}{Chem. Commun.} \textbf{\bibinfo{volume}{50}},
  \bibinfo{pages}{10680} (\bibinfo{year}{2014}),
  \urlprefix\url{http://dx.doi.org/10.1039/C4CC03698K}.

\bibitem[{\citenamefont{Ferrari et~al.}(2018)\citenamefont{Ferrari, Vanbuel,
  Hansen, Lievens, Janssens, and Fielicke}}]{Ferrari}
\bibinfo{author}{\bibfnamefont{P.}~\bibnamefont{Ferrari}},
  \bibinfo{author}{\bibfnamefont{J.}~\bibnamefont{Vanbuel}},
  \bibinfo{author}{\bibfnamefont{K.}~\bibnamefont{Hansen}},
  \bibinfo{author}{\bibfnamefont{P.}~\bibnamefont{Lievens}},
  \bibinfo{author}{\bibfnamefont{E.}~\bibnamefont{Janssens}}, \bibnamefont{and}
  \bibinfo{author}{\bibfnamefont{A.}~\bibnamefont{Fielicke}},
  \bibinfo{journal}{Phys. Rev. A} \textbf{\bibinfo{volume}{98}},
  \bibinfo{pages}{012501} (\bibinfo{year}{2018}),
  \urlprefix\url{https://link.aps.org/doi/10.1103/PhysRevA.98.012501}.

\bibitem[{\citenamefont{Mart\'inez-Guajardo
  et~al.}(2011)\citenamefont{Mart\'inez-Guajardo, Sergeeva, Boldyrev, Heine,
  Ugalde, and Merino}}]{Martinez-Guajardo}
\bibinfo{author}{\bibfnamefont{G.}~\bibnamefont{Mart\'inez-Guajardo}},
  \bibinfo{author}{\bibfnamefont{A.~P.} \bibnamefont{Sergeeva}},
  \bibinfo{author}{\bibfnamefont{A.~I.} \bibnamefont{Boldyrev}},
  \bibinfo{author}{\bibfnamefont{T.}~\bibnamefont{Heine}},
  \bibinfo{author}{\bibfnamefont{J.~M.} \bibnamefont{Ugalde}},
  \bibnamefont{and} \bibinfo{author}{\bibfnamefont{G.}~\bibnamefont{Merino}},
  \bibinfo{journal}{Chem. Commun.} \textbf{\bibinfo{volume}{47}},
  \bibinfo{pages}{6242} (\bibinfo{year}{2011}),
  \urlprefix\url{http://dx.doi.org/10.1039/C1CC10821B}.

\bibitem[{\citenamefont{Moreno et~al.}(2014)\citenamefont{Moreno, Pan, Zeonjuk,
  Islas, Osorio, Martínez-Guajardo, Chattaraj, Heine, and Merino}}]{Moreno}
\bibinfo{author}{\bibfnamefont{D.}~\bibnamefont{Moreno}},
  \bibinfo{author}{\bibfnamefont{S.}~\bibnamefont{Pan}},
  \bibinfo{author}{\bibfnamefont{L.~L.} \bibnamefont{Zeonjuk}},
  \bibinfo{author}{\bibfnamefont{R.}~\bibnamefont{Islas}},
  \bibinfo{author}{\bibfnamefont{E.}~\bibnamefont{Osorio}},
  \bibinfo{author}{\bibfnamefont{G.}~\bibnamefont{Martínez-Guajardo}},
  \bibinfo{author}{\bibfnamefont{P.~K.} \bibnamefont{Chattaraj}},
  \bibinfo{author}{\bibfnamefont{T.}~\bibnamefont{Heine}}, \bibnamefont{and}
  \bibinfo{author}{\bibfnamefont{G.}~\bibnamefont{Merino}},
  \bibinfo{journal}{Chem. Commun.} \textbf{\bibinfo{volume}{50}},
  \bibinfo{pages}{8140} (\bibinfo{year}{2014}),
  \urlprefix\url{http://dx.doi.org/10.1039/C4CC02225D}.

\bibitem[{\citenamefont{Tai et~al.}(2012)\citenamefont{Tai, Ceulemans, and
  Nguyen}}]{Truong}
\bibinfo{author}{\bibfnamefont{T.~B.} \bibnamefont{Tai}},
  \bibinfo{author}{\bibfnamefont{A.}~\bibnamefont{Ceulemans}},
  \bibnamefont{and} \bibinfo{author}{\bibfnamefont{M.~T.}
  \bibnamefont{Nguyen}}, \bibinfo{journal}{Chemistry - A European Journal}
  \textbf{\bibinfo{volume}{18}}, \bibinfo{pages}{4510} (\bibinfo{year}{2012}),
  \eprint{https://chemistry-europe.onlinelibrary.wiley.com/doi/pdf/10.1002/chem.201104064},
  \urlprefix\url{https://chemistry-europe.onlinelibrary.wiley.com/doi/abs/10.1002/chem.201104064}.

\bibitem[{\citenamefont{Fagiani et~al.}(2017)\citenamefont{Fagiani, Song,
  Petkov, Debnath, Gewinner, Schöllkopf, Heine, Fielicke, and
  Asmis}}]{Fagiani}
\bibinfo{author}{\bibfnamefont{M.~R.} \bibnamefont{Fagiani}},
  \bibinfo{author}{\bibfnamefont{X.}~\bibnamefont{Song}},
  \bibinfo{author}{\bibfnamefont{P.}~\bibnamefont{Petkov}},
  \bibinfo{author}{\bibfnamefont{S.}~\bibnamefont{Debnath}},
  \bibinfo{author}{\bibfnamefont{S.}~\bibnamefont{Gewinner}},
  \bibinfo{author}{\bibfnamefont{W.}~\bibnamefont{Schöllkopf}},
  \bibinfo{author}{\bibfnamefont{T.}~\bibnamefont{Heine}},
  \bibinfo{author}{\bibfnamefont{A.}~\bibnamefont{Fielicke}}, \bibnamefont{and}
  \bibinfo{author}{\bibfnamefont{K.~R.} \bibnamefont{Asmis}},
  \bibinfo{journal}{Angewandte Chemie International Edition}
  \textbf{\bibinfo{volume}{56}}, \bibinfo{pages}{501} (\bibinfo{year}{2017}),
  \eprint{https://onlinelibrary.wiley.com/doi/pdf/10.1002/anie.201609766},
  \urlprefix\url{https://onlinelibrary.wiley.com/doi/abs/10.1002/anie.201609766}.

\bibitem[{\citenamefont{Yang et~al.}(2017)\citenamefont{Yang, Jia, Wang, Zhai,
  Man, and Li}}]{Yonggang}
\bibinfo{author}{\bibfnamefont{Y.}~\bibnamefont{Yang}},
  \bibinfo{author}{\bibfnamefont{D.}~\bibnamefont{Jia}},
  \bibinfo{author}{\bibfnamefont{Y.-J.} \bibnamefont{Wang}},
  \bibinfo{author}{\bibfnamefont{H.-J.} \bibnamefont{Zhai}},
  \bibinfo{author}{\bibfnamefont{Y.}~\bibnamefont{Man}}, \bibnamefont{and}
  \bibinfo{author}{\bibfnamefont{S.-D.} \bibnamefont{Li}},
  \bibinfo{journal}{Nanoscale} \textbf{\bibinfo{volume}{9}},
  \bibinfo{pages}{1443} (\bibinfo{year}{2017}),
  \urlprefix\url{http://dx.doi.org/10.1039/C6NR09074E}.

\bibitem[{\citenamefont{Wang et~al.}(2015)\citenamefont{Wang, Zhao, Chen, Zhai,
  and Li}}]{Ying-Jin2015}
\bibinfo{author}{\bibfnamefont{Y.-J.} \bibnamefont{Wang}},
  \bibinfo{author}{\bibfnamefont{X.-Y.} \bibnamefont{Zhao}},
  \bibinfo{author}{\bibfnamefont{Q.}~\bibnamefont{Chen}},
  \bibinfo{author}{\bibfnamefont{H.-J.} \bibnamefont{Zhai}}, \bibnamefont{and}
  \bibinfo{author}{\bibfnamefont{S.-D.} \bibnamefont{Li}},
  \bibinfo{journal}{Nanoscale} \textbf{\bibinfo{volume}{7}},
  \bibinfo{pages}{16054} (\bibinfo{year}{2015}),
  \urlprefix\url{http://dx.doi.org/10.1039/C5NR03732H}.

\bibitem[{\citenamefont{Wang et~al.}(2017{\natexlab{b}})\citenamefont{Wang,
  Guo, and Zhai}}]{Zhai2017}
\bibinfo{author}{\bibfnamefont{Y.-J.} \bibnamefont{Wang}},
  \bibinfo{author}{\bibfnamefont{J.-C.} \bibnamefont{Guo}}, \bibnamefont{and}
  \bibinfo{author}{\bibfnamefont{H.-J.} \bibnamefont{Zhai}},
  \bibinfo{journal}{Nanoscale} \textbf{\bibinfo{volume}{9}},
  \bibinfo{pages}{9310} (\bibinfo{year}{2017}{\natexlab{b}}),
  \urlprefix\url{http://dx.doi.org/10.1039/C7NR03193A}.

\bibitem[{\citenamefont{Ya\~nez et~al.}(2020)\citenamefont{Ya\~nez, Inostroza,
  Usuga-Acevedo, V\'asquez-Espinal, Pino-Rios, Tabilo-Sepulveda, Garza,
  Barroso, Merino, and Tiznado}}]{Osvaldo}
\bibinfo{author}{\bibfnamefont{O.}~\bibnamefont{Ya\~nez}},
  \bibinfo{author}{\bibfnamefont{D.}~\bibnamefont{Inostroza}},
  \bibinfo{author}{\bibfnamefont{B.}~\bibnamefont{Usuga-Acevedo}},
  \bibinfo{author}{\bibfnamefont{A.}~\bibnamefont{V\'asquez-Espinal}},
  \bibinfo{author}{\bibfnamefont{R.}~\bibnamefont{Pino-Rios}},
  \bibinfo{author}{\bibfnamefont{M.}~\bibnamefont{Tabilo-Sepulveda}},
  \bibinfo{author}{\bibfnamefont{J.}~\bibnamefont{Garza}},
  \bibinfo{author}{\bibfnamefont{J.}~\bibnamefont{Barroso}},
  \bibinfo{author}{\bibfnamefont{G.}~\bibnamefont{Merino}}, \bibnamefont{and}
  \bibinfo{author}{\bibfnamefont{W.}~\bibnamefont{Tiznado}},
  \bibinfo{journal}{Theoretical Chemistry Accounts}
  \textbf{\bibinfo{volume}{139}}, \bibinfo{pages}{139} (\bibinfo{year}{2020}),
  \eprint{https://doi.org/10.1007/s00214-020-2548-5},
  \urlprefix\url{https://doi.org/10.1007/s00214-020-2548-5}.

\bibitem[{\citenamefont{Baletto and Ferrando}(2005)}]{Baletto}
\bibinfo{author}{\bibfnamefont{F.}~\bibnamefont{Baletto}} \bibnamefont{and}
  \bibinfo{author}{\bibfnamefont{R.}~\bibnamefont{Ferrando}},
  \bibinfo{journal}{Rev. Mod. Phys.} \textbf{\bibinfo{volume}{77}},
  \bibinfo{pages}{371} (\bibinfo{year}{2005}),
  \urlprefix\url{https://link.aps.org/doi/10.1103/RevModPhys.77.371}.

\bibitem[{\citenamefont{Grigoryan and Springborg}(2019)}]{Grigoryan}
\bibinfo{author}{\bibfnamefont{V.~G.} \bibnamefont{Grigoryan}}
  \bibnamefont{and}
  \bibinfo{author}{\bibfnamefont{M.}~\bibnamefont{Springborg}},
  \bibinfo{journal}{Phys. Chem. Chem. Phys.} \textbf{\bibinfo{volume}{21}},
  \bibinfo{pages}{5646} (\bibinfo{year}{2019}),
  \urlprefix\url{http://dx.doi.org/10.1039/C9CP00123A}.

\bibitem[{\citenamefont{Calvo et~al.}(2001)\citenamefont{Calvo, Doye, and
  Wales}}]{Calvo1}
\bibinfo{author}{\bibfnamefont{F.}~\bibnamefont{Calvo}},
  \bibinfo{author}{\bibfnamefont{J.~P.~K.} \bibnamefont{Doye}},
  \bibnamefont{and} \bibinfo{author}{\bibfnamefont{D.~J.} \bibnamefont{Wales}},
  \bibinfo{journal}{The Journal of Chemical Physics}
  \textbf{\bibinfo{volume}{114}}, \bibinfo{pages}{7312} (\bibinfo{year}{2001}),
  \eprint{https://doi.org/10.1063/1.1359768},
  \urlprefix\url{https://doi.org/10.1063/1.1359768}.

\bibitem[{\citenamefont{Br\'echignac et~al.}(1996)\citenamefont{Br\'echignac,
  Cahuzac, de~Frutos, Keba\"{\i}li, Sarfati, and Akulin}}]{Cahuzac}
\bibinfo{author}{\bibfnamefont{C.}~\bibnamefont{Br\'echignac}},
  \bibinfo{author}{\bibfnamefont{P.}~\bibnamefont{Cahuzac}},
  \bibinfo{author}{\bibfnamefont{M.}~\bibnamefont{de~Frutos}},
  \bibinfo{author}{\bibfnamefont{N.}~\bibnamefont{Keba\"{\i}li}},
  \bibinfo{author}{\bibfnamefont{A.}~\bibnamefont{Sarfati}}, \bibnamefont{and}
  \bibinfo{author}{\bibfnamefont{V.}~\bibnamefont{Akulin}},
  \bibinfo{journal}{Phys. Rev. Lett.} \textbf{\bibinfo{volume}{77}},
  \bibinfo{pages}{251} (\bibinfo{year}{1996}),
  \urlprefix\url{https://link.aps.org/doi/10.1103/PhysRevLett.77.251}.

\bibitem[{\citenamefont{Foster et~al.}(2018)\citenamefont{Foster, Ferrando, and
  Palmer}}]{Foster}
\bibinfo{author}{\bibfnamefont{D.}~\bibnamefont{Foster}},
  \bibinfo{author}{\bibfnamefont{R.}~\bibnamefont{Ferrando}}, \bibnamefont{and}
  \bibinfo{author}{\bibfnamefont{R.}~\bibnamefont{Palmer}},
  \bibinfo{journal}{Nature Communications} \textbf{\bibinfo{volume}{9}},
  \bibinfo{pages}{1323} (\bibinfo{year}{2018}),
  \urlprefix\url{https://doi.org/10.1038/s41467-018-03794-9}.

\bibitem[{\citenamefont{Dzib et~al.}(2019)\citenamefont{Dzib, Cabellos,
  Ortíz-Chi, Pan, Galano, and Merino}}]{Dzib}
\bibinfo{author}{\bibfnamefont{E.}~\bibnamefont{Dzib}},
  \bibinfo{author}{\bibfnamefont{J.~L.} \bibnamefont{Cabellos}},
  \bibinfo{author}{\bibfnamefont{F.}~\bibnamefont{Ortíz-Chi}},
  \bibinfo{author}{\bibfnamefont{S.}~\bibnamefont{Pan}},
  \bibinfo{author}{\bibfnamefont{A.}~\bibnamefont{Galano}}, \bibnamefont{and}
  \bibinfo{author}{\bibfnamefont{G.}~\bibnamefont{Merino}},
  \bibinfo{journal}{International Journal of Quantum Chemistry}
  \textbf{\bibinfo{volume}{119}}, \bibinfo{pages}{e25686}
  (\bibinfo{year}{2019}),
  \eprint{https://onlinelibrary.wiley.com/doi/pdf/10.1002/qua.25686},
  \urlprefix\url{https://onlinelibrary.wiley.com/doi/abs/10.1002/qua.25686}.

\bibitem[{\citenamefont{Vargas-Caamal
  et~al.}(2016{\natexlab{a}})\citenamefont{Vargas-Caamal, Cabellos, Ortiz-Chi,
  Rzepa, Restrepo, and Merino}}]{Vargas-Caamal}
\bibinfo{author}{\bibfnamefont{A.}~\bibnamefont{Vargas-Caamal}},
  \bibinfo{author}{\bibfnamefont{J.~L.} \bibnamefont{Cabellos}},
  \bibinfo{author}{\bibfnamefont{F.}~\bibnamefont{Ortiz-Chi}},
  \bibinfo{author}{\bibfnamefont{H.~S.} \bibnamefont{Rzepa}},
  \bibinfo{author}{\bibfnamefont{A.}~\bibnamefont{Restrepo}}, \bibnamefont{and}
  \bibinfo{author}{\bibfnamefont{G.}~\bibnamefont{Merino}},
  \bibinfo{journal}{Chemistry – A European Journal}
  \textbf{\bibinfo{volume}{22}}, \bibinfo{pages}{2812}
  (\bibinfo{year}{2016}{\natexlab{a}}),
  \eprint{https://chemistry-europe.onlinelibrary.wiley.com/doi/pdf/10.1002/chem.201504016},
  \urlprefix\url{https://chemistry-europe.onlinelibrary.wiley.com/doi/abs/10.1002/chem.201504016}.

\bibitem[{\citenamefont{Shkrebtii et~al.}(2011)\citenamefont{Shkrebtii, Heron,
  Cabellos, Witkowski, Pluchery, Mendoza, and Borensztein}}]{shkrebtii}
\bibinfo{author}{\bibfnamefont{A.}~\bibnamefont{Shkrebtii}},
  \bibinfo{author}{\bibfnamefont{J.}~\bibnamefont{Heron}},
  \bibinfo{author}{\bibfnamefont{J.}~\bibnamefont{Cabellos}},
  \bibinfo{author}{\bibfnamefont{N.}~\bibnamefont{Witkowski}},
  \bibinfo{author}{\bibfnamefont{O.}~\bibnamefont{Pluchery}},
  \bibinfo{author}{\bibfnamefont{B.}~\bibnamefont{Mendoza}}, \bibnamefont{and}
  \bibinfo{author}{\bibfnamefont{Y.}~\bibnamefont{Borensztein}},
  \bibinfo{journal}{MRS Proceedings} \textbf{\bibinfo{volume}{1370}},
  \bibinfo{pages}{mrss11} (\bibinfo{year}{2011}).

\bibitem[{\citenamefont{Goldsmith et~al.}(2019)\citenamefont{Goldsmith,
  Florian, Liu, Gruene, Lyon, Rayner, Fielicke, Scheffler, and
  Ghiringhelli}}]{Goldsmith}
\bibinfo{author}{\bibfnamefont{B.~R.} \bibnamefont{Goldsmith}},
  \bibinfo{author}{\bibfnamefont{J.}~\bibnamefont{Florian}},
  \bibinfo{author}{\bibfnamefont{J.-X.} \bibnamefont{Liu}},
  \bibinfo{author}{\bibfnamefont{P.}~\bibnamefont{Gruene}},
  \bibinfo{author}{\bibfnamefont{J.~T.} \bibnamefont{Lyon}},
  \bibinfo{author}{\bibfnamefont{D.~M.} \bibnamefont{Rayner}},
  \bibinfo{author}{\bibfnamefont{A.}~\bibnamefont{Fielicke}},
  \bibinfo{author}{\bibfnamefont{M.}~\bibnamefont{Scheffler}},
  \bibnamefont{and} \bibinfo{author}{\bibfnamefont{L.~M.}
  \bibnamefont{Ghiringhelli}}, \bibinfo{journal}{Phys. Rev. Materials}
  \textbf{\bibinfo{volume}{3}}, \bibinfo{pages}{016002} (\bibinfo{year}{2019}),
  \urlprefix\url{https://link.aps.org/doi/10.1103/PhysRevMaterials.3.016002}.

\bibitem[{\citenamefont{Ghiringhelli et~al.}(2013)\citenamefont{Ghiringhelli,
  Gruene, Lyon, Rayner, Meijer, Fielicke, and Scheffler}}]{Ghiringhelli_2013}
\bibinfo{author}{\bibfnamefont{L.~M.} \bibnamefont{Ghiringhelli}},
  \bibinfo{author}{\bibfnamefont{P.}~\bibnamefont{Gruene}},
  \bibinfo{author}{\bibfnamefont{J.~T.} \bibnamefont{Lyon}},
  \bibinfo{author}{\bibfnamefont{D.~M.} \bibnamefont{Rayner}},
  \bibinfo{author}{\bibfnamefont{G.}~\bibnamefont{Meijer}},
  \bibinfo{author}{\bibfnamefont{A.}~\bibnamefont{Fielicke}}, \bibnamefont{and}
  \bibinfo{author}{\bibfnamefont{M.}~\bibnamefont{Scheffler}},
  \bibinfo{journal}{New Journal of Physics} \textbf{\bibinfo{volume}{15}},
  \bibinfo{pages}{083003} (\bibinfo{year}{2013}),
  \urlprefix\url{https://doi.org/10.1088%2F1367-2630%2F15%2F8%2F083003}.

\bibitem[{\citenamefont{Schebarchov et~al.}(2018)\citenamefont{Schebarchov,
  Baletto, and Wales}}]{Schebarchov}
\bibinfo{author}{\bibfnamefont{D.}~\bibnamefont{Schebarchov}},
  \bibinfo{author}{\bibfnamefont{F.}~\bibnamefont{Baletto}}, \bibnamefont{and}
  \bibinfo{author}{\bibfnamefont{D.~J.} \bibnamefont{Wales}},
  \bibinfo{journal}{Nanoscale} \textbf{\bibinfo{volume}{10}},
  \bibinfo{pages}{2004} (\bibinfo{year}{2018}),
  \urlprefix\url{http://dx.doi.org/10.1039/C7NR07123J}.

\bibitem[{\citenamefont{Mendoza-Wilson
  et~al.}(2020)\citenamefont{Mendoza-Wilson, Balandrán-Quintana, and
  Cabellos}}]{MENDOZAWILSON2020112912}
\bibinfo{author}{\bibfnamefont{A.~M.} \bibnamefont{Mendoza-Wilson}},
  \bibinfo{author}{\bibfnamefont{R.~R.} \bibnamefont{Balandrán-Quintana}},
  \bibnamefont{and} \bibinfo{author}{\bibfnamefont{J.~L.}
  \bibnamefont{Cabellos}}, \bibinfo{journal}{Computational and Theoretical
  Chemistry} \textbf{\bibinfo{volume}{1186}}, \bibinfo{pages}{112912}
  (\bibinfo{year}{2020}), ISSN \bibinfo{issn}{2210-271X},
  \urlprefix\url{http://www.sciencedirect.com/science/article/pii/S2210271X20302127}.

\bibitem[{\citenamefont{Seitsonen et~al.}(1995)\citenamefont{Seitsonen,
  Laasonen, Nieminen, and Klein}}]{Seitsonen}
\bibinfo{author}{\bibfnamefont{A.~P.} \bibnamefont{Seitsonen}},
  \bibinfo{author}{\bibfnamefont{K.}~\bibnamefont{Laasonen}},
  \bibinfo{author}{\bibfnamefont{R.~M.} \bibnamefont{Nieminen}},
  \bibnamefont{and} \bibinfo{author}{\bibfnamefont{M.~L.} \bibnamefont{Klein}},
  \bibinfo{journal}{The Journal of Chemical Physics}
  \textbf{\bibinfo{volume}{103}}, \bibinfo{pages}{8075} (\bibinfo{year}{1995}),
  \eprint{https://doi.org/10.1063/1.470172},
  \urlprefix\url{https://doi.org/10.1063/1.470172}.

\bibitem[{\citenamefont{Chandrachud et~al.}(2007)\citenamefont{Chandrachud,
  Joshi, and Kanhere}}]{Prachi}
\bibinfo{author}{\bibfnamefont{P.}~\bibnamefont{Chandrachud}},
  \bibinfo{author}{\bibfnamefont{K.}~\bibnamefont{Joshi}}, \bibnamefont{and}
  \bibinfo{author}{\bibfnamefont{D.~G.} \bibnamefont{Kanhere}},
  \bibinfo{journal}{Phys. Rev. B} \textbf{\bibinfo{volume}{76}},
  \bibinfo{pages}{235423} (\bibinfo{year}{2007}),
  \urlprefix\url{https://link.aps.org/doi/10.1103/PhysRevB.76.235423}.

\bibitem[{\citenamefont{Doye and Calvo}(2002)}]{Jonathan}
\bibinfo{author}{\bibfnamefont{J.~P.~K.} \bibnamefont{Doye}} \bibnamefont{and}
  \bibinfo{author}{\bibfnamefont{F.}~\bibnamefont{Calvo}},
  \bibinfo{journal}{The Journal of Chemical Physics}
  \textbf{\bibinfo{volume}{116}}, \bibinfo{pages}{8307} (\bibinfo{year}{2002}),
  \eprint{https://aip.scitation.org/doi/pdf/10.1063/1.1469616},
  \urlprefix\url{https://aip.scitation.org/doi/abs/10.1063/1.1469616}.

\bibitem[{\citenamefont{Li et~al.}(2007)\citenamefont{Li, Jasper, and
  Truhlar}}]{Zhen}
\bibinfo{author}{\bibfnamefont{Z.~H.} \bibnamefont{Li}},
  \bibinfo{author}{\bibfnamefont{A.~W.} \bibnamefont{Jasper}},
  \bibnamefont{and} \bibinfo{author}{\bibfnamefont{D.~G.}
  \bibnamefont{Truhlar}}, \bibinfo{journal}{Journal of the American Chemical
  Society} \textbf{\bibinfo{volume}{129}}, \bibinfo{pages}{14899}
  (\bibinfo{year}{2007}), \bibinfo{note}{pMID: 17994736},
  \eprint{https://doi.org/10.1021/ja073129i},
  \urlprefix\url{https://doi.org/10.1021/ja073129i}.

\bibitem[{\citenamefont{Darby et~al.}(2002)\citenamefont{Darby, Mortimer-Jones,
  Johnston, and Roberts}}]{Darby}
\bibinfo{author}{\bibfnamefont{S.}~\bibnamefont{Darby}},
  \bibinfo{author}{\bibfnamefont{T.~V.} \bibnamefont{Mortimer-Jones}},
  \bibinfo{author}{\bibfnamefont{R.~L.} \bibnamefont{Johnston}},
  \bibnamefont{and} \bibinfo{author}{\bibfnamefont{C.}~\bibnamefont{Roberts}},
  \bibinfo{journal}{The Journal of Chemical Physics}
  \textbf{\bibinfo{volume}{116}}, \bibinfo{pages}{1536} (\bibinfo{year}{2002}),
  \eprint{https://doi.org/10.1063/1.1429658},
  \urlprefix\url{https://doi.org/10.1063/1.1429658}.

\bibitem[{\citenamefont{P.~K.~Doye and J.~Wales}(1998)}]{Doye}
\bibinfo{author}{\bibfnamefont{J.}~\bibnamefont{P.~K.~Doye}} \bibnamefont{and}
  \bibinfo{author}{\bibfnamefont{D.}~\bibnamefont{J.~Wales}},
  \bibinfo{journal}{New J. Chem.} \textbf{\bibinfo{volume}{22}},
  \bibinfo{pages}{733} (\bibinfo{year}{1998}),
  \urlprefix\url{http://dx.doi.org/10.1039/A709249K}.

\bibitem[{\citenamefont{Ohno and Maeda}(2006)}]{Ohno}
\bibinfo{author}{\bibfnamefont{K.}~\bibnamefont{Ohno}} \bibnamefont{and}
  \bibinfo{author}{\bibfnamefont{S.}~\bibnamefont{Maeda}},
  \bibinfo{journal}{The Journal of Physical Chemistry A}
  \textbf{\bibinfo{volume}{110}}, \bibinfo{pages}{8933} (\bibinfo{year}{2006}),
  \bibinfo{note}{pMID: 16836457}, \eprint{https://doi.org/10.1021/jp061149l},
  \urlprefix\url{https://doi.org/10.1021/jp061149l}.

\bibitem[{\citenamefont{Str\o{}m et~al.}(2017)\citenamefont{Str\o{}m, Simon,
  Schnell, Kjelstrup, He, and Bedeaux}}]{Simon}
\bibinfo{author}{\bibfnamefont{B.~A.} \bibnamefont{Str\o{}m}},
  \bibinfo{author}{\bibfnamefont{J.-M.} \bibnamefont{Simon}},
  \bibinfo{author}{\bibfnamefont{S.~K.} \bibnamefont{Schnell}},
  \bibinfo{author}{\bibfnamefont{S.}~\bibnamefont{Kjelstrup}},
  \bibinfo{author}{\bibfnamefont{J.}~\bibnamefont{He}}, \bibnamefont{and}
  \bibinfo{author}{\bibfnamefont{D.}~\bibnamefont{Bedeaux}},
  \bibinfo{journal}{Phys. Chem. Chem. Phys.} \textbf{\bibinfo{volume}{19}},
  \bibinfo{pages}{9016} (\bibinfo{year}{2017}),
  \urlprefix\url{http://dx.doi.org/10.1039/C7CP00874K}.

\bibitem[{\citenamefont{Hill}(2001)}]{Hill2}
\bibinfo{author}{\bibfnamefont{T.~L.} \bibnamefont{Hill}},
  \bibinfo{journal}{Nano Letters} \textbf{\bibinfo{volume}{1}},
  \bibinfo{pages}{159} (\bibinfo{year}{2001}),
  \eprint{https://doi.org/10.1021/nl010009e},
  \urlprefix\url{https://doi.org/10.1021/nl010009e}.

\bibitem[{\citenamefont{Gibbs}(1961)}]{gibbs1961}
\bibinfo{author}{\bibfnamefont{J.}~\bibnamefont{Gibbs}},
  \emph{\bibinfo{title}{Thermodynamics}}, Scientific Papers
  (\bibinfo{publisher}{Dover Publications}, \bibinfo{year}{1961}),
  \urlprefix\url{https://books.google.com.gh/books?id=gPznMrzXfdMC}.

\bibitem[{\citenamefont{Li and Truhlar}(2014)}]{Li-Truhlar}
\bibinfo{author}{\bibfnamefont{Z.~H.} \bibnamefont{Li}} \bibnamefont{and}
  \bibinfo{author}{\bibfnamefont{D.~G.} \bibnamefont{Truhlar}},
  \bibinfo{journal}{Chem. Sci.} \textbf{\bibinfo{volume}{5}},
  \bibinfo{pages}{2605} (\bibinfo{year}{2014}),
  \urlprefix\url{http://dx.doi.org/10.1039/C4SC00052H}.

\bibitem[{\citenamefont{Hill}(1962)}]{Hill}
\bibinfo{author}{\bibfnamefont{T.~L.} \bibnamefont{Hill}},
  \bibinfo{journal}{The Journal of Chemical Physics}
  \textbf{\bibinfo{volume}{36}}, \bibinfo{pages}{3182} (\bibinfo{year}{1962}),
  \eprint{https://doi.org/10.1063/1.1732447},
  \urlprefix\url{https://doi.org/10.1063/1.1732447}.

\bibitem[{\citenamefont{Calvo}(2015)}]{Calvo}
\bibinfo{author}{\bibfnamefont{F.}~\bibnamefont{Calvo}},
  \bibinfo{journal}{Phys. Chem. Chem. Phys.} \textbf{\bibinfo{volume}{17}},
  \bibinfo{pages}{27922} (\bibinfo{year}{2015}),
  \urlprefix\url{http://dx.doi.org/10.1039/C5CP00274E}.

\bibitem[{\citenamefont{Bixon and Jortner}(1989)}]{Bixon}
\bibinfo{author}{\bibfnamefont{M.}~\bibnamefont{Bixon}} \bibnamefont{and}
  \bibinfo{author}{\bibfnamefont{J.}~\bibnamefont{Jortner}},
  \bibinfo{journal}{The Journal of Chemical Physics}
  \textbf{\bibinfo{volume}{91}}, \bibinfo{pages}{1631} (\bibinfo{year}{1989}),
  \eprint{https://doi.org/10.1063/1.457123},
  \urlprefix\url{https://doi.org/10.1063/1.457123}.

\bibitem[{\citenamefont{Kristensen et~al.}(1974)\citenamefont{Kristensen,
  Jensen, and Cotterill}}]{Kristensen}
\bibinfo{author}{\bibfnamefont{W.~D.} \bibnamefont{Kristensen}},
  \bibinfo{author}{\bibfnamefont{E.~J.} \bibnamefont{Jensen}},
  \bibnamefont{and} \bibinfo{author}{\bibfnamefont{R.~M.~J.}
  \bibnamefont{Cotterill}}, \bibinfo{journal}{The Journal of Chemical Physics}
  \textbf{\bibinfo{volume}{60}}, \bibinfo{pages}{4161} (\bibinfo{year}{1974}),
  \eprint{https://doi.org/10.1063/1.1680883},
  \urlprefix\url{https://doi.org/10.1063/1.1680883}.

\bibitem[{\citenamefont{Wales}(1996)}]{Wales925}
\bibinfo{author}{\bibfnamefont{D.~J.} \bibnamefont{Wales}},
  \bibinfo{journal}{Science} \textbf{\bibinfo{volume}{271}},
  \bibinfo{pages}{925} (\bibinfo{year}{1996}), ISSN \bibinfo{issn}{0036-8075},
  \eprint{https://science.sciencemag.org/content/271/5251/925.full.pdf},
  \urlprefix\url{https://science.sciencemag.org/content/271/5251/925}.

\bibitem[{\citenamefont{Jena et~al.}(1992)\citenamefont{Jena, Khanna, and
  Rao}}]{jena1992physics}
\bibinfo{author}{\bibfnamefont{P.}~\bibnamefont{Jena}},
  \bibinfo{author}{\bibfnamefont{S.}~\bibnamefont{Khanna}}, \bibnamefont{and}
  \bibinfo{author}{\bibfnamefont{B.}~\bibnamefont{Rao}},
  \emph{\bibinfo{title}{Physics and Chemistry of Finite Systems: From Clusters
  to Crystals}}, no. \bibinfo{number}{v. 2} in \bibinfo{series}{NATO ASI Series
  : advanced science institutes series: Series C, Mathematical and physical
  sciences} (\bibinfo{publisher}{Kluwer Academic Publishers},
  \bibinfo{year}{1992}), ISBN \bibinfo{isbn}{9780792318163},
  \urlprefix\url{https://books.google.mw/books?id=HDl3xQEACAAJ}.

\bibitem[{\citenamefont{Fox et~al.}(2006)\citenamefont{Fox, Horsfield, and
  Gillan}}]{Fox}
\bibinfo{author}{\bibfnamefont{H.}~\bibnamefont{Fox}},
  \bibinfo{author}{\bibfnamefont{A.~P.} \bibnamefont{Horsfield}},
  \bibnamefont{and} \bibinfo{author}{\bibfnamefont{M.~J.}
  \bibnamefont{Gillan}}, \bibinfo{journal}{The Journal of Chemical Physics}
  \textbf{\bibinfo{volume}{124}}, \bibinfo{pages}{134709}
  (\bibinfo{year}{2006}), \eprint{https://doi.org/10.1063/1.2184313},
  \urlprefix\url{https://doi.org/10.1063/1.2184313}.

\bibitem[{\citenamefont{Beret et~al.}(2011)\citenamefont{Beret, Ghiringhelli,
  and Scheffler}}]{C1FD00027F}
\bibinfo{author}{\bibfnamefont{E.~C.} \bibnamefont{Beret}},
  \bibinfo{author}{\bibfnamefont{L.~M.} \bibnamefont{Ghiringhelli}},
  \bibnamefont{and}
  \bibinfo{author}{\bibfnamefont{M.}~\bibnamefont{Scheffler}},
  \bibinfo{journal}{Faraday Discuss.} \textbf{\bibinfo{volume}{152}},
  \bibinfo{pages}{153} (\bibinfo{year}{2011}),
  \urlprefix\url{http://dx.doi.org/10.1039/C1FD00027F}.

\bibitem[{\citenamefont{Lv et~al.}(2014)\citenamefont{Lv, Xu, Cheng, Chen, and
  Cai}}]{Zhen-Long}
\bibinfo{author}{\bibfnamefont{Z.-L.} \bibnamefont{Lv}},
  \bibinfo{author}{\bibfnamefont{K.}~\bibnamefont{Xu}},
  \bibinfo{author}{\bibfnamefont{Y.}~\bibnamefont{Cheng}},
  \bibinfo{author}{\bibfnamefont{X.-R.} \bibnamefont{Chen}}, \bibnamefont{and}
  \bibinfo{author}{\bibfnamefont{L.-C.} \bibnamefont{Cai}},
  \bibinfo{journal}{The Journal of Chemical Physics}
  \textbf{\bibinfo{volume}{141}}, \bibinfo{pages}{054309}
  (\bibinfo{year}{2014}), \eprint{https://doi.org/10.1063/1.4891721},
  \urlprefix\url{https://doi.org/10.1063/1.4891721}.

\bibitem[{\citenamefont{Malloum et~al.}(2015)\citenamefont{Malloum, Fifen,
  Dhaouadi, Engo, and Jaidane}}]{Malloum}
\bibinfo{author}{\bibfnamefont{A.}~\bibnamefont{Malloum}},
  \bibinfo{author}{\bibfnamefont{J.~J.} \bibnamefont{Fifen}},
  \bibinfo{author}{\bibfnamefont{Z.}~\bibnamefont{Dhaouadi}},
  \bibinfo{author}{\bibfnamefont{S.~G.~N.} \bibnamefont{Engo}},
  \bibnamefont{and} \bibinfo{author}{\bibfnamefont{N.-E.}
  \bibnamefont{Jaidane}}, \bibinfo{journal}{Phys. Chem. Chem. Phys.}
  \textbf{\bibinfo{volume}{17}}, \bibinfo{pages}{29226} (\bibinfo{year}{2015}),
  \urlprefix\url{http://dx.doi.org/10.1039/C5CP03374H}.

\bibitem[{\citenamefont{Malloum et~al.}(2017)\citenamefont{Malloum, Fifen,
  Dhaouadi, Nana~Engo, and Jaidane}}]{Malloum2}
\bibinfo{author}{\bibfnamefont{A.}~\bibnamefont{Malloum}},
  \bibinfo{author}{\bibfnamefont{J.~J.} \bibnamefont{Fifen}},
  \bibinfo{author}{\bibfnamefont{Z.}~\bibnamefont{Dhaouadi}},
  \bibinfo{author}{\bibfnamefont{S.~G.} \bibnamefont{Nana~Engo}},
  \bibnamefont{and} \bibinfo{author}{\bibfnamefont{N.-E.}
  \bibnamefont{Jaidane}}, \bibinfo{journal}{The Journal of Chemical Physics}
  \textbf{\bibinfo{volume}{146}}, \bibinfo{pages}{044305}
  (\bibinfo{year}{2017}), \eprint{https://doi.org/10.1063/1.4974179},
  \urlprefix\url{https://doi.org/10.1063/1.4974179}.

\bibitem[{\citenamefont{Malloum et~al.}(2018)\citenamefont{Malloum, Fifen, and
  Conradie}}]{Malloum3}
\bibinfo{author}{\bibfnamefont{A.}~\bibnamefont{Malloum}},
  \bibinfo{author}{\bibfnamefont{J.~J.} \bibnamefont{Fifen}}, \bibnamefont{and}
  \bibinfo{author}{\bibfnamefont{J.}~\bibnamefont{Conradie}},
  \bibinfo{journal}{The Journal of Chemical Physics}
  \textbf{\bibinfo{volume}{149}}, \bibinfo{pages}{244301}
  (\bibinfo{year}{2018}), \eprint{https://doi.org/10.1063/1.5053172},
  \urlprefix\url{https://doi.org/10.1063/1.5053172}.

\bibitem[{\citenamefont{Fifen and Agmon}(2016)}]{Fifen}
\bibinfo{author}{\bibfnamefont{J.~J.} \bibnamefont{Fifen}} \bibnamefont{and}
  \bibinfo{author}{\bibfnamefont{N.}~\bibnamefont{Agmon}},
  \bibinfo{journal}{Journal of Chemical Theory and Computation}
  \textbf{\bibinfo{volume}{12}}, \bibinfo{pages}{1656} (\bibinfo{year}{2016}),
  \bibinfo{note}{pMID: 26913993},
  \eprint{https://doi.org/10.1021/acs.jctc.6b00038},
  \urlprefix\url{https://doi.org/10.1021/acs.jctc.6b00038}.

\bibitem[{\citenamefont{Gruene et~al.}(2008)\citenamefont{Gruene, Rayner,
  Redlich, van~der Meer, Lyon, Meijer, and Fielicke}}]{Gruene674}
\bibinfo{author}{\bibfnamefont{P.}~\bibnamefont{Gruene}},
  \bibinfo{author}{\bibfnamefont{D.~M.} \bibnamefont{Rayner}},
  \bibinfo{author}{\bibfnamefont{B.}~\bibnamefont{Redlich}},
  \bibinfo{author}{\bibfnamefont{A.~F.~G.} \bibnamefont{van~der Meer}},
  \bibinfo{author}{\bibfnamefont{J.~T.} \bibnamefont{Lyon}},
  \bibinfo{author}{\bibfnamefont{G.}~\bibnamefont{Meijer}}, \bibnamefont{and}
  \bibinfo{author}{\bibfnamefont{A.}~\bibnamefont{Fielicke}},
  \bibinfo{journal}{Science} \textbf{\bibinfo{volume}{321}},
  \bibinfo{pages}{674} (\bibinfo{year}{2008}), ISSN \bibinfo{issn}{0036-8075},
  \eprint{https://science.sciencemag.org/content/321/5889/674.full.pdf},
  \urlprefix\url{https://science.sciencemag.org/content/321/5889/674}.

\bibitem[{\citenamefont{Fielicke et~al.}(2005)\citenamefont{Fielicke, von
  Helden, Meijer, Pedersen, Simard, and Rayner}}]{doi:10.1021/ja0509230}
\bibinfo{author}{\bibfnamefont{A.}~\bibnamefont{Fielicke}},
  \bibinfo{author}{\bibfnamefont{G.}~\bibnamefont{von Helden}},
  \bibinfo{author}{\bibfnamefont{G.}~\bibnamefont{Meijer}},
  \bibinfo{author}{\bibfnamefont{D.~B.} \bibnamefont{Pedersen}},
  \bibinfo{author}{\bibfnamefont{B.}~\bibnamefont{Simard}}, \bibnamefont{and}
  \bibinfo{author}{\bibfnamefont{D.~M.} \bibnamefont{Rayner}},
  \bibinfo{journal}{Journal of the American Chemical Society}
  \textbf{\bibinfo{volume}{127}}, \bibinfo{pages}{8416} (\bibinfo{year}{2005}),
  \bibinfo{note}{pMID: 15941275}, \eprint{https://doi.org/10.1021/ja0509230},
  \urlprefix\url{https://doi.org/10.1021/ja0509230}.

\bibitem[{\citenamefont{Fielicke et~al.}(2004)\citenamefont{Fielicke, Kirilyuk,
  Ratsch, Behler, Scheffler, von Helden, and Meijer}}]{PhysRevLett.93.023401}
\bibinfo{author}{\bibfnamefont{A.}~\bibnamefont{Fielicke}},
  \bibinfo{author}{\bibfnamefont{A.}~\bibnamefont{Kirilyuk}},
  \bibinfo{author}{\bibfnamefont{C.}~\bibnamefont{Ratsch}},
  \bibinfo{author}{\bibfnamefont{J.}~\bibnamefont{Behler}},
  \bibinfo{author}{\bibfnamefont{M.}~\bibnamefont{Scheffler}},
  \bibinfo{author}{\bibfnamefont{G.}~\bibnamefont{von Helden}},
  \bibnamefont{and} \bibinfo{author}{\bibfnamefont{G.}~\bibnamefont{Meijer}},
  \bibinfo{journal}{Phys. Rev. Lett.} \textbf{\bibinfo{volume}{93}},
  \bibinfo{pages}{023401} (\bibinfo{year}{2004}),
  \urlprefix\url{https://link.aps.org/doi/10.1103/PhysRevLett.93.023401}.

\bibitem[{\citenamefont{Sieber et~al.}(2004)\citenamefont{Sieber, Buttet,
  Harbich, F\'elix, Mitri\ifmmode~\acute{c}\else \'{c}\fi{}, and Bona\ifmmode
  \check{c}\else \v{c}\fi{}i\ifmmode \acute{c}\else~\'{c}\fi{}
  Kouteck\'y}}]{Felix}
\bibinfo{author}{\bibfnamefont{C.}~\bibnamefont{Sieber}},
  \bibinfo{author}{\bibfnamefont{J.}~\bibnamefont{Buttet}},
  \bibinfo{author}{\bibfnamefont{W.}~\bibnamefont{Harbich}},
  \bibinfo{author}{\bibfnamefont{C.}~\bibnamefont{F\'elix}},
  \bibinfo{author}{\bibfnamefont{R.}~\bibnamefont{Mitri\ifmmode~\acute{c}\else
  \'{c}\fi{}}}, \bibnamefont{and}
  \bibinfo{author}{\bibfnamefont{V.}~\bibnamefont{Bona\ifmmode \check{c}\else
  \v{c}\fi{}i\ifmmode \acute{c}\else~\'{c}\fi{} Kouteck\'y}},
  \bibinfo{journal}{Phys. Rev. A} \textbf{\bibinfo{volume}{70}},
  \bibinfo{pages}{041201} (\bibinfo{year}{2004}),
  \urlprefix\url{https://link.aps.org/doi/10.1103/PhysRevA.70.041201}.

\bibitem[{\citenamefont{Ji et~al.}(2005)\citenamefont{Ji, Gu, Li, Gong, Li, and
  Wang}}]{Min}
\bibinfo{author}{\bibfnamefont{M.}~\bibnamefont{Ji}},
  \bibinfo{author}{\bibfnamefont{X.}~\bibnamefont{Gu}},
  \bibinfo{author}{\bibfnamefont{X.}~\bibnamefont{Li}},
  \bibinfo{author}{\bibfnamefont{X.}~\bibnamefont{Gong}},
  \bibinfo{author}{\bibfnamefont{J.}~\bibnamefont{Li}}, \bibnamefont{and}
  \bibinfo{author}{\bibfnamefont{L.-S.} \bibnamefont{Wang}},
  \bibinfo{journal}{Angewandte Chemie International Edition}
  \textbf{\bibinfo{volume}{44}}, \bibinfo{pages}{7119} (\bibinfo{year}{2005}),
  \eprint{https://onlinelibrary.wiley.com/doi/pdf/10.1002/anie.200502795},
  \urlprefix\url{https://onlinelibrary.wiley.com/doi/abs/10.1002/anie.200502795}.

\bibitem[{\citenamefont{Wille and Vennik}(1985)}]{Wille_1985}
\bibinfo{author}{\bibfnamefont{L.~T.} \bibnamefont{Wille}} \bibnamefont{and}
  \bibinfo{author}{\bibfnamefont{J.}~\bibnamefont{Vennik}},
  \bibinfo{journal}{Journal of Physics A: Mathematical and General}
  \textbf{\bibinfo{volume}{18}}, \bibinfo{pages}{L419} (\bibinfo{year}{1985}),
  \urlprefix\url{https://doi.org/10.1088%2F0305-4470%2F18%2F8%2F003}.

\bibitem[{\citenamefont{Xu et~al.}(2015)\citenamefont{Xu, Zhao, Liao, and
  Yang}}]{Xu}
\bibinfo{author}{\bibfnamefont{S.-G.} \bibnamefont{Xu}},
  \bibinfo{author}{\bibfnamefont{Y.-J.} \bibnamefont{Zhao}},
  \bibinfo{author}{\bibfnamefont{J.-H.} \bibnamefont{Liao}}, \bibnamefont{and}
  \bibinfo{author}{\bibfnamefont{X.-B.} \bibnamefont{Yang}},
  \bibinfo{journal}{The Journal of Chemical Physics}
  \textbf{\bibinfo{volume}{142}}, \bibinfo{pages}{214307}
  (\bibinfo{year}{2015}), \eprint{https://doi.org/10.1063/1.4922059},
  \urlprefix\url{https://doi.org/10.1063/1.4922059}.

\bibitem[{\citenamefont{Rossi and Ferrando}(2009)}]{Rossi_2009}
\bibinfo{author}{\bibfnamefont{G.}~\bibnamefont{Rossi}} \bibnamefont{and}
  \bibinfo{author}{\bibfnamefont{R.}~\bibnamefont{Ferrando}},
  \bibinfo{journal}{Journal of Physics: Condensed Matter}
  \textbf{\bibinfo{volume}{21}}, \bibinfo{pages}{084208}
  (\bibinfo{year}{2009}),
  \urlprefix\url{https://doi.org/10.1088%2F0953-8984%2F21%2F8%2F084208}.

\bibitem[{\citenamefont{Cheng et~al.}(2009)\citenamefont{Cheng, Feng, Yang, and
  Yang}}]{Cheng}
\bibinfo{author}{\bibfnamefont{L.}~\bibnamefont{Cheng}},
  \bibinfo{author}{\bibfnamefont{Y.}~\bibnamefont{Feng}},
  \bibinfo{author}{\bibfnamefont{J.}~\bibnamefont{Yang}}, \bibnamefont{and}
  \bibinfo{author}{\bibfnamefont{J.}~\bibnamefont{Yang}}, \bibinfo{journal}{The
  Journal of Chemical Physics} \textbf{\bibinfo{volume}{130}},
  \bibinfo{pages}{214112} (\bibinfo{year}{2009}),
  \eprint{https://doi.org/10.1063/1.3152121},
  \urlprefix\url{https://doi.org/10.1063/1.3152121}.

\bibitem[{\citenamefont{Kirkpatrick et~al.}(1983)\citenamefont{Kirkpatrick,
  Gelatt, and Vecchi}}]{kirkpatrick}
\bibinfo{author}{\bibfnamefont{S.}~\bibnamefont{Kirkpatrick}},
  \bibinfo{author}{\bibfnamefont{C.~D.} \bibnamefont{Gelatt}},
  \bibnamefont{and} \bibinfo{author}{\bibfnamefont{M.~P.}
  \bibnamefont{Vecchi}}, \bibinfo{journal}{Science}
  \textbf{\bibinfo{volume}{220}}, \bibinfo{pages}{671} (\bibinfo{year}{1983}).

\bibitem[{\citenamefont{Metropolis et~al.}(1953)\citenamefont{Metropolis,
  Rosenbluth, Rosenbluth, Teller, and Teller}}]{metropolis}
\bibinfo{author}{\bibfnamefont{N.}~\bibnamefont{Metropolis}},
  \bibinfo{author}{\bibfnamefont{A.~W.} \bibnamefont{Rosenbluth}},
  \bibinfo{author}{\bibfnamefont{M.~N.} \bibnamefont{Rosenbluth}},
  \bibinfo{author}{\bibfnamefont{A.~H.} \bibnamefont{Teller}},
  \bibnamefont{and} \bibinfo{author}{\bibfnamefont{E.}~\bibnamefont{Teller}},
  \bibinfo{journal}{J. Chem. Phys.} \textbf{\bibinfo{volume}{21}},
  \bibinfo{pages}{1087} (\bibinfo{year}{1953}),
  \urlprefix\url{http://scitation.aip.org/content/aip/journal/jcp/21/6/10.1063/1.1699114}.

\bibitem[{\citenamefont{Xiang and Gong}(2000)}]{xiang}
\bibinfo{author}{\bibfnamefont{Y.}~\bibnamefont{Xiang}} \bibnamefont{and}
  \bibinfo{author}{\bibfnamefont{X.~G.} \bibnamefont{Gong}},
  \bibinfo{journal}{Phys. Rev. E} \textbf{\bibinfo{volume}{62}},
  \bibinfo{pages}{4473} (\bibinfo{year}{2000}),
  \urlprefix\url{http://link.aps.org/doi/10.1103/PhysRevE.62.4473}.

\bibitem[{\citenamefont{Xiang et~al.}(2013)\citenamefont{Xiang, Gubian,
  Suomela, and Hoeng}}]{yang}
\bibinfo{author}{\bibfnamefont{Y.}~\bibnamefont{Xiang}},
  \bibinfo{author}{\bibfnamefont{S.}~\bibnamefont{Gubian}},
  \bibinfo{author}{\bibfnamefont{B.}~\bibnamefont{Suomela}}, \bibnamefont{and}
  \bibinfo{author}{\bibfnamefont{J.}~\bibnamefont{Hoeng}},
  \bibinfo{journal}{The R Journal} \textbf{\bibinfo{volume}{5}},
  \bibinfo{pages}{13} (\bibinfo{year}{2013}),
  \urlprefix\url{http://journal.r-project.org/archive/2013-1/xiang-gubian-suomela-etal.pdf}.

\bibitem[{\citenamefont{Vlachos et~al.}(1993)\citenamefont{Vlachos, Schmidt,
  and Aris}}]{vlachos}
\bibinfo{author}{\bibfnamefont{D.}~\bibnamefont{Vlachos}},
  \bibinfo{author}{\bibfnamefont{L.}~\bibnamefont{Schmidt}}, \bibnamefont{and}
  \bibinfo{author}{\bibfnamefont{R.}~\bibnamefont{Aris}}, \bibinfo{journal}{Z.
  Phys. D Atom Mol. Cl.} \textbf{\bibinfo{volume}{26}}, \bibinfo{pages}{156}
  (\bibinfo{year}{1993}), ISSN \bibinfo{issn}{0178-7683},
  \urlprefix\url{http://dx.doi.org/10.1007/BF01425649}.

\bibitem[{\citenamefont{Granville et~al.}(1994)\citenamefont{Granville,
  Krivanek, and Rasson}}]{granville}
\bibinfo{author}{\bibfnamefont{V.}~\bibnamefont{Granville}},
  \bibinfo{author}{\bibfnamefont{M.}~\bibnamefont{Krivanek}}, \bibnamefont{and}
  \bibinfo{author}{\bibfnamefont{J.-P.} \bibnamefont{Rasson}},
  \bibinfo{journal}{IEEE Trans. Pattern Anal. Mach. Intell.}
  \textbf{\bibinfo{volume}{16}}, \bibinfo{pages}{652} (\bibinfo{year}{1994}),
  ISSN \bibinfo{issn}{0162-8828}.

\bibitem[{\citenamefont{Saunders}(2004)}]{Saunders}
\bibinfo{author}{\bibfnamefont{M.}~\bibnamefont{Saunders}},
  \bibinfo{journal}{Journal of Computational Chemistry}
  \textbf{\bibinfo{volume}{25}}, \bibinfo{pages}{621} (\bibinfo{year}{2004}),
  \eprint{https://onlinelibrary.wiley.com/doi/pdf/10.1002/jcc.10407},
  \urlprefix\url{https://onlinelibrary.wiley.com/doi/abs/10.1002/jcc.10407}.

\bibitem[{\citenamefont{Saunders}(1987)}]{Saunders2}
\bibinfo{author}{\bibfnamefont{M.}~\bibnamefont{Saunders}},
  \bibinfo{journal}{Journal of the American Chemical Society}
  \textbf{\bibinfo{volume}{109}}, \bibinfo{pages}{3150} (\bibinfo{year}{1987}),
  \eprint{https://doi.org/10.1021/ja00244a051},
  \urlprefix\url{https://doi.org/10.1021/ja00244a051}.

\bibitem[{\citenamefont{Hsu and Lai}(2006)}]{hsu}
\bibinfo{author}{\bibfnamefont{P.~J.} \bibnamefont{Hsu}} \bibnamefont{and}
  \bibinfo{author}{\bibfnamefont{S.~K.} \bibnamefont{Lai}},
  \bibinfo{journal}{J. Chem. Phys.} \textbf{\bibinfo{volume}{124}},
  \bibinfo{eid}{044711} (\bibinfo{year}{2006}),
  \urlprefix\url{http://scitation.aip.org/content/aip/journal/jcp/124/4/10.1063/1.2147159}.

\bibitem[{\citenamefont{Qin et~al.}(2009)\citenamefont{Qin, Lu, Zhao, Zang,
  Wang, and Ho}}]{Wei}
\bibinfo{author}{\bibfnamefont{W.}~\bibnamefont{Qin}},
  \bibinfo{author}{\bibfnamefont{W.-C.} \bibnamefont{Lu}},
  \bibinfo{author}{\bibfnamefont{L.-Z.} \bibnamefont{Zhao}},
  \bibinfo{author}{\bibfnamefont{Q.-J.} \bibnamefont{Zang}},
  \bibinfo{author}{\bibfnamefont{C.~Z.} \bibnamefont{Wang}}, \bibnamefont{and}
  \bibinfo{author}{\bibfnamefont{K.~M.} \bibnamefont{Ho}}, \bibinfo{journal}{J.
  Phys.: Condens. Matter} \textbf{\bibinfo{volume}{21}},
  \bibinfo{pages}{455501} (\bibinfo{year}{2009}),
  \urlprefix\url{http://stacks.iop.org/0953-8984/21/i=45/a=455501}.

\bibitem[{\citenamefont{Goldberg}(1989)}]{goldberg}
\bibinfo{author}{\bibfnamefont{D.~E.} \bibnamefont{Goldberg}},
  \emph{\bibinfo{title}{Genetic Algorithms in Search, Optimization and Machine
  Learning}} (\bibinfo{publisher}{Addison-Wesley Longman Publishing Co., Inc.},
  \bibinfo{address}{Boston, MA, USA}, \bibinfo{year}{1989}),
  \bibinfo{edition}{1st} ed., ISBN \bibinfo{isbn}{0201157675}.

\bibitem[{\citenamefont{Alexandrova and Boldyrev}(2005)}]{alexa}
\bibinfo{author}{\bibfnamefont{A.~N.} \bibnamefont{Alexandrova}}
  \bibnamefont{and} \bibinfo{author}{\bibfnamefont{A.~I.}
  \bibnamefont{Boldyrev}}, \bibinfo{journal}{J. Chem. Theory Comput.}
  \textbf{\bibinfo{volume}{1}}, \bibinfo{pages}{566} (\bibinfo{year}{2005}),
  \eprint{http://pubs.acs.org/doi/pdf/10.1021/ct050093g},
  \urlprefix\url{http://pubs.acs.org/doi/abs/10.1021/ct050093g}.

\bibitem[{\citenamefont{Alexandrova et~al.}(2004)\citenamefont{Alexandrova,
  Boldyrev, Fu, Yang, Wang, and Wang}}]{alexandrova}
\bibinfo{author}{\bibfnamefont{A.~N.} \bibnamefont{Alexandrova}},
  \bibinfo{author}{\bibfnamefont{A.~I.} \bibnamefont{Boldyrev}},
  \bibinfo{author}{\bibfnamefont{Y.-J.} \bibnamefont{Fu}},
  \bibinfo{author}{\bibfnamefont{X.}~\bibnamefont{Yang}},
  \bibinfo{author}{\bibfnamefont{X.-B.} \bibnamefont{Wang}}, \bibnamefont{and}
  \bibinfo{author}{\bibfnamefont{L.-S.} \bibnamefont{Wang}},
  \bibinfo{journal}{J. Chem. Phys.} \textbf{\bibinfo{volume}{121}},
  \bibinfo{pages}{5709} (\bibinfo{year}{2004}),
  \urlprefix\url{http://scitation.aip.org/content/aip/journal/jcp/121/12/10.1063/1.1783276}.

\bibitem[{\citenamefont{Alexandrova}(2010)}]{alexan}
\bibinfo{author}{\bibfnamefont{A.~N.} \bibnamefont{Alexandrova}},
  \bibinfo{journal}{J. Phys. Chem. A} \textbf{\bibinfo{volume}{114}},
  \bibinfo{pages}{12591} (\bibinfo{year}{2010}),
  \urlprefix\url{http://pubs.acs.org/doi/abs/10.1021/jp1092543}.

\bibitem[{\citenamefont{Harding et~al.}(2006)\citenamefont{Harding, Mackenzie,
  and Walsh}}]{harding}
\bibinfo{author}{\bibfnamefont{D.}~\bibnamefont{Harding}},
  \bibinfo{author}{\bibfnamefont{S.~R.} \bibnamefont{Mackenzie}},
  \bibnamefont{and} \bibinfo{author}{\bibfnamefont{T.~R.} \bibnamefont{Walsh}},
  \bibinfo{journal}{J. Phys. Chem. B} \textbf{\bibinfo{volume}{110}},
  \bibinfo{pages}{18272} (\bibinfo{year}{2006}),
  \eprint{http://pubs.acs.org/doi/pdf/10.1021/jp062603o},
  \urlprefix\url{http://pubs.acs.org/doi/abs/10.1021/jp062603o}.

\bibitem[{\citenamefont{Wales and Doye}(1997)}]{wales}
\bibinfo{author}{\bibfnamefont{D.~J.} \bibnamefont{Wales}} \bibnamefont{and}
  \bibinfo{author}{\bibfnamefont{J.~P.~K.} \bibnamefont{Doye}},
  \bibinfo{journal}{J. Phys. Chem. A} \textbf{\bibinfo{volume}{101}},
  \bibinfo{pages}{5111} (\bibinfo{year}{1997}),
  \eprint{http://pubs.acs.org/doi/pdf/10.1021/jp970984n},
  \urlprefix\url{http://pubs.acs.org/doi/abs/10.1021/jp970984n}.

\bibitem[{\citenamefont{Mondal et~al.}(2016)\citenamefont{Mondal, Cabellos,
  Pan, Osorio, Torres-Vega, Tiznado, Restrepo, and Merino}}]{Mondal}
\bibinfo{author}{\bibfnamefont{S.}~\bibnamefont{Mondal}},
  \bibinfo{author}{\bibfnamefont{J.~L.} \bibnamefont{Cabellos}},
  \bibinfo{author}{\bibfnamefont{S.}~\bibnamefont{Pan}},
  \bibinfo{author}{\bibfnamefont{E.}~\bibnamefont{Osorio}},
  \bibinfo{author}{\bibfnamefont{J.~J.} \bibnamefont{Torres-Vega}},
  \bibinfo{author}{\bibfnamefont{W.}~\bibnamefont{Tiznado}},
  \bibinfo{author}{\bibfnamefont{A.}~\bibnamefont{Restrepo}}, \bibnamefont{and}
  \bibinfo{author}{\bibfnamefont{G.}~\bibnamefont{Merino}},
  \bibinfo{journal}{Phys. Chem. Chem. Phys.} \textbf{\bibinfo{volume}{18}},
  \bibinfo{pages}{11909} (\bibinfo{year}{2016}),
  \urlprefix\url{http://dx.doi.org/10.1039/C6CP00671J}.

\bibitem[{\citenamefont{Ravell et~al.}(2018)\citenamefont{Ravell, Jalife,
  Barroso, Orozco-Ic, Hernandez-Juarez, Ortiz-Chi, Pan, Cabellos, and
  Merino}}]{Ravell}
\bibinfo{author}{\bibfnamefont{E.}~\bibnamefont{Ravell}},
  \bibinfo{author}{\bibfnamefont{S.}~\bibnamefont{Jalife}},
  \bibinfo{author}{\bibfnamefont{J.}~\bibnamefont{Barroso}},
  \bibinfo{author}{\bibfnamefont{M.}~\bibnamefont{Orozco-Ic}},
  \bibinfo{author}{\bibfnamefont{G.}~\bibnamefont{Hernandez-Juarez}},
  \bibinfo{author}{\bibfnamefont{F.}~\bibnamefont{Ortiz-Chi}},
  \bibinfo{author}{\bibfnamefont{S.}~\bibnamefont{Pan}},
  \bibinfo{author}{\bibfnamefont{J.~L.} \bibnamefont{Cabellos}},
  \bibnamefont{and} \bibinfo{author}{\bibfnamefont{G.}~\bibnamefont{Merino}},
  \bibinfo{journal}{Chemistry - An Asian Journal}
  \textbf{\bibinfo{volume}{13}}, \bibinfo{pages}{1467} (\bibinfo{year}{2018}),
  \eprint{https://onlinelibrary.wiley.com/doi/pdf/10.1002/asia.201800261},
  \urlprefix\url{https://onlinelibrary.wiley.com/doi/abs/10.1002/asia.201800261}.

\bibitem[{\citenamefont{Pan et~al.}(2014)\citenamefont{Pan, Moreno, Cabellos,
  Romero, Reyes, Merino, and Chattaraj}}]{Sudip}
\bibinfo{author}{\bibfnamefont{S.}~\bibnamefont{Pan}},
  \bibinfo{author}{\bibfnamefont{D.}~\bibnamefont{Moreno}},
  \bibinfo{author}{\bibfnamefont{J.~L.} \bibnamefont{Cabellos}},
  \bibinfo{author}{\bibfnamefont{J.}~\bibnamefont{Romero}},
  \bibinfo{author}{\bibfnamefont{A.}~\bibnamefont{Reyes}},
  \bibinfo{author}{\bibfnamefont{G.}~\bibnamefont{Merino}}, \bibnamefont{and}
  \bibinfo{author}{\bibfnamefont{P.~K.} \bibnamefont{Chattaraj}},
  \bibinfo{journal}{The Journal of Physical Chemistry A}
  \textbf{\bibinfo{volume}{118}}, \bibinfo{pages}{487} (\bibinfo{year}{2014}),
  \bibinfo{note}{pMID: 24199587}, \eprint{https://doi.org/10.1021/jp409941v},
  \urlprefix\url{https://doi.org/10.1021/jp409941v}.

\bibitem[{\citenamefont{Cui et~al.}(2015)\citenamefont{Cui, Ding, Cabellos,
  Osorio, Islas, Restrepo, and Merino}}]{Cui}
\bibinfo{author}{\bibfnamefont{Z.-h.} \bibnamefont{Cui}},
  \bibinfo{author}{\bibfnamefont{Y.-h.} \bibnamefont{Ding}},
  \bibinfo{author}{\bibfnamefont{J.~L.} \bibnamefont{Cabellos}},
  \bibinfo{author}{\bibfnamefont{E.}~\bibnamefont{Osorio}},
  \bibinfo{author}{\bibfnamefont{R.}~\bibnamefont{Islas}},
  \bibinfo{author}{\bibfnamefont{A.}~\bibnamefont{Restrepo}}, \bibnamefont{and}
  \bibinfo{author}{\bibfnamefont{G.}~\bibnamefont{Merino}},
  \bibinfo{journal}{Phys. Chem. Chem. Phys.} \textbf{\bibinfo{volume}{17}},
  \bibinfo{pages}{8769} (\bibinfo{year}{2015}),
  \urlprefix\url{http://dx.doi.org/10.1039/C4CP05707D}.

\bibitem[{\citenamefont{Vargas-Caamal
  et~al.}(2016{\natexlab{b}})\citenamefont{Vargas-Caamal, Pan, Ortiz-Chi,
  Cabellos, Boto, Contreras-Garcia, Restrepo, Chattaraj, and
  Merino}}]{Vargas-Caamal2}
\bibinfo{author}{\bibfnamefont{A.}~\bibnamefont{Vargas-Caamal}},
  \bibinfo{author}{\bibfnamefont{S.}~\bibnamefont{Pan}},
  \bibinfo{author}{\bibfnamefont{F.}~\bibnamefont{Ortiz-Chi}},
  \bibinfo{author}{\bibfnamefont{J.~L.} \bibnamefont{Cabellos}},
  \bibinfo{author}{\bibfnamefont{R.~A.} \bibnamefont{Boto}},
  \bibinfo{author}{\bibfnamefont{J.}~\bibnamefont{Contreras-Garcia}},
  \bibinfo{author}{\bibfnamefont{A.}~\bibnamefont{Restrepo}},
  \bibinfo{author}{\bibfnamefont{P.~K.} \bibnamefont{Chattaraj}},
  \bibnamefont{and} \bibinfo{author}{\bibfnamefont{G.}~\bibnamefont{Merino}},
  \bibinfo{journal}{Phys. Chem. Chem. Phys.} \textbf{\bibinfo{volume}{18}},
  \bibinfo{pages}{550} (\bibinfo{year}{2016}{\natexlab{b}}),
  \urlprefix\url{http://dx.doi.org/10.1039/C5CP05956A}.

\bibitem[{\citenamefont{Cui et~al.}(2017)\citenamefont{Cui, Vassilev-Galindo,
  Luis~Cabellos, Osorio, Orozco, Pan, Ding, and Merino}}]{Cui2}
\bibinfo{author}{\bibfnamefont{Z.-h.} \bibnamefont{Cui}},
  \bibinfo{author}{\bibfnamefont{V.}~\bibnamefont{Vassilev-Galindo}},
  \bibinfo{author}{\bibfnamefont{J.}~\bibnamefont{Luis~Cabellos}},
  \bibinfo{author}{\bibfnamefont{E.}~\bibnamefont{Osorio}},
  \bibinfo{author}{\bibfnamefont{M.}~\bibnamefont{Orozco}},
  \bibinfo{author}{\bibfnamefont{S.}~\bibnamefont{Pan}},
  \bibinfo{author}{\bibfnamefont{Y.-h.} \bibnamefont{Ding}}, \bibnamefont{and}
  \bibinfo{author}{\bibfnamefont{G.}~\bibnamefont{Merino}},
  \bibinfo{journal}{Chem. Commun.} \textbf{\bibinfo{volume}{53}},
  \bibinfo{pages}{138} (\bibinfo{year}{2017}),
  \urlprefix\url{http://dx.doi.org/10.1039/C6CC08273D}.

\bibitem[{\citenamefont{Vargas-Caamal et~al.}(2015)\citenamefont{Vargas-Caamal,
  Ortiz-Chi, Moreno, Restrepo, Merino, and Cabellos}}]{Vargas-Caamal2015}
\bibinfo{author}{\bibfnamefont{A.}~\bibnamefont{Vargas-Caamal}},
  \bibinfo{author}{\bibfnamefont{F.}~\bibnamefont{Ortiz-Chi}},
  \bibinfo{author}{\bibfnamefont{D.}~\bibnamefont{Moreno}},
  \bibinfo{author}{\bibfnamefont{A.}~\bibnamefont{Restrepo}},
  \bibinfo{author}{\bibfnamefont{G.}~\bibnamefont{Merino}}, \bibnamefont{and}
  \bibinfo{author}{\bibfnamefont{J.~L.} \bibnamefont{Cabellos}},
  \bibinfo{journal}{Theoretical Chemistry Accounts}
  \textbf{\bibinfo{volume}{134}}, \bibinfo{pages}{16} (\bibinfo{year}{2015}),
  \eprint{https://doi.org/10.1007/s00214-015-1615-9},
  \urlprefix\url{https://doi.org/10.1007/s00214-015-1615-9}.

\bibitem[{\citenamefont{Fl\'orez et~al.}(2016)\citenamefont{Fl\'orez, Acelas,
  Ibargüen, Mondal, Cabellos, Merino, and Restrepo}}]{Florez}
\bibinfo{author}{\bibfnamefont{E.}~\bibnamefont{Fl\'orez}},
  \bibinfo{author}{\bibfnamefont{N.}~\bibnamefont{Acelas}},
  \bibinfo{author}{\bibfnamefont{C.}~\bibnamefont{Ibargüen}},
  \bibinfo{author}{\bibfnamefont{S.}~\bibnamefont{Mondal}},
  \bibinfo{author}{\bibfnamefont{J.~L.} \bibnamefont{Cabellos}},
  \bibinfo{author}{\bibfnamefont{G.}~\bibnamefont{Merino}}, \bibnamefont{and}
  \bibinfo{author}{\bibfnamefont{A.}~\bibnamefont{Restrepo}},
  \bibinfo{journal}{RSC Adv.} \textbf{\bibinfo{volume}{6}},
  \bibinfo{pages}{71913} (\bibinfo{year}{2016}),
  \urlprefix\url{http://dx.doi.org/10.1039/C6RA15059D}.

\bibitem[{\citenamefont{Frisch et~al.}(2009)\citenamefont{Frisch, Trucks,
  Schlegel, Scuseria, Robb, Cheeseman, Scalmani, Barone, Mennucci, Petersson
  et~al.}}]{gauss}
\bibinfo{author}{\bibfnamefont{M.~J.} \bibnamefont{Frisch}},
  \bibinfo{author}{\bibfnamefont{G.~W.} \bibnamefont{Trucks}},
  \bibinfo{author}{\bibfnamefont{H.~B.} \bibnamefont{Schlegel}},
  \bibinfo{author}{\bibfnamefont{G.~E.} \bibnamefont{Scuseria}},
  \bibinfo{author}{\bibfnamefont{M.~A.} \bibnamefont{Robb}},
  \bibinfo{author}{\bibfnamefont{J.~R.} \bibnamefont{Cheeseman}},
  \bibinfo{author}{\bibfnamefont{G.}~\bibnamefont{Scalmani}},
  \bibinfo{author}{\bibfnamefont{V.}~\bibnamefont{Barone}},
  \bibinfo{author}{\bibfnamefont{B.}~\bibnamefont{Mennucci}},
  \bibinfo{author}{\bibfnamefont{G.~A.} \bibnamefont{Petersson}},
  \bibnamefont{et~al.}, \emph{\bibinfo{title}{{Gaussian 09, Revision B.01}}}
  (\bibinfo{year}{2009}).

\bibitem[{\citenamefont{McQuarrie and A}(1975)}]{mcquarrie1975statistical}
\bibinfo{author}{\bibfnamefont{D.}~\bibnamefont{McQuarrie}} \bibnamefont{and}
  \bibinfo{author}{\bibfnamefont{M.}~\bibnamefont{A}},
  \emph{\bibinfo{title}{Statistical Mechanics}}, Chemistry Series
  (\bibinfo{publisher}{Harper \& Row}, \bibinfo{year}{1975}),
  \urlprefix\url{https://books.google.com.bo/books?id=PANRAAAAMAAJ}.

\bibitem[{\citenamefont{Hill}(1986)}]{hill1986introduction}
\bibinfo{author}{\bibfnamefont{T.}~\bibnamefont{Hill}},
  \emph{\bibinfo{title}{An Introduction to Statistical Thermodynamics}},
  Addison-Wesley series in chemistry (\bibinfo{publisher}{Dover Publications},
  \bibinfo{year}{1986}), ISBN \bibinfo{isbn}{9780486652429},
  \urlprefix\url{https://books.google.com.vc/books?id=0fNItAEACAAJ}.

\bibitem[{\citenamefont{Teague}(2003)}]{Teague}
\bibinfo{author}{\bibfnamefont{S.~J.} \bibnamefont{Teague}},
  \bibinfo{journal}{Nature Reviews Drug Discovery}
  \textbf{\bibinfo{volume}{2}}, \bibinfo{pages}{527} (\bibinfo{year}{2003}),
  \urlprefix\url{https://doi.org/10.1038/nrd1129}.

\bibitem[{\citenamefont{An et~al.}(2006)\citenamefont{An, Bulusu, Gao, and
  Zeng}}]{Bulusu}
\bibinfo{author}{\bibfnamefont{W.}~\bibnamefont{An}},
  \bibinfo{author}{\bibfnamefont{S.}~\bibnamefont{Bulusu}},
  \bibinfo{author}{\bibfnamefont{Y.}~\bibnamefont{Gao}}, \bibnamefont{and}
  \bibinfo{author}{\bibfnamefont{X.~C.} \bibnamefont{Zeng}},
  \bibinfo{journal}{The Journal of Chemical Physics}
  \textbf{\bibinfo{volume}{124}}, \bibinfo{pages}{154310}
  (\bibinfo{year}{2006}), \eprint{https://doi.org/10.1063/1.2187003},
  \urlprefix\url{https://doi.org/10.1063/1.2187003}.

\bibitem[{\citenamefont{Shortle}(2003)}]{Shortle}
\bibinfo{author}{\bibfnamefont{D.}~\bibnamefont{Shortle}},
  \bibinfo{journal}{Computational and Theoretical Chemistry}
  \textbf{\bibinfo{volume}{12}}, \bibinfo{pages}{1298} (\bibinfo{year}{2003}),
  ISSN \bibinfo{issn}{1469-896X},
  \urlprefix\url{https://pubmed.ncbi.nlm.nih.gov/12761401}.

\bibitem[{\citenamefont{Kubicki and Watts}(2019)}]{Kubicki_2019}
\bibinfo{author}{\bibfnamefont{J.}~\bibnamefont{Kubicki}} \bibnamefont{and}
  \bibinfo{author}{\bibfnamefont{H.}~\bibnamefont{Watts}},
  \bibinfo{journal}{Minerals} \textbf{\bibinfo{volume}{9}},
  \bibinfo{pages}{141} (\bibinfo{year}{2019}), ISSN \bibinfo{issn}{2075-163X},
  \urlprefix\url{http://dx.doi.org/10.3390/min9030141}.

\bibitem[{\citenamefont{Adamo and Barone}(1999)}]{Adamo}
\bibinfo{author}{\bibfnamefont{C.}~\bibnamefont{Adamo}} \bibnamefont{and}
  \bibinfo{author}{\bibfnamefont{V.}~\bibnamefont{Barone}},
  \bibinfo{journal}{The Journal of Chemical Physics}
  \textbf{\bibinfo{volume}{110}}, \bibinfo{pages}{6158} (\bibinfo{year}{1999}),
  \eprint{https://doi.org/10.1063/1.478522},
  \urlprefix\url{https://doi.org/10.1063/1.478522}.

\bibitem[{\citenamefont{Weigend and Ahlrichs}(2005)}]{Weigend}
\bibinfo{author}{\bibfnamefont{F.}~\bibnamefont{Weigend}} \bibnamefont{and}
  \bibinfo{author}{\bibfnamefont{R.}~\bibnamefont{Ahlrichs}},
  \bibinfo{journal}{Phys. Chem. Chem. Phys.} \textbf{\bibinfo{volume}{7}},
  \bibinfo{pages}{3297} (\bibinfo{year}{2005}),
  \urlprefix\url{http://dx.doi.org/10.1039/B508541A}.

\bibitem[{\citenamefont{Grimme et~al.}(2010)\citenamefont{Grimme, Antony,
  Ehrlich, and Krieg}}]{Grimme}
\bibinfo{author}{\bibfnamefont{S.}~\bibnamefont{Grimme}},
  \bibinfo{author}{\bibfnamefont{J.}~\bibnamefont{Antony}},
  \bibinfo{author}{\bibfnamefont{S.}~\bibnamefont{Ehrlich}}, \bibnamefont{and}
  \bibinfo{author}{\bibfnamefont{H.}~\bibnamefont{Krieg}},
  \bibinfo{journal}{The Journal of Chemical Physics}
  \textbf{\bibinfo{volume}{132}}, \bibinfo{pages}{154104}
  (\bibinfo{year}{2010}), \eprint{https://doi.org/10.1063/1.3382344},
  \urlprefix\url{https://doi.org/10.1063/1.3382344}.

\bibitem[{\citenamefont{Pan et~al.}(2008)\citenamefont{Pan, Li, and
  Wang}}]{Li-Li}
\bibinfo{author}{\bibfnamefont{L.-L.} \bibnamefont{Pan}},
  \bibinfo{author}{\bibfnamefont{J.}~\bibnamefont{Li}}, \bibnamefont{and}
  \bibinfo{author}{\bibfnamefont{L.-S.} \bibnamefont{Wang}},
  \bibinfo{journal}{The Journal of Chemical Physics}
  \textbf{\bibinfo{volume}{129}}, \bibinfo{pages}{024302}
  (\bibinfo{year}{2008}), \eprint{https://doi.org/10.1063/1.2948405},
  \urlprefix\url{https://doi.org/10.1063/1.2948405}.

\bibitem[{\citenamefont{Shoji et~al.}(2011)\citenamefont{Shoji, Matsuo,
  Hashizume, Gutmann, Fueno, Tanaka, and Tamao}}]{Shoji}
\bibinfo{author}{\bibfnamefont{Y.}~\bibnamefont{Shoji}},
  \bibinfo{author}{\bibfnamefont{T.}~\bibnamefont{Matsuo}},
  \bibinfo{author}{\bibfnamefont{D.}~\bibnamefont{Hashizume}},
  \bibinfo{author}{\bibfnamefont{M.~J.} \bibnamefont{Gutmann}},
  \bibinfo{author}{\bibfnamefont{H.}~\bibnamefont{Fueno}},
  \bibinfo{author}{\bibfnamefont{K.}~\bibnamefont{Tanaka}}, \bibnamefont{and}
  \bibinfo{author}{\bibfnamefont{K.}~\bibnamefont{Tamao}},
  \bibinfo{journal}{Journal of the American Chemical Society}
  \textbf{\bibinfo{volume}{133}}, \bibinfo{pages}{11058}
  (\bibinfo{year}{2011}), \bibinfo{note}{pMID: 21711029},
  \eprint{https://doi.org/10.1021/ja203333j},
  \urlprefix\url{https://doi.org/10.1021/ja203333j}.

\bibitem[{\citenamefont{Zhai et~al.}(2003)\citenamefont{Zhai, Alexandrova,
  Birch, Boldyrev, and Wang}}]{Birch}
\bibinfo{author}{\bibfnamefont{H.-J.} \bibnamefont{Zhai}},
  \bibinfo{author}{\bibfnamefont{A.~N.} \bibnamefont{Alexandrova}},
  \bibinfo{author}{\bibfnamefont{K.~A.} \bibnamefont{Birch}},
  \bibinfo{author}{\bibfnamefont{A.~I.} \bibnamefont{Boldyrev}},
  \bibnamefont{and} \bibinfo{author}{\bibfnamefont{L.-S.} \bibnamefont{Wang}},
  \bibinfo{journal}{Angewandte Chemie International Edition}
  \textbf{\bibinfo{volume}{42}}, \bibinfo{pages}{6004} (\bibinfo{year}{2003}),
  \eprint{https://onlinelibrary.wiley.com/doi/pdf/10.1002/anie.200351874},
  \urlprefix\url{https://onlinelibrary.wiley.com/doi/abs/10.1002/anie.200351874}.

\bibitem[{\citenamefont{Moezzi et~al.}(1992)\citenamefont{Moezzi, Olmstead, and
  Power}}]{Moezzi}
\bibinfo{author}{\bibfnamefont{A.}~\bibnamefont{Moezzi}},
  \bibinfo{author}{\bibfnamefont{M.~M.} \bibnamefont{Olmstead}},
  \bibnamefont{and} \bibinfo{author}{\bibfnamefont{P.~P.} \bibnamefont{Power}},
  \bibinfo{journal}{Journal of the American Chemical Society}
  \textbf{\bibinfo{volume}{114}}, \bibinfo{pages}{2715} (\bibinfo{year}{1992}),
  \eprint{https://doi.org/10.1021/ja00033a054},
  \urlprefix\url{https://doi.org/10.1021/ja00033a054}.

\bibitem[{\citenamefont{Zhou et~al.}(2002)\citenamefont{Zhou, Tsumori, Li, Fan,
  Andrews, and Xu}}]{Zhou}
\bibinfo{author}{\bibfnamefont{M.}~\bibnamefont{Zhou}},
  \bibinfo{author}{\bibfnamefont{N.}~\bibnamefont{Tsumori}},
  \bibinfo{author}{\bibfnamefont{Z.}~\bibnamefont{Li}},
  \bibinfo{author}{\bibfnamefont{K.}~\bibnamefont{Fan}},
  \bibinfo{author}{\bibfnamefont{L.}~\bibnamefont{Andrews}}, \bibnamefont{and}
  \bibinfo{author}{\bibfnamefont{Q.}~\bibnamefont{Xu}},
  \bibinfo{journal}{Journal of the American Chemical Society}
  \textbf{\bibinfo{volume}{124}}, \bibinfo{pages}{12936}
  (\bibinfo{year}{2002}), \bibinfo{note}{pMID: 12405806},
  \eprint{https://doi.org/10.1021/ja026257+},
  \urlprefix\url{https://doi.org/10.1021/ja026257+}.

\bibitem[{\citenamefont{Feixas et~al.}(2011)\citenamefont{Feixas, Matito,
  Poater, and Sola}}]{Feixas}
\bibinfo{author}{\bibfnamefont{F.}~\bibnamefont{Feixas}},
  \bibinfo{author}{\bibfnamefont{E.}~\bibnamefont{Matito}},
  \bibinfo{author}{\bibfnamefont{J.}~\bibnamefont{Poater}}, \bibnamefont{and}
  \bibinfo{author}{\bibfnamefont{M.}~\bibnamefont{Sola}}, \bibinfo{journal}{The
  Journal of Physical Chemistry A} \textbf{\bibinfo{volume}{115}},
  \bibinfo{pages}{13104} (\bibinfo{year}{2011}), \bibinfo{note}{pMID:
  21932863}, \eprint{https://doi.org/10.1021/jp205152n},
  \urlprefix\url{https://doi.org/10.1021/jp205152n}.

\bibitem[{\citenamefont{Szabo et~al.}(2005)\citenamefont{Szabo, Kovacs, and
  Frenking}}]{Szabo}
\bibinfo{author}{\bibfnamefont{A.}~\bibnamefont{Szabo}},
  \bibinfo{author}{\bibfnamefont{A.}~\bibnamefont{Kovacs}}, \bibnamefont{and}
  \bibinfo{author}{\bibfnamefont{G.}~\bibnamefont{Frenking}},
  \bibinfo{journal}{Zeitschrift für anorganische und allgemeine Chemie}
  \textbf{\bibinfo{volume}{631}}, \bibinfo{pages}{1803} (\bibinfo{year}{2005}),
  \eprint{https://onlinelibrary.wiley.com/doi/pdf/10.1002/zaac.200500183},
  \urlprefix\url{https://onlinelibrary.wiley.com/doi/abs/10.1002/zaac.200500183}.

\bibitem[{\citenamefont{Geudtner et~al.}(2012)\citenamefont{Geudtner,
  Calaminici, Carmona-Esp\'indola, del Campo, Dom\'inguez-Soria, Moreno,
  Gamboa, Goursot, K\"oster, Reveles et~al.}}]{demon2k}
\bibinfo{author}{\bibfnamefont{G.}~\bibnamefont{Geudtner}},
  \bibinfo{author}{\bibfnamefont{P.}~\bibnamefont{Calaminici}},
  \bibinfo{author}{\bibfnamefont{J.}~\bibnamefont{Carmona-Esp\'indola}},
  \bibinfo{author}{\bibfnamefont{J.~M.} \bibnamefont{del Campo}},
  \bibinfo{author}{\bibfnamefont{V.~D.} \bibnamefont{Dom\'inguez-Soria}},
  \bibinfo{author}{\bibfnamefont{R.~F.} \bibnamefont{Moreno}},
  \bibinfo{author}{\bibfnamefont{G.~U.} \bibnamefont{Gamboa}},
  \bibinfo{author}{\bibfnamefont{A.}~\bibnamefont{Goursot}},
  \bibinfo{author}{\bibfnamefont{A.~M.} \bibnamefont{K\"oster}},
  \bibinfo{author}{\bibfnamefont{J.~U.} \bibnamefont{Reveles}},
  \bibnamefont{et~al.}, \bibinfo{journal}{WIREs Computational Molecular
  Science} \textbf{\bibinfo{volume}{2}}, \bibinfo{pages}{548}
  (\bibinfo{year}{2012}),
  \eprint{https://onlinelibrary.wiley.com/doi/pdf/10.1002/wcms.98},
  \urlprefix\url{https://onlinelibrary.wiley.com/doi/abs/10.1002/wcms.98}.

\bibitem[{\citenamefont{Li et~al.}(2015)\citenamefont{Li, Wang, Fu, and
  Hou}}]{Li-Min}
\bibinfo{author}{\bibfnamefont{M.}~\bibnamefont{Li}},
  \bibinfo{author}{\bibfnamefont{J.}~\bibnamefont{Wang}},
  \bibinfo{author}{\bibfnamefont{B.}~\bibnamefont{Fu}}, \bibnamefont{and}
  \bibinfo{author}{\bibfnamefont{Q.}~\bibnamefont{Hou}}, \bibinfo{journal}{AIP
  Advances} \textbf{\bibinfo{volume}{5}}, \bibinfo{pages}{127131}
  (\bibinfo{year}{2015}), \eprint{https://doi.org/10.1063/1.4939137},
  \urlprefix\url{https://doi.org/10.1063/1.4939137}.

\bibitem[{\citenamefont{Spickermann}(2011)}]{spickermann2011entropies}
\bibinfo{author}{\bibfnamefont{C.}~\bibnamefont{Spickermann}},
  \emph{\bibinfo{title}{Entropies of Condensed Phases and Complex Systems: A
  First Principles Approach}}, Springer Theses (\bibinfo{publisher}{Springer
  Berlin Heidelberg}, \bibinfo{year}{2011}), ISBN
  \bibinfo{isbn}{9783642157363},
  \urlprefix\url{https://books.google.com.mx/books?id=19Mc2WWxSQQC}.

\end{thebibliography}
\newpage 
\appendix
\section{Average Be-B Bond Length}
\label{appendix:a}
\begin{figure}[ht]
  \begin{center}
  \includegraphics[scale=1.0]{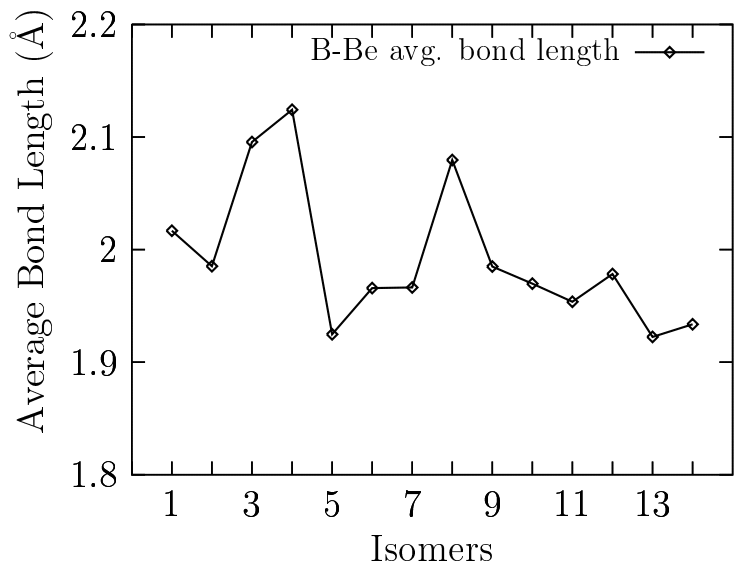} 
  \caption{(Color online.)  The average Be-B bond length a function of the number of isomers. The isomers are arranged in energy, from the lowest- (1) to the highest-energy structure (14). The coaxial triple-layered structures with C$_s$ and C$_{2v}$ symmetries (3 and 4) have the highest average bond lengths of 2.09, and 2.12~\AA,~respectively.}
  \label{cubic}
  \end{center}
\end{figure}
\newpage 
\section{Energetic Ordering According to CCSD(T) Energies}
\label{appendix:b}
\begin{figure}[ht]
  \begin{center}
  \includegraphics[scale=0.60]{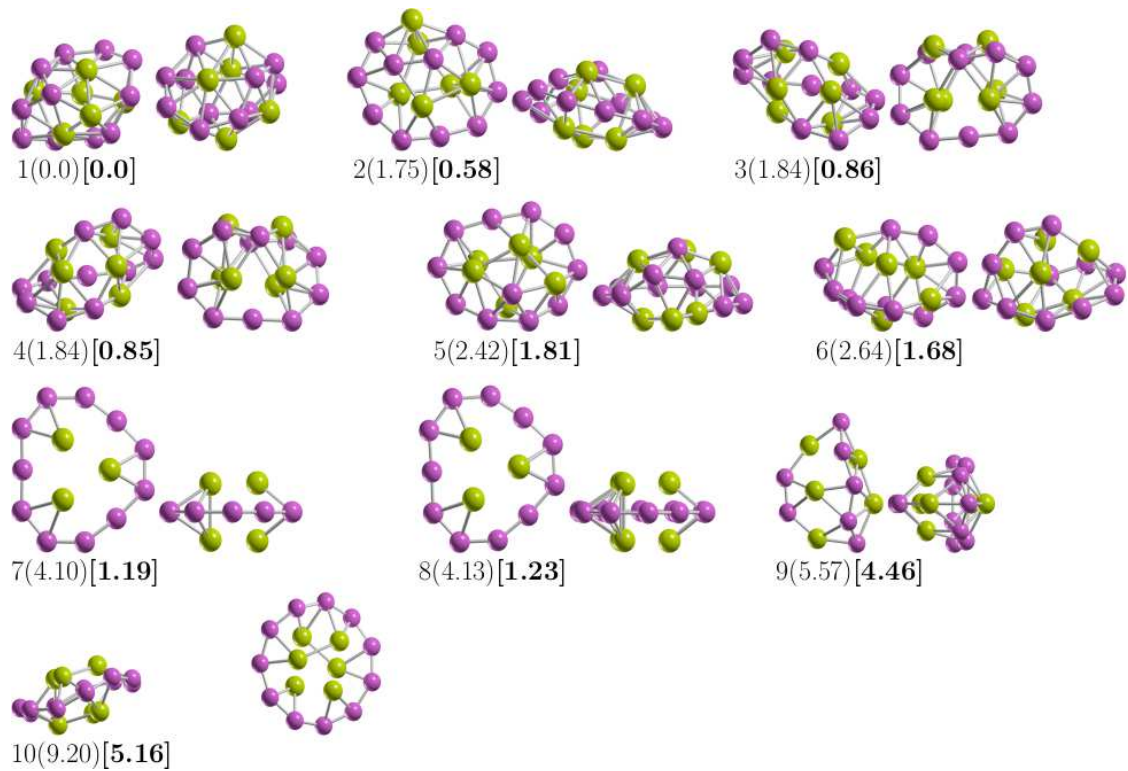}
  \caption{(Color online.)The most important energy isomers shown in two orientations, rotated 90 degrees up to plane paper and front. Single-point relative CCSDT energies in kcal/mol (in parentheses) and single point relative CCSDT energies considering the ZPE energies (in square brackets). Single-point relative CCSDT energies agree with those reported previously~\cite{Osvaldo}. The purple- and green-colored spheres represent the boron and beryllium atoms, respectively.}
  \label{ccsdt_geo}
  \end{center}
\end{figure}
The low-energy structures computed at single point CCSD(T) level of theory.
\newpage
\section{XYZ atomic coordiantes}
\label{appendix:c}
\begin{verbatim}
 17    
0.000000000         a1.out
B      -0.809816000      1.579400000      0.643022000
Be      0.596573000     -0.814845000      2.010115000
B       0.051091000      1.760054000     -0.791777000
B       1.580974000      1.309936000     -0.966076000
B      -2.070825000      0.461623000      0.890950000
B      -0.603948000      0.506270000      1.857829000
B       2.628206000      0.531768000     -0.066162000
Be      0.883805000      0.850245000      0.737351000
Be     -0.903703000     -1.025277000      0.582096000
B      -2.628548000     -0.531967000     -0.178140000
B      -1.569501000     -1.073309000     -1.210345000
B      -0.048006000     -1.595631000     -1.037654000
Be      1.608400000     -0.779034000     -1.006380000
Be      0.006693000      0.158977000     -1.774212000
Be     -1.600348000      0.919041000     -0.888479000
B       2.151414000     -0.709603000      0.737654000
B       0.845823000     -1.685827000      0.392307000
17    
1.756121500         a2.out
Be      0.871471000     -2.240509000      0.377306000
Be     -1.189569000      0.091524000     -1.070090000
Be     -1.553783000     -0.035182000      1.029073000
Be      0.353918000      0.927628000      1.191148000
B      -2.388865000     -1.286223000     -0.191009000
Be      0.902109000      0.576938000     -1.026034000
B       2.138354000      1.405375000      0.186226000
B       2.104451000     -1.290639000     -0.677320000
B      -0.695071000      2.012979000     -0.103576000
B      -2.937178000      0.167009000     -0.293782000
B      -0.966176000     -1.813381000      0.224983000
B       0.854208000      2.267959000      0.001131000
B       1.541700000     -0.519470000      0.705420000
B      -2.190898000      1.507723000     -0.161535000
B       2.776437000      0.070546000     -0.362169000
Be      0.251303000     -1.389407000     -1.254458000
B       0.054679000     -0.866672000      1.274075000
17    
1.840865375         a3.out
B      -1.398792000     -2.376827000     -0.608428000
B       1.860269000      1.944117000      0.832486000
B       0.000000000     -2.314681000     -1.284177000
B       1.398792000      2.376827000     -0.608428000
B      -1.860269000     -1.944117000      0.832486000
B       1.198910000      1.039826000      1.873576000
Be     -0.846279000     -0.686255000     -1.602453000
B       0.000000000      2.314681000     -1.284177000
Be      0.026437000     -1.647024000      0.486714000
Be      1.524481000      0.000032000      0.331634000
B       0.978975000     -1.100162000     -1.127179000
Be     -0.026437000      1.647024000      0.486714000
B      -1.198910000     -1.039826000      1.873576000
B      -0.978975000      1.100162000     -1.127179000
Be     -1.524481000     -0.000032000      0.331634000
Be      0.846279000      0.686255000     -1.602453000
B       0.000000000      0.000000000      1.882013000
17    
1.846625825         gibss_0002.out
Be      0.000000000      1.524670000      0.325548000
Be      0.000000000     -1.524670000      0.325548000
Be     -1.646433000      0.020113000      0.488496000
Be      1.646433000     -0.020113000      0.488496000
B       1.104502000     -0.972620000     -1.129209000
B      -1.104502000      0.972620000     -1.129209000
B       0.000000000      0.000000000      1.882165000
Be     -0.681645000     -0.847320000     -1.608316000
Be      0.681645000      0.847320000     -1.608316000
B       1.033017000      1.204123000      1.875487000
B      -1.033017000     -1.204123000      1.875487000
B      -2.314461000     -0.012187000     -1.282308000
B       2.314461000      0.012187000     -1.282308000
B      -1.936501000     -1.869964000      0.837138000
B       1.936501000      1.869964000      0.837138000
B       2.363598000      1.412169000     -0.606771000
B      -2.363598000     -1.412169000     -0.606771000
17    
2.420995400         a5.out
Be     -1.462704000     -0.014249000      0.967655000
B       2.728694000      0.030013000     -0.619010000
B      -2.146695000     -1.259141000     -0.336682000
B       2.196858000      1.375491000      0.031415000
B      -0.880491000     -1.823626000      0.448212000
B       1.934511000     -1.299414000     -0.736505000
B      -0.694414000      1.931806000      0.025551000
Be      0.874964000      0.501056000     -0.992441000
B      -2.883951000      0.123283000     -0.317897000
Be      1.826138000     -0.480368000      0.902917000
Be      0.408897000      0.951482000      1.344252000
B       0.772439000     -1.870483000      0.277861000
B      -2.220982000      1.503653000     -0.070210000
Be     -0.113249000     -1.338601000     -1.218981000
B       0.106145000     -0.876434000      1.446090000
B       0.872097000      2.199862000      0.022620000
Be     -1.264310000      0.336919000     -1.217707000
17    
2.642791550         a6.out
B       1.528679000      1.701700000      0.509569000
B      -0.117021000     -0.964914000     -1.615283000
B      -1.648819000     -0.707237000     -1.161047000
Be     -0.302031000     -1.449462000      0.175321000
B       2.641512000     -0.596366000     -0.644848000
Be     -1.789419000      1.565781000      0.309923000
Be      1.619805000     -0.697074000      0.999108000
B      -1.341198000     -0.925353000      1.696499000
B       1.342610000     -1.332495000     -1.175056000
Be     -0.127111000      0.623224000      1.414746000
Be     -0.940410000      0.987997000     -1.391875000
Be      1.010995000      0.511028000     -0.864484000
B      -2.628326000      0.474501000     -0.801661000
B       0.139389000     -1.127775000      1.950483000
B       2.710593000      0.706957000      0.147077000
B       0.040832000      1.912171000      0.081513000
B      -2.245715000     -0.374385000      0.498565000
17    
4.109553975         gibss_0000.out
Be     -0.597631000      1.124894000      1.052561000
Be     -0.597631000      1.124894000     -1.052561000
B      -2.238140000     -1.547111000      0.000000000
B       1.559153000      1.828212000      0.000000000
Be     -0.597631000     -1.125067000      1.051152000
B       0.289604000      2.707478000      0.000000000
B      -2.209386000      0.001996000      0.000000000
B      -1.233725000      2.689543000      0.000000000
B      -2.233185000      1.551289000      0.000000000
Be      1.279363000      0.001246000     -1.086218000
Be      1.279363000      0.001246000      1.086218000
B      -1.242591000     -2.688862000      0.000000000
Be     -0.597631000     -1.125067000     -1.051152000
B       2.675935000      0.757041000      0.000000000
B       0.280383000     -2.711048000      0.000000000
B       2.670570000     -0.762843000      0.000000000
B       1.546820000     -1.827411000      0.000000000
17    
4.139184525         a8.out
B       0.000000000      1.549129000      2.235698000
B       0.000000000      1.827820000     -1.553909000
B       0.000000000     -0.759893000     -2.673713000
B       0.000000000     -2.707671000     -0.285066000
B       0.000000000     -1.549129000      2.235698000
B       0.000000000     -1.827820000     -1.553909000
B       0.000000000      0.759893000     -2.673713000
B       0.000000000      2.707671000     -0.285066000
B       0.000000000      2.688931000      1.237960000
B       0.000000000      0.000000000      2.207900000
B       0.000000000     -2.688931000      1.237960000
Be      1.052108000      1.124714000      0.598858000
Be      1.085258000      0.000000000     -1.278866000
Be      1.052108000     -1.124714000      0.598858000
Be     -1.052108000      1.124714000      0.598858000
Be     -1.085258000      0.000000000     -1.278866000
Be     -1.052108000     -1.124714000      0.598858000
17    
5.572858875         gibss_0004.out
Be      1.784070000     -0.881154000      0.000000000
Be     -1.178349000      1.635847000      0.000000000
Be      1.064755000      1.033056000      0.000000000
Be     -0.937588000     -0.323485000      1.174575000
B      -2.310757000      0.242038000      0.000000000
B       0.892147000     -0.081631000     -1.612675000
B       0.352235000      1.385244000      1.718054000
Be     -0.937588000     -0.323485000     -1.174575000
Be     -0.817189000     -2.391480000      0.000000000
B       0.892147000     -0.081631000      1.612675000
B       0.621579000     -1.643571000      1.185587000
B      -2.180768000     -1.259499000      0.000000000
B       0.352235000      1.385244000     -1.718054000
B       0.285545000      2.649292000      0.765420000
B       1.006026000     -2.600648000      0.000000000
B       0.621579000     -1.643571000     -1.185587000
B       0.285545000      2.649292000     -0.765420000
17    
9.200317150         gibss_0003.out
Be     -1.140416000      0.103343000      1.097260000
Be      1.140416000     -0.103343000      1.097260000
B       1.806969000      1.963089000     -0.386217000
B      -1.542626000     -1.932366000      1.133568000
Be      0.000000000      1.577206000      0.008758000
B      -0.638597000     -1.316233000      2.228919000
B       1.542626000      1.932366000      1.133568000
B       0.000000000      0.000000000      2.724881000
B       0.638597000      1.316233000      2.228919000
Be      0.000000000     -1.577206000      0.008758000
Be     -1.148876000      0.082815000     -1.070831000
B       1.176139000      1.697602000     -1.775334000
Be      1.148876000     -0.082815000     -1.070831000
B      -1.806969000     -1.963089000     -0.386217000
B       0.326002000      0.702995000     -2.591527000
B      -1.176139000     -1.697602000     -1.775334000
B      -0.326002000     -0.702995000     -2.591527000
17    
11.507201675        gibss_0006.out
Be      2.136962000      0.466122000     -0.136821000
Be     -2.136962000     -0.466122000     -0.136821000
Be      0.448793000      1.453418000      0.501781000
Be     -0.448793000     -1.453418000      0.501781000
B      -1.013219000     -1.183852000     -1.416467000
B       1.013219000      1.183852000     -1.416467000
B      -1.516619000     -0.303767000      1.526878000
Be      0.858236000     -0.661623000     -1.331966000
Be     -0.858236000      0.661623000     -1.331966000
B       1.516619000      0.303767000      1.526878000
B       0.000000000      0.000000000      1.865341000
B       0.000000000      2.333754000     -1.105419000
B       0.000000000     -2.333754000     -1.105419000
B       1.546339000     -1.193439000      0.921284000
B      -1.546339000      1.193439000      0.921284000
B      -1.178288000      2.314234000     -0.085342000
B       1.178288000     -2.314234000     -0.085342000
17    
12.741801650        gibss_0005.out
Be     -0.967864000     -0.003324000     -2.025249000
Be     -0.977869000     -0.908311000      1.005310000
Be      0.977869000      0.908311000      1.005310000
Be      0.967864000      0.003324000     -2.025249000
Be     -1.631200000      0.684878000     -0.239010000
B       0.932498000     -0.979840000      1.526020000
B      -0.932498000      0.979840000      1.526020000
Be      1.631200000     -0.684878000     -0.239010000
B       1.725675000      1.166405000     -0.793865000
B       0.000000000     -1.306175000     -1.105711000
B       0.000000000      1.306175000     -1.105711000
B       0.440133000     -2.107280000      0.385744000
B      -1.725675000     -1.166405000     -0.793865000
B      -0.440133000      2.107280000      0.385744000
B       0.995287000      2.407450000     -0.164110000
B      -0.995287000     -2.407450000     -0.164110000
B       0.000000000      0.000000000      2.318162000
17    
25.066610725        gibss_0007.out
Be     -1.277429000      1.666744000      0.000000000
B      -1.541731000     -2.135511000      0.000000000
Be     -0.626280000     -1.277275000      1.405826000
B      -0.006859000     -2.441749000      0.000000000
Be     -0.626280000     -1.277275000     -1.405826000
B       1.212038000     -1.565687000      0.808025000
B       0.037720000      1.982302000      1.348142000
B      -0.811599000      0.576246000      1.455720000
B       1.777627000     -0.269016000      0.000000000
Be      0.926136000      0.082001000     -1.658674000
Be      0.926136000      0.082001000      1.658674000
B       0.037720000      1.982302000     -1.348142000
B       1.212038000     -1.565687000     -0.808025000
Be     -1.878824000     -0.366759000      0.000000000
B      -0.811599000      0.576246000     -1.455720000
B       0.324901000      2.740038000      0.000000000
B       0.614978000      0.992966000      0.000000000
17    
36.617310700        gibss_0008.out
Be     -0.000618000      1.265154000     -1.244622000
Be      0.000618000     -1.265154000     -1.244622000
B      -0.768950000     -0.000057000     -2.360234000
Be      1.577107000      0.000458000      1.121963000
B       0.001596000      1.217530000      1.359563000
B      -1.321661000      1.600756000      0.207885000
B       1.321661000     -1.600756000      0.207885000
Be     -1.852297000     -0.000057000     -0.867758000
B       1.322470000      1.601879000      0.207441000
B      -1.322470000     -1.601879000      0.207441000
B       0.768950000      0.000057000     -2.360234000
B       0.000000000      0.000000000      2.355612000
Be     -1.577107000     -0.000458000      1.121963000
Be      1.852297000      0.000057000     -0.867758000
B       0.000000000      2.439167000      0.199873000
B      -0.001596000     -1.217530000      1.359563000
B       0.000000000     -2.439167000      0.199873000 
\end{verbatim}
\typeout{get arXiv to do 4 passes: Label(s) may have changed. Rerun}
\end{document}